\documentclass[particles, review, accept, pdftex, moreauthors]{Definitions/mdpi}
\firstpage{57}
\pubvolume{6}
\issuenum{1}
\articlenumber{0}
\pubyear{2023}
\copyrightyear{2023}
\datereceived{2022 Nov 21}
\daterevised{2022 Dec 18} 
\dateaccepted{2022 Dec 19}
\datepublished{2023 Jan 11}
\hreflink{https://doi.org/10.3390/particles6010004} 

\usepackage{wasysym}

\Title{$\,$\\[-5ex]\hspace*{\fill}{\normalsize{\sf\emph{Preprint no}.\ NJU-INP 066/22}}\\[1.25ex]
Emergence of Hadron Mass and Structure}

\TitleCitation{Emergence of Hadron Mass and Structure}

\Author{Minghui Ding $^{1,\ddagger}$\orcidA{}, Craig D.\ Roberts $^{2,3,\ddagger}$\orcidB{} and Sebastian M.\ Schmidt $^{1,4,\ddagger}$\orcidC{}\,*}

\AuthorNames{Minghui Ding, Craig D.\ Roberts and Sebastian M.\ Schmidt}

\AuthorCitation{Ding, M.; Roberts, C.D.; Schmidt, S.M.}

\address{%
$^{1}$ \quad Helmholtz-Zentrum Dresden-Rossendorf, Bautzner Landstra{\ss}e 400, D-01328 Dresden, Germany\\
$^{2}$ \quad School of Physics, Nanjing University, Nanjing, Jiangsu 210093, China\\
$^{3}$ \quad Institute for Nonperturbative Physics, Nanjing University, Nanjing, Jiangsu 210093, China\\
$^{4}$ \quad RWTH Aachen University, III. Physikalisches Institut B, D-52074 Aachen, Germany
}

\corres{Correspondence:\\
\qquad m.ding@hzdr.de (M.\ Ding);
cdroberts@nju.edu.cn (C.\,D.\ Roberts);
s.schmidt@hzdr.de (S.\,M.\ Schmidt)}

\secondnote{These authors contributed equally to this work.}

\abstract{
Visible matter is characterised by a single mass scale; namely, the proton mass.  The proton's existence and structure are supposed to be described by quantum chromodynamics (QCD); yet, absent Higgs boson couplings, chromodynamics is scale invariant.  Thus, if the Standard Model is truly a part of the theory of Nature, then the proton mass is an emergent feature of QCD; and emergent hadron mass (EHM) must provide the basic link between theory and observation.  Nonperturbative tools are necessary if such connections are to be made; and in this context, we sketch recent progress in the application of continuum Schwinger function methods to an array of related problems in hadron and particle physics.  Special emphasis is given to the three pillars of EHM -- namely, the running gluon mass, process-independent effective charge, and running quark mass; their role in stabilising QCD; and their measurable expressions in a diverse array of observables.
}


\usepackage[mathscr,scaled=1.15]{urwchancal}
\DeclareFontFamily{OT1}{pzc}{}
\DeclareFontShape{OT1}{pzc}{m}{it}%
{<-> s * [1.15] pzcmi7t}{}
\DeclareMathAlphabet{\mathpzc}{OT1}{pzc}{m}{it}

\begin{document}





\begin{footnotesize}

\tableofcontents

\end{footnotesize}

\section{Introduction}
Our Universe exists; and even the small part that we occupy contains much which might be considered miraculous.  Nevertheless, science typically assumes that the Universe's evolution can be explained by some collection of equations -- even a single equation, perhaps, which replaces distinct theories of many things with a single theory of everything.  Choosing not to approach that frontier, then, within the current paradigm, the Standard Model of particle physics (SM) is given a central role; and it must account for a huge array of observable phenomena.  Herein, we focus on one especially important aspect, \emph{viz}.\ the fact that the mass of the vast bulk of visible material in the Universe is explained as soon as one understands why the proton is absolutely stable and how it comes to possess a mass $m_p \approx 1\,$GeV.  In elucidating this connection, we will argue that the theory of strong interactions may deliver far more than was originally asked of it.

We have evidently supposed that quantum gauge field theory is the correct paradigm for understanding Nature.  In this connection, it is important to note that, in our tangible Universe, time and space give us four noncompact dimensions.
Consider, therefore, that quantum gauge field theories in $D\neq 4$ dimensions are characterised by an explicit, intrinsic mass-scale: the basic couplings generated by minimal substitution are mass-dimensioned and set the scale for all calculated quantities.  For $D>4$, such theories manifest uncontrollable ultraviolet divergences, making them of little physical use.  In contrast, for $D<4$, they are super-convergent, but are afflicted with a hierarchy problem, \emph{viz}.\ dynamical mass-generation effects are typically very small when compared with the theory's explicit scale \cite{Appelquist:1981vg, Appelquist:1986fd, Bashir:2008fk, Bashir:2009fv, Braun:2014wja}.
Hence, perhaps unsurprisingly, $D=4$ is a critical point.
Removing Higgs boson couplings, the classical gauge theory elements of the SM are scale-invariant.  Taking the step to quantum theories, they are all (at least perturbatively) renormalisable; and that procedure introduces a mass scale.  As we have noted, the scale for visible matter is $m_{\rm Nature} \approx m_p \approx 1\,$GeV.  However, the size of this scale is not determined by the theory; so, whence does it come?  Further, how much tolerance does Nature give us?  Is the Universe habitable when $m_{\rm Nature} \to (1 \pm \delta) m_{\rm Nature}$, with $\delta = 0.1$ or $0.2$, etc.?  It is comforting to imagine that our (ultimate?) theory of Nature will answer these questions, but the existence of such a theory is not certain.

Returning to concrete issues, strong interactions within the SM are described by quantum chromodynamics (QCD).  Therefore, consider the classical Lagrangian density that serves as the starting point on the road to QCD:
\begin{subequations}
\label{QCDdefine}
\begin{align}
{\mathpzc L}_{\rm QCD} & = \sum_{{\mathpzc f}=u,d,s,\ldots}
\bar{q}_{\mathpzc f} [\gamma\cdot\partial
    + i g \tfrac{1}{2} \lambda^a\gamma\cdot A^a+ m_{\mathpzc f}] q_{\mathpzc f}
    + \tfrac{1}{4} G^a_{\mu\nu} G^a_{\mu\nu},\\
%
%
\label{gluonSI}
G^a_{\mu\nu} & = \partial_\mu A^a_\nu + \partial_\nu A^a_\mu -
g f^{abc}A^b_\mu A^c_\nu,
\end{align}
\end{subequations}
where $\{q_{\mathpzc f}\,|\,{\mathpzc f}=u,d,s,c,b,t\}$ are fields associated with the six known flavours of quarks; $\{m_{\mathpzc f}\}$ are their current-masses, generated by the Higgs boson;
$\{A_\mu^a\,|\,a=1,\ldots,8\}$ represent the gluon fields, whose matrix structure is encoded in $\{\tfrac{1}{2}\lambda^a\}$, the generators of SU$(3)$ in the fundamental representation; and $g$ is the \emph{unique} QCD coupling, using which one conventionally defines $\alpha = g^2/[4\pi]$.  As remarked above, if one removes Higgs boson couplings into QCD, so that $\{m_{\mathpzc f}\equiv 0 \}$ in Eq.\,\eqref{QCDdefine}, then the classical action associated with this Lagrangian is scale invariant.  A scale invariant theory cannot produce compact bound states; indeed, scale invariant theories do not support dynamics, only kinematics \cite{Roberts:2016vyn}.  So if Eq.\,\eqref{QCDdefine} is really capable of explaining, amongst other things, the proton's mass, size, and stability, then remarkable features must emerge via the process of defining \underline{quantum} chromodynamics.

This point is placed in stark relief when one appreciates that the gluon and quark fields used to express the one-line Lagrangian of QCD are \underline{not} the degrees-of-freedom measured in detectors.  This is an empirical manifestation of confinement.  Amongst other things, a solution of QCD will reveal the meaning of confinement, predict the observable states, and explain how they are built from the Lagrangian's gluon and quark partons.  But the search for a solution presumes that QCD is actually a theory.  Effective theories are tools for use in obtaining a realistic description of phenomena perceived at a given scale.  A true theory must be rigorously defined at all scales and unify phenomena perceived at vastly different energies.  If QCD really is a well-defined quantum field theory, then it may serve as a paradigm for physics far beyond the SM.

Having raised this possibility, then it is appropriate to provide a working definition of ``well-defined'' in relation to quantum field theory.  Aspects of the mathematical problem are discussed elsewhere \cite{GJ81, SE82}.  Herein, we consider that a quantum (gauge) field theory is well-defined if its ultraviolet renormalisation can be accomplished with a finite number of renormalisation constants, $\{Z_j | j=1,\ldots, N\}$, $N \lesssim 10$,\footnote{Here, the value ``10'' is arbitrary.  More generally, the number should be small enough to ensure that predictive power is not lost through a need to fit too many renormalised observables to measured quantities.}
all of which can (\emph{a}) be computed nonperturbatively and (\emph{b}) remain bounded real numbers as any regularisation scale is removed.  Further, that the renormalisation of ultraviolet divergences is sufficient to ensure that any/all infrared divergences are eliminated, \emph{i.e}., the theory is infrared complete.

Quantum electrodynamics (QED) is not well-defined owing to the existence of a Landau pole in the far ultraviolet  (see, \emph{e.g}.\ Ref.\,\cite[Ch.\,13]{IZ80} and Refs.\,\cite{Rakow:1990jv, Gockeler:1994ci, Reenders:1999bg, Kizilersu:2014ela}).  Furthermore, weak interactions are essentially perturbative because the inclusion of the Higgs scalar-boson introduces an enormous infrared scale that suppresses all nonperturbative effects; moreover, the Higgs boson mass is quadratically divergent, making the theory non-renormalisable.

On the other hand, as we will explain herein, it is beginning to seem increasingly likely that QCD satisfies the tests listed above; hence, is the first well-defined quantum field theory that humanity has developed.  QCD may thus stand alone as an internally consistent theory, so that after quantisation of Eq.\,\eqref{QCDdefine}, with nothing further added, it is a genuinely predictive mathematical framework for the explanation of natural phenomena.

We have used a Euclidean metric and consistent Dirac matrices in writing Eq.\,\eqref{QCDdefine} because if there is any hope of arriving at a rigorous definition of QCD, then it is by formulating the theory in Euclidean space.  There are many reasons for adopting this perspective.  Amongst the most significant being the fact that a lattice-regularisation of the theory is only possible in Euclidean space, where one can use the action associated with Eq.\,\eqref{QCDdefine} to define a probability measure \cite[Sec.\,2.1]{Roberts:2000aa}.  Notably, a choice must be made because any ``Wick rotation'' between Minkowski space and Euclidean space is a purely formal exercise, whose validity is only guaranteed for perturbative calculations \cite{Krein:1990sf, Roberts:1994dr}.  If QCD really does (somehow) explain the emergence of hadron mass and structure, then nonlinear, nonperturbative dynamics must be crucial.  Consequently, one cannot assume that any of the requirements necessary to mathematically justify a Wick rotation are satisfied when calculating and summing the necessarily infinite collection of processes associated with a given experimental observable.

One concrete example may serve to illustrate the point.  Both continuum and lattice analyses of the gluon two-point Schwinger function (often called the Euclidean-space gluon propagator) yield a result whose analytic properties are very different from those one would obtain in perturbation theory at any finite-order \cite{Binosi:2019ecz}.  As a consequence, the Minkowski space gluon gap equation that is obtained from the Euclidean form via the standard transcriptions used to implement the Wick rotation \cite[Sec.\,2.3]{Roberts:1994dr}, whilst being similar in appearance, cannot possess the same solutions.  Thus, to avoid confusion, one should begin with all such equations formulated in Euclidean space, where the solutions determined have a direct and unambiguous connection with results obtained using numerical simulations of the lattice regularised theory.  Anything else is an unnecessary and potentially misleading pretence.  Furthermore, only those Schwinger functions corresponding to observable quantities need have a continuation to Minkowski space and that can be accomplished following standard notions from constructive field theory \cite[Secs.\,3, 4]{Krein:1990sf},  \cite[Sec.\,2.3]{Roberts:1994dr}.

We proceed then by supposing that QCD is defined by the Euclidean space generating functional built using the Lagrangian density in Eq.\,\eqref{QCDdefine}.  Here, a new choice presents itself.  One might attempt to solve the thus quantised theory using a lattice regularisation \cite{Wilson:1974sk, Wilson:2004de}.  Lattice-regularised QCD (lQCD) is a popular framework, which, owing to growth in computer power and algorithm improvements, is becoming more effective -- see, \emph{e.g}., Ref.\,\cite{Lattice2021}.
On the other hand, continuum Schwinger function methods (CSMs) are also available \cite{Roberts:1994dr, Roberts:2000aa, Maris:2003vk, Roberts:2007ji, Chang:2011vu, Bashir:2012fs, Roberts:2012sv}.  Much has been achieved using this approach, especially during the past decade \cite{Roberts:2015lja, Horn:2016rip, Eichmann:2016yit, Burkert:2017djo, Fischer:2018sdj, Qin:2020rad} and particularly in connection with elucidating the origins and wide-ranging expressions of emergent hadron mass (EHM) \cite{Roberts:2020udq, Roberts:2020hiw, Roberts:2021xnz, Roberts:2021nhw, Binosi:2022djx, Papavassiliou:2022wrb}.  It is upon those advances that we focus herein.

\begin{table}[t]
\caption{\label{massbudget}
Mass budgets of a collection of hadrons, with each panel ordered according to the contribution from Higgs boson couplings into QCD (HB), and including the component which is entirely unrelated to the Higgs (EHM) and that arising from constructive interference between these two mass sources (EHM+HB).  (Separation at $\zeta = 2\,GeV$, produced using information from Refs.\,\cite{Flambaum:2005kc, RuizdeElvira:2017stg, Aoki:2019cca, Roberts:2021nhw, Workman:2022ynf}.)
 }
\begin{center}
\begin{tabular*}
{\hsize}
{
l@{\extracolsep{0ptplus1fil}}
|r@{\extracolsep{0ptplus1fil}}
r@{\extracolsep{0ptplus1fil}}
r@{\extracolsep{0ptplus1fil}}}\hline
& \multicolumn{3}{c}{mass fraction (\%)}\\
hadron\,(mass/GeV)$\ $ & HB & EHM+HB & EHM \\\hline
$p\,(0.938)\ $ & $1\ $ & $6\ $ & $93\ $\\
$\rho\,(0.775)\ $ & $1\ $ & $2\ $ & $97\ $\\
$D^\ast\,(2.010)\ $ & $63\ $ & $30\ $ & $7\ $\\
$B^\ast\,(5.325)\ $ & $78\ $ & $21\ $ & $1\ $\\\hline
$\pi\,(0.140)\ $ & $5\ $ & $95\ $ & $0\ $\\
$K\,(0.494)\ $ & $20\ $ & $80\ $ & $0\ $\\
$D\,(1.870)\ $ & $68\ $ & $32\ $ & $0\ $\\
$B\,(5.279)\ $ & $79\ $ & $21\ $ & $0\ $\\\hline
\end{tabular*}
\end{center}
\end{table}

\section{Hadron Mass Budgets}
There is one generally recognised mass generating mechanism in the SM; namely, that associated with Higgs boson couplings \cite{Englert:2014zpa, Higgs:2014aqa}.  Insofar as QCD is concerned, there are six distinct such couplings, each of which generates the current-mass of a different quark flavour.  Those current-quark masses exhibit a remarkable hierarchy of scales, ranging from an electron-like size for the $u$, $d$ quarks up to a value five-orders-of-magnitude larger for the $t$ quark \cite[page~32]{Workman:2022ynf}.  Faced with such discordance, we choose to begin our discussion of mass by considering the proton and its closest relatives, \emph{viz}.\ the $\pi$- and $\rho$-mesons.

The proton is defined as the lightest state constituted from the valence-quark combination $u+u+d$.  The $\pi^+$ is a pseudoscalar meson built from $u+\bar d$ valence quarks and the $\rho^+$ is its kindred vector meson partner: in quark models, the $\pi$ and $\rho$ are identified as $^1S_0$ and $^3S_1$ states, respectively \cite[Sec.\,63]{Workman:2022ynf}.  Table~\ref{massbudget} presents a breakdown of the masses of these states into three contributions: the simplest to count is that associated with the Higgs-generated current-masses of the valence-quarks (HB); the least well understood is that part which has no connection with the Higgs boson (EHM); and the remainder is that arising from constructive interference between these two sources of mass (EHM+HB).

\begin{figure}[!t]
\hspace*{-1ex}\begin{tabular}{ll}
{\sf A} & {\sf B} \\[-2ex]
\includegraphics[clip, width=0.46\textwidth]{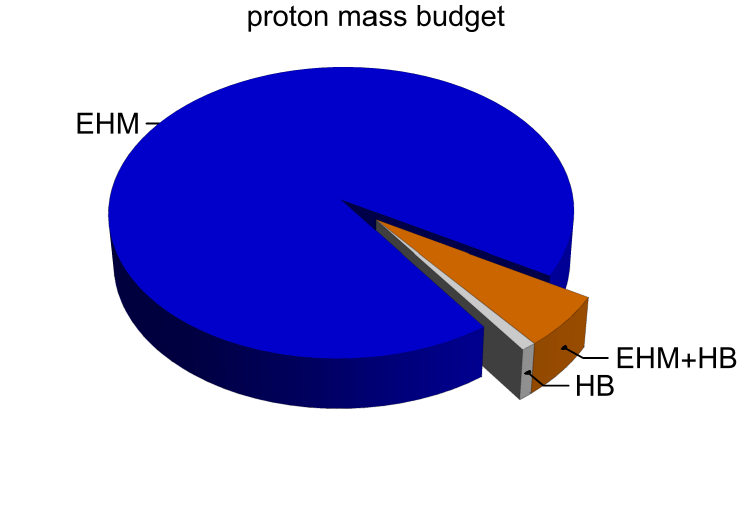} &
\includegraphics[clip, width=0.46\textwidth]{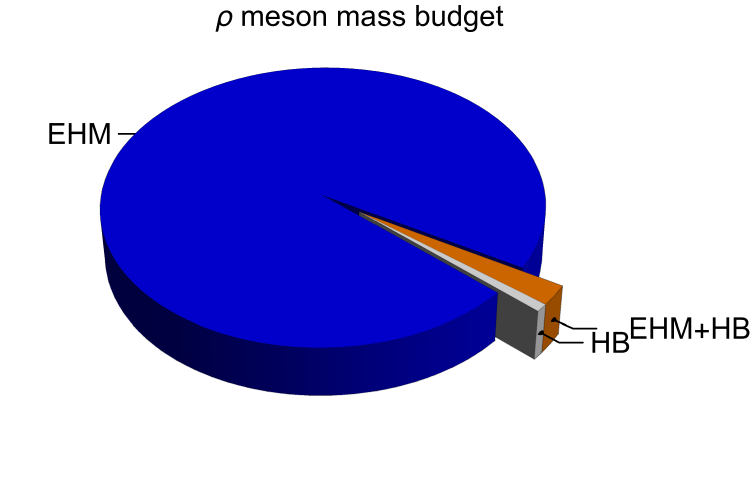}
\end{tabular}

\hspace*{-1ex}\begin{tabular}{ll}
{\sf C} & {\sf D} \\[-2ex]
\includegraphics[clip, width=0.5\textwidth]{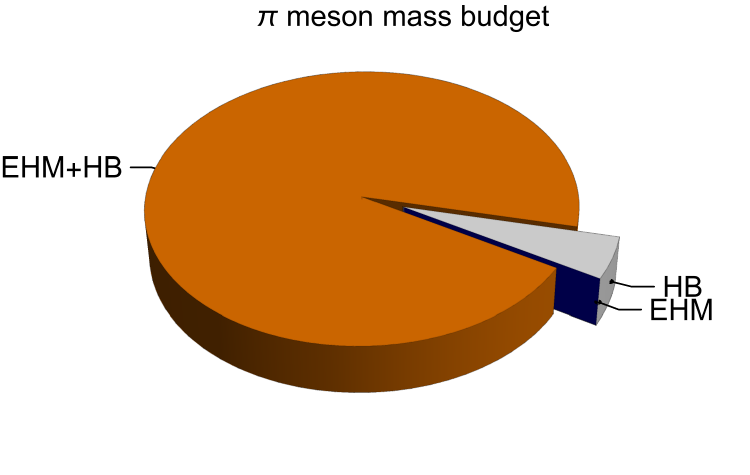}&
\hspace*{-1em}\includegraphics[clip, width=0.5\textwidth]{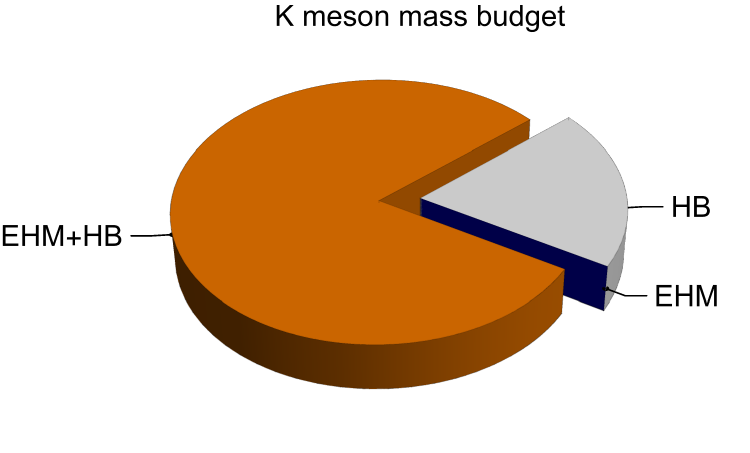}
\end{tabular}
\caption{\label{Fmassbudget}
Mass budgets:
\mbox{\sf A}\,--\,proton;
\mbox{\sf B}\,--\,$\rho$-meson;
\mbox{\sf C}\,--\,pion; and
\mbox{\sf D}\,--\,kaon.
Each is drawn using a Poincar\'e invariant decomposition and the numerical values listed in Table~\ref{massbudget}.
(Separation at $\zeta = 2\,GeV$, calculated using information from Refs.\,\cite{Flambaum:2005kc, RuizdeElvira:2017stg, Aoki:2019cca, Workman:2022ynf}.)
 }
\end{figure}

The information listed in rows 1, 2, 5, 6, of Table~\ref{massbudget} is represented pictorially in Fig.\,\ref{Fmassbudget}: plainly, there are significant differences between the upper and lower panels.
Regarding the proton and $\rho$-meson, the HB-alone component of their masses is just 1\% in each case.  Notwithstanding that, their masses are large, and remain so even in the absence of Higgs boson couplings into QCD, \emph{i.e}., in the chiral limit.  This overwhelmingly dominant component is a manifestation of EHM in the SM.  It produces roughly 95\% of the measured mass.  Evidently, baryons and vector mesons are similar in these respects.

Conversely and yet still owing to EHM via its dynamical chiral symmetry breaking (DCSB) corollary, the pion is massless in the chiral limit -- it is the SM's Nambu-Goldstone (NG) mode \cite{Nambu:1960tm, Goldstone:1961eq, GellMann:1968rz, Casher:1974xd, Brodsky:2009zd, Brodsky:2010xf, Chang:2011mu, Brodsky:2012ku, Cloet:2013jya, Horn:2016rip}.  Returning to the quark model picture, the only difference between $\rho$- and $\pi$-mesons is a spin-flip: in the $\rho$, the constituent quark spins are aligned, whereas they are antialigned in the $\pi$.  Yet, their mass budgets are fundamentally different: Fig.\,\ref{Fmassbudget}B \emph{cf}.\  Fig.\,\ref{Fmassbudget}C.  An inability to explain this difference is a conspicuous failure of quark models: whilst it is easy to obtain a satisfactory mass for the $\rho$, a low-mass pion can only be obtained by fine-tuning the quark model's potential.  Nature, however, doesn't fine-tune the pion: in the absence of Higgs boson couplings, it is massless irrespective of the size of $m_\rho$ and, in fact, the mass of any other hadron.

The kaon mass budget is also drawn -- see Fig.\,\ref{Fmassbudget}D.  In the chiral limit, then, like the $\pi$, the $K$-meson is a NG boson.  However, with realistic values of Higgs boson couplings into QCD, the $s$ quark current-mass is approximately $27$-times the average of the $u$ and $d$ current-masses \cite{Workman:2022ynf}: $2 m_s \approx 27 (m_u+m_d)$.  Consequently, the HB wedge in Fig.\,\ref{Fmassbudget}D accounts for 20\% of $m_K$.  The remaining 80\% is generated by constructive EHM+HB interference.  It follows that comparisons between $\pi$ and $K$ properties present good opportunities for studying Higgs boson modulation of EHM, because the HB mass fraction is four-times larger in kaons than in pions.  Moreover, the array of images in Fig.\,\ref{Fmassbudget} highlight that additional, complementary information can be obtained from comparisons between baryons/vector-mesons and the array of kindred pseudoscalar mesons.  For instance, studies of spectra (Sec.\,\ref{SecSpectroscopy}), transitions between vector-mesons and pseudoscalar mesons (Sec.\,\ref{SecTransition}), and comparative analyses of proton and pion parton distribution functions (DFs -- see Sec.\,\ref{SecDFs}).  In all cases, predominantly EHM systems on one hand are contrasted/overlapped with final states that possess varying degrees of EHM+HB interference.

These observations highlight that EHM -- whatever it is -- can be accessed via experiment.  The task for theory is to identify and explain its source, then elucidate a broad range of observable consequences so that the origins and explanations can be validated.


\section{Gluons and the Emergence of Mass}
\label{SecGluonMass}
The requirement of gauge invariance ensures that the Higgs boson does not couple to gluons and precludes any other means of generating an explicit mass term for the gluon fields in Eq.\,\eqref{QCDdefine}.  Consequently, it is widely believed that gluons are massless; and this is recorded by the Particle Data Group (PDG) \cite[page\,25]{Workman:2022ynf}.  (We stress that gluon partons \underline{are} massless.)

In QCD, this ``gauge invariance'' statement is properly translated into a property of the two-point gluon Schwinger function.  Namely, using the class of covariant gauges as an illustrative tool, characterised by a gauge fixing parameter $\xi$, the inverse of the gluon two point function can be expressed in terms of a gluon vacuum polarisation (or self energy):
\begin{equation}
D_{\mu\nu}^{-1}(k) = \delta_{\mu\nu} k^2 - k_\mu k_\nu (1-\xi) + \Pi_{\mu\nu}(k)
=:  \,_0D_{\mu\nu}^{-1}(k) + \Pi_{\mu\nu}(k) + \xi k_\mu k_\nu\,,
\end{equation}
where $k$ is the gluon momentum.
(Regarding $\xi$, common choices in perturbation theory are $\xi=0,1$, \emph{viz}.\ Landau and Feynman gauges, respectively.)
Gauge invariance (BRST symmetry of the quantised theory \cite[Ch.\,II]{Pascual:1984zb}) is expressed in the following Slavnov-Taylor identity \cite{Taylor:1971ff, Slavnov:1972fg}:
\begin{equation}
\label{gPiSTI}
k_\mu \Pi_{\mu\nu}(k) = 0 = \Pi_{\mu\nu}(k) k_\nu\,.
\end{equation}
This restrictive, yet generous, constraint states that interactions cannot affect the four-longitudinal component of the gluon two-point function, but leaves room for modifications of the propagation characteristics of the three four-transverse degrees-of-freedom.

Equation~\eqref{gPiSTI} means
\begin{equation}
\Pi_{\mu\nu}(k) = [\delta_{\mu\nu} k^2 - k_\mu k_\nu ] \Pi(k^2)=: T_{\mu\nu}^k k^2 \Pi(k^2)\,,
\end{equation}
where $\Pi(k^2)$ is the dimensionless gluon self energy; hence, the gauge invariance constraint entails
\begin{equation}
\label{GluonSF}
D_{\mu\nu}(k) = T_{\mu\nu}^k \frac{1}{k^2 [ 1 +\Pi(k^2)] }
+ \xi \frac{k_\mu k_\nu}{k^4}
=: T_{\mu\nu}^k \overline{D}(k^2) + \xi \frac{k_\mu k_\nu}{k^4} \,.
\end{equation}
This is the propagator of a massless vector-boson \emph{unless}
\begin{equation}
\label{Eqgluonmass}
\Pi(k^2) \stackrel{k^2\simeq 0}{=} \frac{m_J^2}{k^2},
\end{equation}
in the event of which both the dressed-gluon acquires a mass and all symmetry constraints are preserved.  That Eq.\,\eqref{Eqgluonmass} is possible in an interacting quantum gauge field theory was first shown in a study of two-dimensional QED \cite{Schwinger:1962tn, Schwinger:1962tp} and the phenomenon is now known as the Schwinger mechanism of gauge boson mass generation.  Three-dimensional QED supports a similar outcome \cite{Appelquist:1988sr, Maris:1996zg, Bashir:2008fk, Bashir:2009fv, Braun:2014wja}, as does $D=3$ QCD \cite{Aguilar:2010zx, Cornwall:2015lna}; but, as already noted above, there is a difference between both these examples and QCD.  Namely, whereas the Lagrangian couplings in $D<4$ theories carry a mass dimension, which explicitly breaks scale invariance, this is not the case for $D=4$ chromodynamics.

\begin{figure}[!t]
\centering
\includegraphics[clip, width=0.62\textwidth]{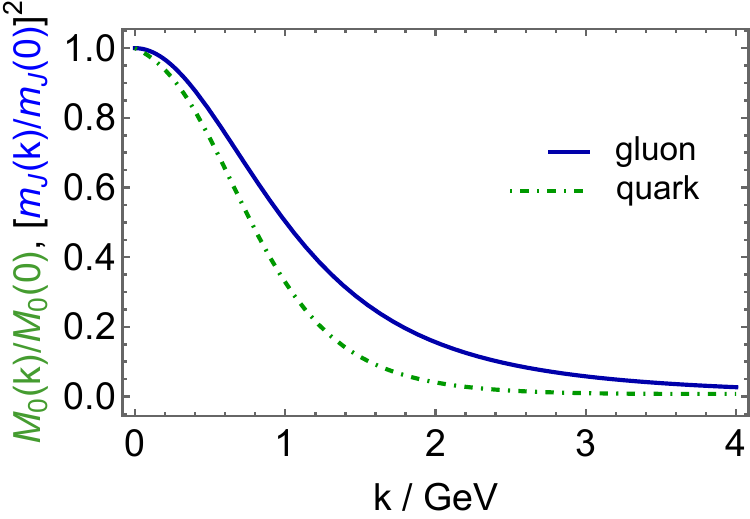}
\caption{\label{Fmasses}
Renormalisation group invariant dressed-gluon mass function (solid blue curve) calculated, following the method in Ref.\,\cite{Aguilar:2019uob}, from a gluon two-point function obtained using the lQCD configurations in Refs.\,\cite{Blum:2014tka, Boyle:2015exm, Boyle:2017jwu}.  The mass-squared curve is plotted, normalised by its $k=0$ value, and compared with the kindred chiral-limit dressed-quark mass function drawn from Ref.\,\cite{Binosi:2016wcx} (dot-dashed green curve).  It is this pair of curves that is $1/k^2$-suppressed in the ultraviolet, each with additional logarithmic corrections.
}
\end{figure}

The existence of a Schwinger mechanism in QCD was first conjectured forty years ago \cite{Cornwall:1981zr}.  The idea has subsequently been explored and refined \cite{Aguilar:2008xm, Boucaud:2008ky, Binosi:2009qm, Boucaud:2011ug, Aguilar:2015bud}, so that today a detailed picture is emerging, which unifies both the gauge and matter sectors \cite{Binosi:2014aea}.  The dynamical origin of the QCD Schwinger mechanism and its intimate connection with nonperturbative dynamics in the three-gluon vertex are elucidated elsewhere \cite{Binosi:2022djx, Papavassiliou:2022wrb}.  This is an area of continuing research, where synergies between continuum and lattice QCD are being exploited \cite{Aguilar:2021uwa, Pinto-Gomez:2022brg}.  For our purposes, it is sufficient to know that Eq.\,\eqref{Eqgluonmass} is realised in QCD.  Indeed, owing to their self-interactions, gluon partons transmogrify into gluon quasiparticles whose propagation characteristics are determined by a momentum-dependent mass function.  That mass function is power-law suppressed in the ultraviolet -- hence, invisible in perturbation theory; yet large at infrared momenta, being characterised by a renormalisation point independent value  \cite{Cui:2019dwv}:
\begin{equation}
\label{gluonmass}
m_0 = 0.43(1)\,{\rm GeV}.
\end{equation}
The renormalisation group invariant (RGI) gluon mass function is drawn in Fig.\,\ref{Fmasses}.

Before closing this section, it is worth stressing the importance of Poincar\'e covariance in modern physics.%
\footnote{
When working with a Euclidean formulation, as we do, Poincar\'e covariance maps straightforwardly into Euclidean covariance, \emph{viz}.\ valid Schwinger functions must transform covariantly under O$(4)$ rotations and linear translations in $\mathbb R^4$.  Owing to the simplicity of this connection, we avoid transliteration and speak of Poincar\'e covariance and invariance throughout.
}
If one chooses to formulate a problem in quantum field theory using a scheme that does not ensure Poincar\'e invariance of physical quantities, then artificial or ``pseudodynamical'' effects are typically encountered \cite{Brodsky:2022fqy}.  In connection with gauge theory Schwinger functions, Poin\-car\'e covariance very effectively limits the nature and number of independent amplitudes that are required for a complete representation.  In contrast, analyses and quantisation procedures which violate Poincar\'e covariance engender a rapid proliferation in the number of such functions.  For instance, the  covariant-gauge gluon two-point function in Eq.\,\eqref{GluonSF} is fully specified by one scalar function; whereas in the class of axial gauges, two unconnected functions are required and unphysical, kinematic singularities appear in the associated tensors \cite{West:1982gg, Brown:1988bn}.  This is why covariant gauges are normally employed for concrete calculations in both continuum and lattice-regularised QCD.  In fact, Landau gauge, \emph{i.e}., $\xi = 0$ in Eq.\,\eqref{GluonSF}, is often used because, amongst other things, it is a fixed point of the renormalisation group \cite[Ch.\,IV]{Pascual:1984zb} and implemented readily in lQCD \cite{Cucchieri:2009kk}.  We typically refer to Landau gauge results herein.  Naturally, gauge covariance of Schwinger functions ensures that expressions of EHM in physical observables are independent of the gauge used for their elucidation.

Equation~\eqref{gluonmass} is the cleanest expression of EHM in Nature, being truly a manifestation of mass emerging from \emph{nothing}: infinitely many massless gluon partons fuse together so that, to all intents and purposes, they behave as coherent quasiparticle fields with a long-wavelength mass which is almost half that of the proton.  The implications of this result are enormous and far-reaching, including, \emph{e.g}., key steps toward elimination of the problem of Gribov ambiguities \cite{Gao:2017uox}, which were long thought to prevent a rigorous definition of QCD.

\section{Process-Independent Effective Charge}
\label{SecPICharge}
In classical field theories, couplings and masses are constants.  Typically, this is also true in quantum mechanics models of strong interaction phenomena.  However, it is not the case in renormalisable quantum gauge field theories, as highlighted by the Gell-Mann--Low effective-charge/running-coupling in QED \cite{GellMann:1954fq}, which is a textbook case \cite[Ch.\,13.1]{IZ80}.

A highlight of twentieth century physics was the realisation that QCD in particular, and non-Abelian gauge theories in general, express asymptotic freedom \cite{Politzer:2005kc, Wilczek:2005az, Gross:2005kv}, \cite[Ch.\,7.1]{Pickering:1984tk}, \emph{i.e}., the feature that the interaction between charge-carriers in the theory becomes weaker as $k^2$, the momentum-squared characterising the scattering process, becomes larger.  Analysed perturbatively at one-loop order in the modified minimal subtraction renormalisation scheme, $\overline{MS}$, the QCD running coupling takes the form
\begin{equation}
\label{EqRunningCoupling}
\alpha_{\overline{MS}}(k^2) = \frac{\gamma_m \pi}{\ln k^2/\Lambda_{\rm QCD}^2}\,,
\end{equation}
where $\gamma_m = 12/[33 -  2 n_f]$, with $n_f$ the number of quark flavours whose mass does not exceed $k^2$, and $\Lambda_{\rm QCD}\sim 0.2\,$GeV is the RGI mass-parameter that sets the scale for perturbative analyses.

Asymptotic freedom comes with a ``flip side'', which came to be known as \emph{infrared slavery} \cite[Sec.\,3.1.2]{Marciano:1977su}.  Namely, beginning with some $k^2 \gg \Lambda_{\rm QCD}^2$, then the interaction strength grows as $k^2$ is reduced, with the coupling diverging at $k^2=\Lambda_{\rm QCD}^2$.  (This is the Landau pole.)  Qualitatively, this statement is true at any finite order in perturbation theory: whilst the value of $\Lambda_{\rm QCD}$ changes somewhat, the divergence remains.  In concert with the area law demonstrated in Ref.\,\cite{Wilson:1974sk}, which entails that the potential between any two \underline{infinitely massive} colour sources grows linearly with their separation, many practitioners were persuaded that the complex dynamical phenomenon of confinement could simply be explained by an unbounded potential that grows with parton separation.  As we shall see, that is not the case, but the notion is persistent.

Given the character of QCD's perturbative running coupling, two big questions arise:
\begin{enumerate}[label=(\emph{\alph*}), ref=(\alph*)]
\item does QCD possess a unique, nonperturbatively well-defined and calculable effective charge, \emph{viz}.\ a veritable analogue of QED's Gell-Mann--Low running coupling; \label{alphaa}
and
\item does Eq.\,\eqref{EqRunningCoupling} express the large-$k^2$ behaviour of that charge? \label{alphab}
\end{enumerate}
If both questions can be answered in the affirmative, then great strides have been made toward verifying that QCD is truly a theory.

Following roughly forty years of two practically disjoint research efforts, one focused on QCD's gauge sector \cite{Fischer:2006ub, Boucaud:2011ug, Aguilar:2015bud} and another on its matter sector \cite{Maris:2003vk, Chang:2011vu, Bashir:2012fs, Roberts:2012sv}, a key step on the path to answering these questions was taken in Ref.\,\cite{Binosi:2014aea}.  The two distinct efforts were designated therein as the top-down approach -- \emph{ab initio} computation of the interaction via direct analyses of gauge-sector gap equations; and the bottom-up scheme -- inferring the interaction by describing data within a well-defined truncation of those matter sector equations that are relevant to bound-state properties.   Reference~\cite{Binosi:2014aea} showed that the top-down and bottom-up approaches are unified when the RGI running-interaction predicted by then-contemporary analyses of QCD's gauge sector is used to explain ground-state hadron observables using nonperturbatively-improved truncations of the matter sector bound-state equations.  The first such truncation was introduced in Ref.\,\cite{Chang:2009zb}.

\begin{figure}[!t]
\centering
\includegraphics[clip, width=0.7\textwidth]{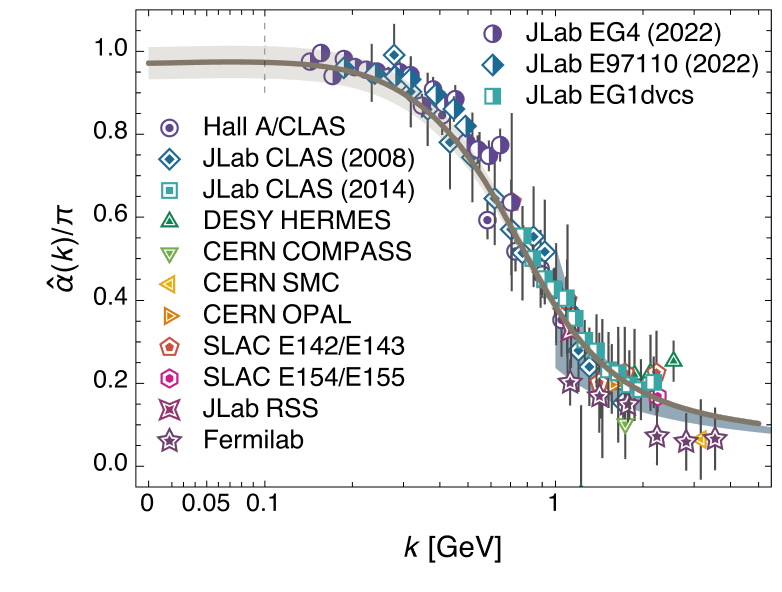}
\caption{\label{Falpha}
Process-independent effective charge, $\hat{\alpha}(k)/\pi$, obtained by combining modern results from continuum and lattice analyses of QCD's gauge sector \cite{Cui:2019dwv}.
%
%
Existing data on the process-dependent charge $\alpha_{g_1}$ \cite{Deur:2016tte, Deur:2022msf}, defined via the Bjorken sum rule, is shown for comparison -- see Refs.\,\cite{%
Deur:2005cf, Deur:2008rf, Deur:2014vea, Deur:2022msf,
Ackerstaff:1997ws, Ackerstaff:1998ja, Airapetian:1998wi, Airapetian:2002rw, Airapetian:2006vy,
Kim:1998kia,
Alexakhin:2006oza, Alekseev:2010hc, Adolph:2015saz,
Anthony:1993uf, Abe:1994cp, Abe:1995mt, Abe:1995dc, Abe:1995rn, Anthony:1996mw, Abe:1997cx, Abe:1997qk, Abe:1997dp, Abe:1998wq, Anthony:1999py, Anthony:1999rm, Anthony:2000fn, Anthony:2002hy}.
(Image courtesy of D.\,Binosi.)}
\end{figure}

It was a short walk from this point to a realisation \cite{Binosi:2016nme} that in QCD, by means of the pinch-technique \cite{Pilaftsis:1996fh, Binosi:2009qm, Cornwall:2010upa} and background field method \cite{Abbott:1981ke}, one can define and calculate a unique, process-independent (PI) and RGI analogue of the Gell-Mann--Low effective charge, now denoted $\hat\alpha(k^2)$.  The analysis was refined in Ref.\,\cite{Cui:2019dwv}, which combined modern results from continuum analyses of QCD's gauge sector and lQCD configurations generated with three domain-wall fermions at the physical pion mass \cite{Blum:2014tka, Boyle:2015exm, Boyle:2017jwu} to obtain a parameter-free prediction of $\hat\alpha(k^2)$.  The resulting charge is drawn in Fig.\,\ref{Falpha}.  It is reliably interpolated by writing
\begin{align}
\label{Eqhatalpha}
\hat{\alpha}(k^2) & = \frac{\gamma_m \pi}{\ln\left[{\mathpzc K}^2(k^2)/\Lambda_{\rm QCD}^2\right]}
\,,\; {\mathpzc K}^2(y=k^2) = \frac{a_0^2 + a_1 y + y^2}{b_0 + y}\,,
\end{align}
%
with (in GeV$^2$): $a_0=0.104(1)$, $a_1=0.0975$, $b_0=0.121(1)$.  The curve was obtained using a momentum-subtraction renormalisation scheme: $\Lambda_{\rm QCD}= 0.52\,$GeV when $n_f=4$.

Notably, $\hat{\alpha}(k^2)$ is PI and RGI in any gauge; but it is sufficient to know $\hat{\alpha}(k^2)$ in Landau gauge, $\xi=0$ in Eq.\,\eqref{GluonSF}, which is the choice both for easiest calculation and the result in Eq.\,\eqref{Eqhatalpha}.  This is because $\hat{\alpha}(k^2)$ is form invariant under gauge transformations, as may be shown using identities discussed elsewhere \cite{Binosi:2013cea}, and gauge covariance ensures that any such transformations can be absorbed into the Schwinger functions of the quasiparticles whose interactions are described by $\hat{\alpha}(k^2)$ \cite{Aslam:2015nia}.

The following physical features of $\hat{\alpha}$ deserve to be highlighted because they expose a great deal about QCD.
\begin{description}
\label{alphaPIlist}
\item[Absence of a Landau pole.]
Whereas the perturbative running coupling, \emph{e.g}., Eq.\,\eqref{EqRunningCoupling}, diverges at $k^2 = \Lambda_{\rm QCD}^2$, revealing the Landau pole, the PI charge is a smooth function on $k^2 \geq 0$: the Landau pole is eliminated owing to the appearance of a gluon mass scale, Eq.\,\eqref{gluonmass}.

Implicit in the ``screening function'', ${\mathpzc K}(k^2)$, is a screening mass,
\begin{equation}
\zeta_{\cal H} = {\mathpzc K}(k^2=\Lambda^2_{\rm QCD}) \approx 1.4\,\Lambda_{\rm QCD} < m_p,
\end{equation}
at which point the perturbative coupling would diverge but the PI coupling passes through an inflection point on its way to saturation.
On $\surd k^2 \lesssim \zeta_{\cal H}$, the PI charge enters a new domain, upon which the running slows, practically ceasing on $\surd k^2 \leq m_0/2$ so that QCD is once again effectively a conformal theory and the charge saturates to a constant infrared value $\hat\alpha(k^2=0) = \pi \times 0.97(4)$.  This value is a prediction: within 3(4)\%, the coupling saturates to a value of $\pi$ at $k^2=0$.  It is not yet known whether this proximity to $\pi$ has any deeper significance in Nature, but a potential explanation is provided in the next bullet.

These features emphasise the role of EHM as expressed in Eq.\,\eqref{gluonmass}: the existence of $m_0 \approx m_p/2$ guarantees that long wavelength gluons are screened, so play no dynamical role.
Consequently, $\zeta_H$ marks the boundary between soft/nonperturbative and hard/perturbative physics.  It is therefore a natural choice for the ``hadron scale'', \emph{viz}.\ the renormalisation scale at which valence quasiparticle degrees-of-freedom should be used to formulate and solve hadron bound-state problems \cite{Cui:2019dwv}.  Implementing that notion, then those quasiparticles carry all hadron properties at $\zeta=\zeta_{\cal H}$.  This approach is today being used to good effect in the prediction of hadron parton distribution functions (DFs) -- see Sec.\,\ref{SecDFs} and Refs.\,\cite{Ding:2019qlr, Ding:2019lwe, Cui:2020dlm, Cui:2020tdf, Han:2020vjp, Chang:2021utv, Xie:2021ypc, Raya:2021zrz, Cui:2021mom, Cui:2022bxn, Chang:2022jri, Lu:2022cjx, dePaula:2022pcb}.

\item[Match with the Bjorken process-dependent charge.]
The theory of process \emph{dependent} (PD) charges was introduced in Ref.\,\cite{Grunberg:1982fw, Grunberg:1989xf}: ``\ldots \emph{to each physical quantity depending on a single scale variable is associated an effective charge, whose corresponding St\"uckelberg -- Peterman -- Gell-Mann--Low function is identified as the proper object on which perturbation theory applies}."  PD charges have since been widely canvassed \cite{Dokshitzer:1998nz, Prosperi:2006hx, Deur:2016tte}.

One of the most fascinating things about the PI running coupling is highlighted by its comparison with the data in Fig.\,\ref{Falpha}, which express measurements of the PD effective charge, $\alpha_{g_1}(k^2)$, defined via the Bjorken sum rule \cite{Bjorken:1966jh, Bjorken:1969mm}.  The charge calculated in Ref.\,\cite{Cui:2019dwv} is an essentially PI charge.  There are no parameters; and, \emph{prima facie}, no reason to expect that it should match $\alpha_{g_1}(k^2)$.  The almost precise agreement is a discovery, given more weight by new results on $\alpha_{g_1}(k^2)$ \cite{Deur:2022msf}, which now reach into the conformal window at infrared momenta.

Mathematically, at least part of the explanation lies in the fact that the Bjorken sum rule is an isospin non-singlet relation, which eliminates many dynamical contributions that might distinguish between the two charges.  It is known that the two charges are not identical; yet, equally, on any domain for which perturbation theory is valid, the charges are nevertheless very much alike:
\begin{equation}
\frac{\alpha_{g_1}(k^2)}{\hat\alpha(k^2)} \stackrel{k^2 \gg m_0^2}{=} 1+\frac{1}{20} \alpha_{\overline{\rm MS}}(k^2)\,,
\end{equation}
where $\alpha_{\overline{\rm MS}}$ is given in Eq.\,\eqref{EqRunningCoupling}.  At the $c$ quark current-mass, the ratio is $1.007$, \emph{i.e}., indistinguishable from unity insofar as currently achievable precision is concerned.
At the other extreme, in the far infrared, the Bjorken charge saturates to $\alpha_{g_1}(k^2=0)=\pi$; hence,
\begin{equation}
\frac{\alpha_{g_1}(k^2)}{\hat\alpha(k^2)} \stackrel{k^2 \ll m_0^2}{=} 1.03(4)\,.
\end{equation}

Evidently, the PD charge determined from the Bjorken sum rule is, for practical intents and purposes, indistinguishable from the PI charge generated by QCD's gauge-sector dynamics \cite{Binosi:2016nme, Cui:2019dwv}.

\item[Infrared completion.] Being process independent, $\hat\alpha(k^2)$ serves numerous purposes and unifies many observables.  It is therefore a good candidate for that long-sought running coupling which describes QCD's effective charge at all accessible momentum scales \cite{Dokshitzer:1998nz}, from the deep infrared to the far ultraviolet, and at all scales in between, without any modification.

    Significantly, the properties of $\hat\alpha(k^2)$ support a conclusion that QCD is actually a theory, \emph{viz}.\ a well-defined $D=4$ quantum gauge field theory.  QCD therefore emerges as a viable tool for use in moving beyond the SM by giving substructure to particles that today seem elementary.  A good example was suggested long ago; namely, perhaps all spin-$J=0$ bosons may be \cite{Schwinger:1962tp} ``\ldots \emph{secondary dynamical manifestations of strongly coupled primary fermion fields and vector gauge fields} \ldots'.  Adopting this position, the SM's Higgs boson might also be composite; in which case, \emph{inter alia}, the quadratic divergence of Higgs boson mass corrections would be eliminated.

\end{description}
Qualitatively equivalent remarks have been developed using light-front holographic models of QCD based on anti-de Sitter/conformal field theory (AdS/CFT) duality \cite{Brodsky:2003px, Brodsky:2010ur}.

Returning to the two questions posed following Eq.\,\eqref{EqRunningCoupling} in items \ref{alphaa}, \ref{alphab}, it is now apparent that they are answerable in the affirmative: QCD does possess a unique, nonperturbatively well-defined and calculable effective charge whose large-$k^2$ behaviour connects smoothly with that in Eq.\,\eqref{EqRunningCoupling}.  These facts provide strong support for the view that QCD is a well-defined 4D quantum gauge field theory.

\section{Confinement}
\label{SecConfinement}
Confinement is much discussed but little understood.  In large part, both these things stem from the absence of a clear, agreed definition of confinement.  With certainty, it is only known that nothing with quantum numbers matching those of the gluon or quark fields in Eq.\,\eqref{QCDdefine} has ever reached a detector.

An interpretation of confinement is included in the official description of the Yang-Mills Millennium Problem \cite{millennium:2006}.  The simpler background statement is worth repeating: \\[0.5ex]
\hspace*{0.05\textwidth}\parbox[t]{0.9\textwidth}{``\emph{Quantum Yang-Mills theory is now the foundation of most of elementary particle theory, and its predictions have been tested at many experimental laboratories, but its mathematical foundation is still unclear.  The successful use of Yang-Mills theory to describe the strong interactions of elementary particles depends on a subtle quantum mechanical property called the `mass gap': the quantum particles have positive masses, even though the classical waves travel at the speed of light.  This property has been discovered by physicists from experiment and confirmed by computer simulations, but it still has not been understood from a theoretical point of view.  Progress in establishing the existence of the Yang-Mills theory and a mass gap will require the introduction of fundamental new ideas both in physics and in mathematics}.''}
\vspace*{1ex}

The formulation of this problem focuses entirely on quenched-QCD, \emph{i.e}., QCD without quarks; so, its solution is not directly relevant to our Universe.  Confinement in pure quantum SU$(3)$ gauge theory and in QCD proper are probably very different because the pion exists and is unnaturally light on the hadron scale \cite{Horn:2016rip}.  On the other hand, the remarks concerning the emergence of a ``mass gap'' relate directly to Fig.\,\ref{Fmasses} and Eq.\,\eqref{gluonmass} herein.  Whilst these properties of QCD may be considered proven by the canons of theoretical physics, such arguments do not meet the standards of mathematical physics and constructive field theory because they involve input from numerical analyses of QCD Schwinger functions.  Hereafter, therefore, we will continue within the theoretical physics perspective.

As noted above, a mechanism for the total confinement of infinitely massive charge sources has been identified in the lattice-regularised treatment of quantum field theories using compact representations of Abelian or non-Abelian gauge fields \cite{Wilson:1974sk}, \emph{viz}.\ the area law $\equiv$ linear source-antisource potential.  However, no treatment of the continuum meson bound-state problem has yet been able to demonstrate how such an area law emerges as the masses of the meson's valence degrees-of-freedom grow to infinity.

In the era of infrared slavery, it was widely assumed that some sort of nonperturbatively improved one-gluon exchange could simultaneously produce asymptotic freedom and a linearly rising potential between quarks; and many models were developed with just such features \cite{Eichten:1978tg, Richardson:1978bt, Buchmuller:1980su, Godfrey:1985xj, Lucha:1991vn}.  However, as highlighted by our discussion of QCD's effective charge, ongoing developments in the study of mesons, using rigorous treatments of the Schwinger functions involved, do not support this picture of confinement via dressed-one-gluon-exchange.  The path to an area law is far more complex.

\begin{figure}[!t]
\centering
\hspace*{-1ex}\begin{tabular}{lcl}
{\sf A} & \hspace*{5em} & {\sf B} \\[-0ex]
\vspace*{+3.5ex} & &  \\
\includegraphics[clip, width=0.2\textwidth]{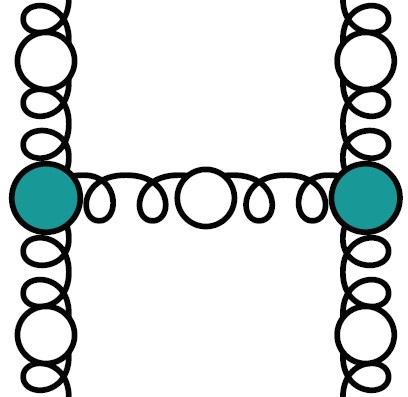} & \hspace*{5em} & \\
\vspace*{-24.7ex} & \hspace*{5em} &  \\
& \hspace*{5em} &
\includegraphics[clip, width=0.3\textwidth]{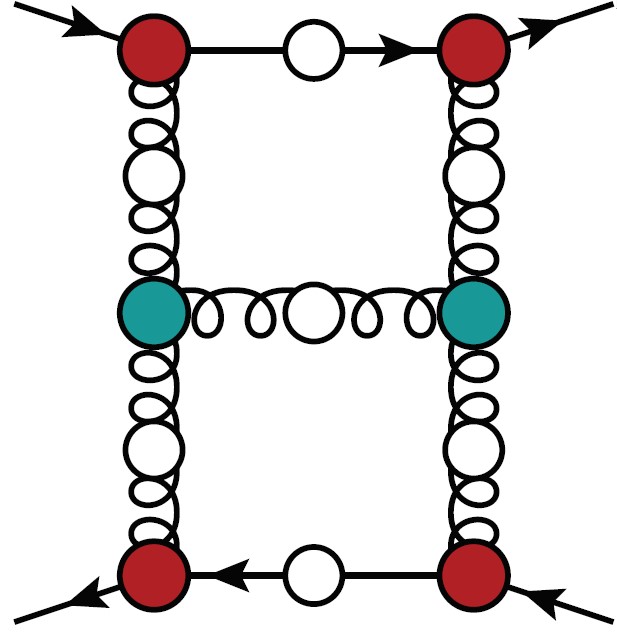}
\end{tabular}
\caption{\label{Hdiagram}
Panel \mbox{\sf A}\,--\,primitive gluon $H$-diagram.
Panel \mbox{\sf B}\,--\,one-$H$-diagram contribution to quark+antiquark scattering.
Legend.
Dressed-gluon two-point function -- spring with open circle insertion;
dressed-quark two-point function -- straight line with open circle insertion;
blue-filled circle at $3$-gluon junction -- dressed $3$-gluon vertex;
and
red-filled circle at gluon-quark junction -- dressed gluon-quark.
Repeated insertion of $H$-diagrams within $H$-diagrams, exploiting both gluon-quark and gluon self-couplings, leads to a plaquette-like area-filling structure, reminiscent of the planar summation of elementary squares in Ref.\,\cite{Wilson:1974sk}.
 }
\end{figure}

One direction that deserves exploration is connected with the gluon ``$H$-diagrams'' drawn in Ref.\,\cite[Fig.\,8]{Binosi:2016rxz} and reproduced in Fig.\,\ref{Hdiagram}A.  Imagine a valence-quark and -anti\-quark scattering via such a process, as drawn in Fig.\,\ref{Hdiagram}B; then keep adding $H$-diagrams within $H$-diagrams, exploiting both gluon-quark and gluon self-couplings.
Such $H$-diagram scattering processes produce an infrared divergence in the perturbative computation of a static quark potential \cite{Smirnov:2009fh}, \emph{viz}.\ a contribution that exhibits unbounded growth as the source-antisource separation increases.  Nonperturbatively, that divergence is tamed because the effective charge saturates -- Fig.\,\ref{Falpha}.  On the other hand, there are infinitely many such contributions; and in the limit of static valence degrees-of-freedom, the entire unbounded sum of planar $H$-diagrams is contracted to a point-connection of infinitely dense fisherman's-net/spider's-web diagrams on both the source and antisource.
It is conceivable that the confluence of these effects could yield the long-sought area law via the Bethe-Salpeter equation \cite{Binosi:2016rxz, Brodsky:2015oia}.

Real-world QCD, however, is characterised by light degrees-of-freedom: $u$ and $d$ quarks with electron-size current masses; and $s$ quarks with mass roughly one order-of-magnitude larger, so still much less than $m_p$.  Pions and kaons are constituted from such valence degrees-of-freedom and these mesons are light.  In fact, the pion has a lepton like mass \cite{Workman:2022ynf}: $m_\pi \approx m_\mu$, where $m_\mu$ is the mass of the $\mu$ lepton.  Owing to the presence of such degrees-of-freedom, light-particle annihilation and creation effects are essentially nonperturbative in QCD.  Consequently, despite continuing dedicated efforts \cite{Brodsky:2014yha, Reinhardt:2017pyr, Hoyer:2021adf}, it has thus far proved impossible to either define or calculate a static quantum mechanical potential between two light quarks.

This may be illustrated by apprehending that a potential which increases with separation can be described by a flux tube extending between the source and antisource.  As the source-antisource separation increases, so does the potential energy stored in the flux tube.  However, it can only increase until the stored energy matches that required to produce a particle+antiparticle pair of the theory's lightest asymptotic states -- in QCD, a $\pi^+ \pi^-$ pair.  Numerical simulations of lQCD reveal \cite{Bali:2005fu, Prkacin:2005dc} that once the energy exceeds this critical value, the flux tube then dissolves along its entire length, leaving two isolated colour-singlet systems.  Given that $m_\pi = 0.14\,$GeV, then this disintegration must occur at source+antisource centre-of-mass separation $r\approx (1/3)\,$fm \cite{Chang:2009ae}, which is well within the interior of any hadron.  This example assumes that the source and antisource are static.  The situation is even more complex for real, dynamical quarks.
Thus, at least in the $u$, $d$, $s$ quark sector, confinement is manifested in features of Schwinger functions that are far more subtle than can be captured in typical potential models.

One non-static, \emph{i.e}., dynamical, picture of confinement has emerged from studies of the analytic properties of the two-point Schwinger functions associated with propagation of coloured gluon and quark quasiparticles -- see, \emph{e.g}., Fig.\,\ref{Fmasses}.  The development of this perspective may be traced back to a beginning almost forty years ago \cite{Munczek:1983dx, Stingl:1985hx, Zwanziger:1989mf, Krein:1990sf, Burden:1991gd, Gribov:1999ui}.  It has subsequently been carefully explored \cite{Burden:1991gd, Stingl:1994nk, Roberts:2007ji, Dudal:2008sp, Brodsky:2012ku, Qin:2013ufa, Lucha:2016vte, Gao:2017uox, Binosi:2019ecz, Dudal:2019gvn, Fischer:2020xnb, Roberts:2020hiw}; and in this connection one may profitably observe that only Schwinger functions which satisfy the axiom of reflection positivity \cite{Osterwalder:1973dx, Osterwalder:1974tc, GJ81} can be connected with states that appear in the Hilbert space of observables.

The axioms referred to here are those first presented in Refs.\,\cite{Osterwalder:1973dx, Osterwalder:1974tc} and subsequently modified in \cite{GJ81}, which identify the properties of Schwinger functions that are necessary and sufficient to ensure equivalence between the formulation of a given quantum field theory in Euclidean and Minkowski space.  (A contemporary literature compilation is presented elsewhere \cite{Dedushenko:2022zwd}.)  In effect, this means that all and only those Schwinger functions which satisfy the five Osterwalder-Schrader axioms possess connections with elements in the Hilbert space of physical states.  Regarding strong interactions, all physical states are colour singlets.  Consequently, for QCD to be the theory of strong interactions, all its colour-singlet Schwinger functions must satisfy the Osterwalder-Schrader axioms; equally, all its colour-nonsinglet functions must violate at least one.

Reflection positivity is a severe constraint.  It requires that the Fourier transform of the momentum-space Schwinger function, treated as a function of analytic, Poincar\'e-invariant arguments, is a positive-definite function.  To illustrate, consider the gluon Schwinger function in Eq.\,\eqref{GluonSF}.  A massless partonic gluon is described by $\overline{D}(k^2) = 1/k^2$; and the $4D$ Fourier transform of this function is
\begin{equation}
\label{GJ1}
\int \frac{d^4 k}{(2\pi)^4} {\rm e}^{i k\cdot x} \frac{1}{k^2} = \frac{1}{4\pi^2 x^2} > 0 \; \forall x^2>0 \,.
\end{equation}

More generally regarding two-point functions, \emph{viz}.\ those connected with propagation of elementary excitations in QCD, reflection positivity is satisfied if, and only if, the Schwinger function has a K\"all\'en-Lehmann representation.  Returning to the gluon Schwinger function in Eq.\,\eqref{GluonSF}, this means one must be able to write
\begin{equation}
\label{GJ2}
\overline{D}(k^2) = \int_0^\infty d\zeta \frac{\rho(\varsigma)}{k^2+\varsigma^2} \,, \quad \rho(\varsigma) > 0 \; \forall \varsigma > 0.
\end{equation}
Plainly, $\rho(\varsigma)=\delta(\varsigma)$ yields $\overline{D}(k^2) =1/k^2$, \emph{i.e}., the two-point function for a bare gluon parton.  Hence, according to Eqs.\,\eqref{GJ1}, \eqref{GJ2}, absent dressing, the gluon parton could appear in the Hilbert space of physical states.

It is important to observe that any function which satisfies Eq.\,\eqref{GJ2} is positive definite itself.  Moreover, given Eq.\,\eqref{GJ2},
\begin{equation}
\label{inflexion}
sgn\left( [\frac{d}{dk^2}]^n \,\overline{D}(k^2)\right) = (-1)^n\,;
\end{equation}
consequently, \emph{inter alia}, treated as a function of the analytic, Poincar\'e-invariant variable $k^2$, no function with a K\"all\'en-Lehmann representation of the form written in Eq.\,\eqref{GJ2} can possess an inflection point.  Conversely, any function that exhibits an inflection point or, more generally, has a second derivative which changes sign, must violate the axiom of reflection positivity \cite{Roberts:2007ji}; hence, the associated excitation cannot appear in the Hilbert space of observables.

Take another step, and consider the following configuration space Schwinger function ($\tau=x_4$, $\ell = k_4$)
\begin{equation}
\label{EqDelta}
\Delta(\tau) = \int d^3 x  \int \frac{d^4 k}{(2\pi)^4} {\rm e}^{i k\cdot x} \overline{D}(k^2)
= \frac{1}{\pi} \int_{0}^\infty d \ell \cos (\ell \tau) \overline{D}(\ell^2)\,.
\end{equation}
Suppose that interactions generate a constant mass for the gluon parton, so that $\overline{D}(k^2) = 1/(k^2+\mu^2)$.  Does that trigger confinement?  The answer is "no" because this Schwinger function has a spectral representation with $\rho(\varsigma) = \delta(\varsigma^2 - \mu^2)$; Eq.\,\eqref{inflexion} is satisfied; and so is positivity:
\begin{equation}
\Delta(\tau) = \frac{1}{2\mu} {\rm e}^{-\mu \tau}\,.
\end{equation}

\begin{figure}[!t]
\centering
\includegraphics[clip, width=0.6\textwidth]{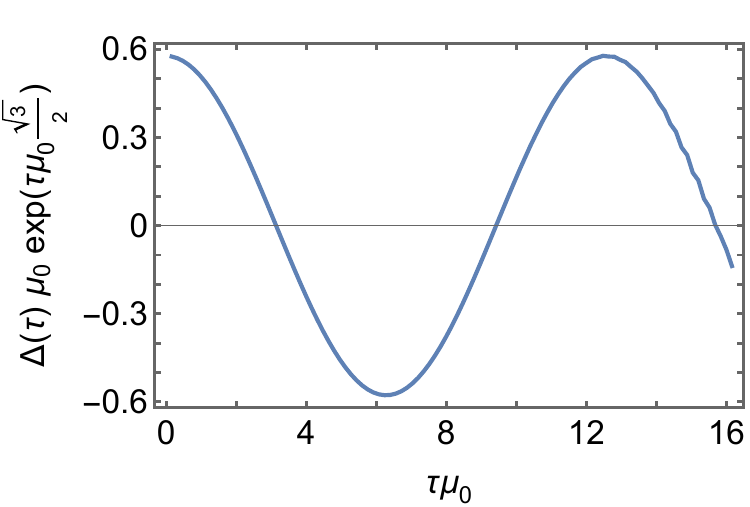}
\caption{\label{FSzero}
$\Delta(\tau) \mu_0  \exp(\mu_0 \tau \tfrac{\surd 3}{2})$ computed from the Schwinger function in Eq.\,\eqref{ModelGluon} using Eq.\,\eqref{EqDelta}.
}
\end{figure}

Suppose instead that interactions produce a momentum-dependent mass-squared function like that in Fig.\,\ref{Fmasses}, which is $1/k^2$ suppressed in the ultraviolet:
\begin{equation}
m_J^2(k^2) = \frac{\mu_0^4}{k^2 + \mu_0^2}
\Rightarrow \overline{D}(k^2) = \frac{k^2 + \mu_0^2}{k^2 (k^2+\mu_0^2)+\mu_0^4}\,.
\label{ModelGluon}
\end{equation}
The mass function itself is a monotonically decreasing, concave-up function; yet, in this case, the Schwinger function has an inflection point at $k^2 = 0.53 \mu_0^2$.  Hence, it does not have a K\"all\'en-Lehmann representation; so, the associated excitation cannot appear in the Hilbert space of observables.  Furthermore, evaluation of the configuration space Schwinger function defined by Eq.\,\eqref{EqDelta} yields \cite{Gao:2017uox}
\begin{equation}
\label{DeltaSimple}
\Delta(\tau) = \frac{1}{\mu_0} {\rm e}^{-\tau \mu_0 \tfrac{\surd 3}{2}}  \cos \frac{\mu_0 \tau}{2}
=: \Delta_{\rm p}(\tau)  \cos \frac{\mu_0 \tau}{2}\,,
\end{equation}
using which the curve in Fig.\,\ref{FSzero} is drawn: plainly, the configuration space Schwinger function violates reflection positivity.
(Notably, the algebraic calculation of\emph{\emph{}} $\Delta(\tau)$ is often difficult and not always possible; so, uniform positivity of the second-derivative, Eq.\,\eqref{inflexion}, is a much quicker means of testing for reflection positivity.  Nevertheless, when it can be obtained, an explicit form of $\Delta(\tau)$ does provide additional insights.)

It is interesting to extend Eq.\,\eqref{ModelGluon} using $m_J^2(k^2) = \alpha \mu_0^4/[k^2 + \mu_0^2]$, in which case
\begin{equation}
\label{Dk2RGZ}
\overline{D}(k^2=\mu_0^2 y) =\frac{1}{\mu_0^2 } \frac{1+y}{y^2 + y + \alpha}=: \frac{1}{\mu_0^2 } {\mathpzc d}(y)\,.
\end{equation}
This type of Schwinger function lies within the so-called ``refined Gribov-Zwanziger'' class \cite{Dudal:2008sp}.  For $\alpha > \tfrac{1}{2}$, as in the example above, the function in Eq.\,\eqref{Dk2RGZ} exhibits an inflection point at some $y>0$; and when $\alpha\in(\tfrac{1}{4},\tfrac{1}{2})$, the inflection point is found at a location $y\in (-\tfrac{1}{2},0)$.  With $\alpha\in(0,\tfrac{1}{4})$, on the other hand, the function in Eq.\,\eqref{Dk2RGZ} separates into a sum of two terms:
\begin{equation}
{\mathpzc d}(y) = \frac{\mathpzc n_1}{y+{\mathpzc p}_1} - \frac{\mathpzc n_2}{y+{\mathpzc p}_2}\,,
\end{equation}
where ${\mathpzc p}_{1,2} \in \mathbb R$ and $sgn({\mathpzc p}_1/{\mathpzc p}_2)=+1$.  In this case, there is no inflection point; nevertheless, the second derivative does change sign, switching $\pm\infty \leftrightarrow \mp\infty$ as $y$ passes through the pole locations.  Consequently, any excitation whose propagation is described by a Schwinger function obtained with $\alpha > 0$ in Eq.\,\eqref{Dk2RGZ} cannot appear in the Hilbert space of observable states.

Inserting Eq.\,\eqref{Dk2RGZ} into Eq.\,\eqref{EqDelta}, one finds, after some careful algebra  \cite{Gao:2017uox} (${\mathpzc a}=\alpha^{1/4}$):
\begin{align}
\Delta(\tau) &=
\frac{1}{2\mu_0 {\mathpzc a} s_{\varphi/2} }{\rm e}^{-\tau \mu_0 {\mathpzc a} c_{\varphi/2} } \nonumber \\
& \quad \times \left[
(1+\tfrac{1}{{\mathpzc a}^2}) s_{\varphi/2} \cos( \tau \mu_0 {\mathpzc a} s_{\varphi/2} )
-(1-\tfrac{1}{{\mathpzc a}^2}) c_{\varphi/2} \sin( \tau \mu_0 {\mathpzc a} s_{\varphi/2} )
\right]\,,
\label{Deltatau}
\end{align}
where
$c_{\varphi/2}=[\tfrac{1}{2} + \tfrac{1}{4 {\mathpzc a}^2}]^{1/2}$,
$s_{\varphi/2}=[1-c_{\varphi/2}^2]^{1/2}$.
(Eq.\,\eqref{DeltaSimple} is a special case of this result.)
Equation~\eqref{Deltatau} reveals that the distance from $\tau=0$ of the first zero in the configuration space Schwinger function, $\tau_z$, increases with decreasing $\alpha$, \emph{i.e}., as the infrared value of the gluon mass-squared function is reduced.  Thus, confinement would practically be lost if $\tau_z$ were to become much greater than $\pi/\mu_0$.
Considering realistic gluon two-point functions, however, one finds $\mu_0 \approx (3/4) m_p$, $\alpha \approx 1$ \cite{Gao:2017uox}; so, $\tau_z \approx 1\,$fm, expressing a natural confinement length-scale.
(It is worth observing that if $\tau_z$ were instead measured on the \AA\ scale, then the notion of confinement would be lost because modern detectors are able to directly image targets of this size \cite{PhysRevLett.110.213001}.)

This discussion is readily summarised.  Owing to complex nonlinear dynamics in QCD, gluon and quark partons acquire momentum-dependent mass functions, as a consequence of which they emerge as quasiparticles whose propagation characteristics are described by two-point Schwinger functions that are incompatible with reflection positivity.
Normally, the dynamical generation of \underline{running} masses is alone sufficient to ensure this outcome.
It follows that the dressed-gluons and -quarks cannot appear in the Hilbert space of physical states.  In this sense, they are confined.  The associated confinement length-scale is $\tau_z \approx 1\,$fm.
It is worth stressing that the use of such two-point functions in the calculation of colour-singlet matrix elements ensures the absence of coloured particle+antiparticle production thresholds \cite{Bhagwat:2002tx}, thereby securing the empirical expression of real-QCD confinement.

Considering these quasiparticle Schwinger functions further, one may also define a parton \emph{persistence} or \emph{fragmentation} length, $\tau_{\rm F}$, as the scale whereat the deviation of the Schwinger function from parton-like behaviour is $50$\%: $\Delta(\tau_{\rm F})/\Delta_{\rm p}(\tau_{\rm F}) =0.5$.  Referring to Eq.\,\eqref{DeltaSimple}, one reads $\tau_{\rm F} = (2/3) \tau_z$.  This result is also found using realistic gluon two-point functions \cite{Gao:2017uox}.  (The value 50\% is merely a reasonable choice.  At this level, 30\% would also be acceptable, in which case $\tau_{\rm F}\to \tau_{\rm F}^\prime  = (1/2) \tau_z$.)

A physical picture of dynamical confinement now becomes apparent \cite{Stingl:1994nk}.  Namely, once a gluon or quark parton is produced, it begins to propagate in spacetime; but after traversing a spacetime distance characterised by $\tau_{\rm F}$, interactions occur, causing the parton to lose its identity, sharing it with others.  Ultimately, combining the effects on this parton with similar impacts on those produced along with it, a countable infinity of partons (a parton cloud) is produced, from which detectable colour-singlet final states coalesce.
This train of events is the physics expressed in parton fragmentation functions (FFs) \cite{Field:1977fa}.  Such distributions describe how the QCD partons in Eq.\,\eqref{QCDdefine}, generated in a high-energy event and almost massless in perturbation theory, transform into a shower of massive hadrons, \emph{viz}.\ they describe how hadrons with mass emerge from practically massless partons.  It is natural, therefore, to view FFs as the cleanest expression of dynamical confinement in QCD.
Furthermore, in the neighbourhood of their common boundary of support, DFs and FFs are related by crossing symmetry \cite{Gribov:1971zn}: FFs are timelike analogues of DFs.  Hence, an understanding of FFs and their deep connection with DFs can deliver fundamental insights into EHM.
This picture of parton propagation, hadronisation and confinement -- of DFs and FFs -- can be tested in experiments at modern and planned facilities \cite{Brodsky:2015aia, Denisov:2018unj, Aguilar:2019teb, Brodsky:2020vco, Chen:2020ijn, Anderle:2021wcy, Arrington:2021biu, Aoki:2021cqa, Quintans:2022utc}.
A pressing demand on theory is delivery of predictions for FFs before such experiments are completed so as, for instance, to guide development of facilities and detectors.  As yet, however, there are no realistic computations of FFs.  In fact, even a formulation of this problem remains uncertain.

Before moving on, it is worth reiterating that confinement means different things to different people.  Whilst some see confinement only in an area law for Wilson loops \cite{Wilson:1974sk}, our perspective stresses a dynamical picture, in which dynamically driven changes in the analytic structure of coloured Schwinger functions ensures the absence of colour-carrying objects from the Hilbert space of observable states.  In time, perhaps, as strong QCD is better understood, it may be found that these two realisations are connected?  The only certain thing is the necessity to keep an open mind on this subject.

\section{Spectroscopy}
\label{SecSpectroscopy}
Insofar as the spectrum of hadrons is concerned, results from nonrelativistic or somewhat relativised quark models \cite{Capstick:2000qj, Giannini:2015zia, Plessas:2015mpa} are still often cited as benchmarks.  Indeed, a standard reference \cite[Sec.\,63]{Workman:2022ynf} includes the following assertions: ``\emph{The spectrum of baryons and mesons exhibits a high degree of regularity.  The organizational principle which best categorizes this regularity is encoded in the quark model.  All descriptions of strongly interacting states use the language of the quark model}.''
This is despite the fact that neither the ``quarks'' nor the potentials in quark models have been shown to possess any mathematical link with Eq.\,\eqref{QCDdefine} -- rigorous or otherwise; and, furthermore, the orbital angular momentum and spin used to label quark model states are not Poincar\'e-invariant (observable) quantum numbers.

In step with improvements in computer performance, lQCD is delivering interesting results for hadron spectra \cite{Dudek:2010wm, Edwards:2011jj}, amongst which one may highlight indications for the existence of hybrid and exotic hadrons \cite{HadronSpectrum:2012gic, Dudek:2012ag, Ryan:2020iog, Woss:2020ayi}.  Continuum studies in quantum field theory are lagging behind owing in part to impediments placed by the character of the Bethe-Salpeter equation; primarily the fact that it is impossible to write the complete Bethe-Salpeter kernel in a closed form.

A systematic approach to truncating the integral equations associated with bound-state problems in QCD was introduced almost thirty years ago \cite{Munczek:1994zz, Bender:1996bb}.  Amongst other things, the scheme highlighted the importance of preserving continuous and discrete symmetries when formulating bound-state problems; enabled proof of Goldberger-Treiman identities and the Gell-Mann--Oakes--Renner relation in QCD \cite{Maris:1997hd, Maris:1997tm}; and opened the door to symmetry-preserving, Poincar\'e-invariant predictions of hadron observables, including elastic and transition form factors and DFs \cite{Maris:2003vk, Chang:2011vu, Bashir:2012fs, Roberts:2012sv, Eichmann:2016yit, Chen:2018rwz, Ding:2018xwy, Eichmann:2019bqf, Xu:2019ilh, Xu:2021mju, Yao:2020vef, Yao:2021pyf, Yao:2021pdy, Ding:2019lwe}.  Some of the more recent developments are sketched below.

An issue connected with the leading-order (RL - rainbow-ladder) term in the truncation scheme of Refs.\,\cite{Munczek:1994zz, Bender:1996bb} is that it only serves well for those ground-state hadrons which possess little rest-frame orbital angular momentum, $L$, between the dressed valence constituents \cite{Holl:2004fr, Holl:2005vu, Fischer:2009jm, Krassnigg:2009zh, Qin:2011dd, Qin:2011xq, Blank:2011ha, Hilger:2014nma, Fischer:2014cfa, Eichmann:2016hgl, Qin:2019hgk}.  This limitation can be traced to its inability to realistically express impacts of EHM on hadron observables, a weakness that is not overcome at any finite order of elaboration \cite{Fischer:2009jm}.  Improved schemes, which express EHM in the kernels, have been identified \cite{Chang:2009zb, Chang:2011ei, Binosi:2014aea, Williams:2015cvx, Binosi:2016rxz, Qin:2020jig}.  They have shown promise in applications to ground-state mesons constituted from $u$, $d$ valence quarks and/or antiquarks.  However, that is a small subset of the hadron spectrum; so, a recent extension to the spectrum and decay constants of $u$, $d$, $s$ meson ground- and first-excited states is welcome \cite{Xu:2022kng}.

Returning to quark models, it was long ago claimed \cite{Godfrey:1985xj} ``\ldots \emph{that all mesons -- from the pion to the upsilon -- can be described in a unified framework}.''  The context for this assertion was a model potential built using one-gluon-like exchange combined with an infrared-slavery ``confinement'' term that increases linearly with colour-source separation.  The basic mass-scales in such potential models are set by the constituent quark masses; and one might draw a qualitative link between those scales and the far-infrared values of the momentum-dependent dressed-quark running-masses \cite[Fig.\,2.5]{Roberts:2021nhw}: $M_{u,d}(0)\simeq 0.41\,$GeV, $M_s(0) \simeq 0.53\,$GeV.  Thereafter, mass-splittings and level orderings are arranged by tuning details of the potential.  Such a procedure can be quantitatively efficacious; however, it is qualitatively incorrect.  This is readily seen by recalling the Gell-Mann--Oakes--Renner relation \cite{GellMann:1968rz, Maris:1997hd, Qin:2014vya}: $m_\pi^2 \propto \hat m$, where $\hat m$ is Nature's explicit source of chiral symmetry breaking, generated by Higgs boson couplings to quarks in the SM.  Such behaviour is impossible in a potential model \cite{Roberts:2012sv, Horn:2016rip}, but natural in the CSM treatment of bound-states -- see, \emph{e.g}., Refs.\,\cite{Maris:1997hd, Holl:2004fr}, \cite[Fig.\,3.3]{Maris:2003vk}, \cite[Fig.\,7]{Bhagwat:2004hn}, \cite[Fig.\,1A]{Chang:2009zb}.  Thus, whilst potential models might deliver a fit to hadron spectra, they do not provide an explanation.

\begin{figure*}[!t]
\centerline{%
\includegraphics[clip, width=0.95\textwidth]{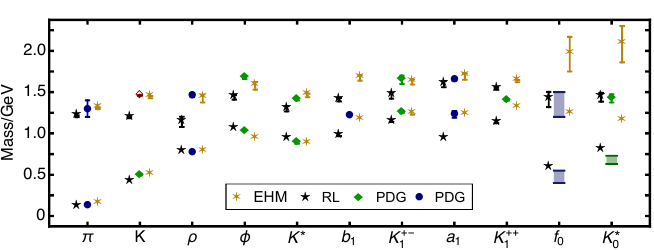}}
\caption{\label{figAllstrange}
Empirical spectrum \cite[PDG Summary Tables]{Workman:2022ynf}:
blue circles (bars) -- $u$, $d$ states; and green diamonds (bar) -- systems with $s$ and/or $\bar s$ quarks.
Little is known about $K(1460)$, which is therefore drawn as an open red diamond.
Gold six-pointed stars -- spectrum of low-lying $u$, $d$, $s$ mesons predicted by the EHM-improved Bethe-Salpeter kernel developed in Ref.\,\cite{Xu:2022kng};
and black five-pointed stars -- same spectrum computed using RL truncation.
}
\end{figure*}

That such challenges are surmounted when using CSMs to treat hadron bound-state problems is further exemplified in Ref.\,\cite{Xu:2022kng}, which adapts a novel scheme for including EHM effects in the Bethe-Salpeter kernel \cite{Qin:2020jig} to simultaneously treat ground- and first-excited-states of $u$, $d$, $s$ quarks.  As revealed in Fig.\,\ref{figAllstrange}, the empirical spectrum displays some curious features, \emph{e.g}.:
consistent with quark mass-scale counting, $m_\rho < m_{K^\ast}$, but this ordering is reversed for the first excitations of these states;
the first excited state of the $\pi$ is lighter than that of the $\rho$, but the ordering is switched for $K$, $K^\ast$;
and all axialvector mesons are nearly degenerate, with the larger mass of the $s$ quarks appearing to have little or no impact.
In delivering the first symmetry-preserving analysis of this collection of states to employ an EHM-improved kernel, Ref.\,\cite{Xu:2022kng} supplies fresh insights into the dynamical foundations of the properties of lighter-quark mesons.

In order to sketch that effort, we note that, using CSMs, the dressed propagator (two-point Schwinger function) for a quark with flavour $g$ is obtained as the solution of the following gap equation:
\begin{subequations}
\label{gendseN}
\begin{align}
S_g^{-1}(k) & = i\gamma\cdot k \, A_g(k^2) + B_g(k^2) = [i\gamma\cdot k + M_g(k^2)]/Z_g(k^2) \,,  \\
&  = Z_2 \,(i\gamma\cdot k + m_g) + \Sigma_g(k)\,,\\
\Sigma(k)& =  Z_1 \int^\Lambda_{dq}\!\! g^2 D_{\mu\nu}(k-q)\frac{\lambda^a}{2}\gamma_\mu S_g(q) \frac{\lambda^a}{2} \Gamma_\nu^g(k,p) ,
\end{align}
\end{subequations}
where $M_g(k^2)$ is RGI;
\begin{equation}
\label{DressedGluon}
D_{\mu\nu}(k) = \Delta(k^2) \,_0D_{\mu\nu}(k) \,;
\end{equation}
$\Gamma_\nu^g$ is the quark-gluon vertex;
$\int^\Lambda_{dq}$ denotes a Poincar\'e invariant regularisation of the four-dimensional integral, with $\Lambda$ the regularisation mass-scale;
and $Z_{1,2}(\zeta^2,\Lambda^2)$ are, respectively, the vertex and quark wave-function renormalisation constants, with $\zeta$ the renormalisation point.  (In such applications, when renormalisation is necessary, a mass-independent scheme is important, as discussed elsewhere \cite{Chang:2008ec}.)

It was anticipated almost forty years ago \cite{Singh:1985sg, Bicudo:1998qb}, and confirmed more recently \cite{Chang:2010hb, Bashir:2011dp, Binosi:2016wcx, Kizilersu:2021jen}, that EHM engenders a large anomalous chromomagnetic moment (ACM) for the lighter quarks; and with development of the first EHM-improved kernels, it was shown that such an ACM has a big impact on the $u$, $d$ meson spectrum \cite{Chang:2011ei, Williams:2015cvx, Qin:2020jig}.

The aim in Ref.\,\cite{Xu:2022kng} was to extend Refs.\,\cite{Chang:2011ei, Williams:2015cvx, Qin:2020jig} and highlight additional impacts of an ACM on the spectrum of mesons constituted from $u$, $d$, $s$ quarks.  An ACM emerges as a feature of the dressed-gluon-quark vertex, a three-point Schwinger function.  Its character and impacts can be exposed by writing $(l=k-q)$
\begin{equation}
\label{EqVertexGap}
\Gamma_\nu^g(q,k)  = \gamma_\nu + \tau_\nu(l)\,,\; \tau_\nu(l) = \eta \kappa(l^2)\sigma_{l\nu} \,,
\end{equation}
$\sigma_{l\nu} = \sigma_{\rho\nu} l_\rho$, $\kappa(l^2) = (1/\omega)\exp{(-l^2/\omega^2)}$.
Here, $\eta>0$ is the strength of the ACM term, $\tau_\nu(l)$; and it is assumed that the vertex is flavour-independent, which is a sound approximation for the lighter quarks \cite{Bhagwat:2004hn, Williams:2014iea}.
When considering Eq.\,\eqref{EqVertexGap}, one might remark that the complete gluon-quark vertex is far more complicated -- potentially containing twelve distinct terms -- and, in QCD, $\kappa(l^2)$ is power-law suppressed in the ultraviolet.  Notwithstanding these things, illustrative purposes are well served by Eq.\,\eqref{EqVertexGap}.

ACM effects are most immediately felt by the dressed-quark propagator.  The presence of an ACM in the kernel of Eq.\,\eqref{gendseN} increases positive EHM-induced feedback on dynamical mass generation.  Consequently, as shown elsewhere \cite{Binosi:2016wcx}, realistic values of the dressed-quark mass at infrared momenta are achieved using the PI effective charge in Fig.\,\ref{Falpha}.  Such an outcome requires tuning when using the PI charge in a rainbow truncation of the gap equation; in fact, DCSB cannot be guaranteed in that case \cite{Binosi:2014aea}.

Following Refs.\,\cite{Chang:2009zb, Chang:2010hb, Qin:2020jig} in continuing to emphasise clarity over numerical complexity, Ref.\,\cite{Xu:2022kng} also simplified the kernel in Eq.\,\eqref{gendseN}, writing
\begin{equation}
\label{GapKernel}
g^2 \Delta(k^2) = 4\pi \hat\alpha(0) \frac{D(\eta)}{\omega^4}  {\rm e}^{-k^2/\omega^2}\,,
\end{equation}
where $\omega =0.8\,$GeV, a value matching that suggested by analyses of QCD's gauge sector \cite{Binosi:2014aea, Cui:2019dwv}, and $D(\eta)= D_{\rm RL} (1+0.27 \eta)/(1+1.47 \eta)$, with $\omega D_{\rm RL} = (1.286\,{\rm GeV})^3$ chosen to achieve $m_\rho = 0.77\,$GeV in RL truncation.  The $\eta$-dependence of $D(\eta)$ was fixed \emph{a posteriori} by requiring that $m_\rho$ remain unchanged as $\eta$ is increased.  Since $\eta > 0 $ adds EHM strength to the gap equation's kernel, then $D$ must become smaller as $\eta$ grows in order to maintain a fixed value of $m_\rho$.  Following this procedure, $m_\rho$ becomes the benchmark against which all ACM-induced changes are measured.

It is worth noting that when one identifies $(g^2/[4\pi])\Delta(k^2=0) = 1/\mu_0^2$, then $\mu_0=0.39\,$GeV in RL truncation and $\mu_0 = 0.57\,$GeV at $\eta=1.2$.  So, the interaction specified by Eq.\,\eqref{GapKernel} is consistent with gluon mass generation as described in Sec.\,\ref{SecGluonMass}.
On the other hand, the large-$k^2$ behaviour of Eq.\,\eqref{GapKernel} does not respect the renormalisation group flow of QCD.  This would be an issue if one were using it, \emph{e.g}., to calculate hadron form factors at large $Q^2$ \cite{Chen:2018rwz, Ding:2018xwy, Xu:2019ilh, Xu:2021mju}, where $Q^2$ is momentum transfer squared, or parton distribution functions and amplitudes near the endpoints of their support domains \cite{Ding:2019qlr, Ding:2019lwe, Cui:2020dlm, Cui:2020tdf}.%
\footnote{This is well-known and explains why the truncated interaction in Eq.\,\eqref{GapKernel} was not used for any of the calculations described in Secs.\,\ref{SecBaryonWaveFunctions}\,--\,\ref{SecDFs} below.  All those studies are based on interactions which at least preserve QCD's ultraviolet power-law behaviour, where more has not yet been achieved -- Secs.\,\ref{SecBaryonWaveFunctions}, \ref{SecBaryonFormFactors}; and also the one-loop logarithmic improvement, when the necessary algorithms are already available -- Secs.\,\ref{SecFormFactors}, \ref{SecTransition}, \ref{SecDFs}.
}
However, it is far less important when calculating global, integrated properties, like hadron masses.  In such applications, Eq.\,\eqref{GapKernel} is satisfactory.  Indeed, good results can even be obtained using a symmetry-preserving treatment of a momentum-independent interaction \cite{Yin:2021uom, Xu:2021iwv, Gutierrez-Guerrero:2021rsx}.  A key merit of Eq.\,\eqref{GapKernel} lies in its elimination of the need for renormalisation, which simplifies analyses without materially affecting relevant results.

The scheme introduced in Ref.\,\cite{Qin:2020jig} provides a direct route from any reasonable set of gap equation elements to closed-form Bethe-Salpeter kernels for meson bound-state problems.  Thus having specified physically reasonable gap equations via Eqs.\,\eqref{EqVertexGap}, \eqref{GapKernel}, Ref.\,\cite{Xu:2022kng} adapted that scheme to arrive at Bethe-Salpeter equations for each of the mesons identified in Fig.\,\ref{figAllstrange}, obtaining solutions in all cases.  The image compares experiment with RL truncation results, also calculated in Ref.\,\cite{Xu:2022kng}, and predictions obtained using the EHM-improved kernel.%
\footnote{Dressed-quark propagators form an important part of the kernels of all bound-state equations.  As on-shell meson masses increase, poles in those propagators enter the complex-plane integration domain sampled by the Bethe-Salpeter equation \cite{Maris:1997tm}.  For such cases -- here, meson excited states -- a direct on-shell solution cannot be obtained using simple algorithms.  So, to obtain the masses of those mesons, Ref.\,\cite{Xu:2022kng} used an extrapolation procedure based on Pad\'e approximants.  This is the origin of the uncertainty bar on the CSM predictions.}

It is first worth mentioning the RL truncation mass predictions in Fig.\,\ref{figAllstrange}.  On the whole, the mean absolute relative difference, $\overline{\mbox{ard}}$, between RL results and central experimental values is $13(8)$\%.  This is tolerable; but there is substantial scatter and there are many qualitative discrepancies.

In contrast, compared with central experimental values, the EHM-improved masses in Fig.\,\ref{figAllstrange} agree at the level of $\overline{\mbox{ard}}=2.9(2.7)$\%.  This is a factor of $4.6$ improvement over the RL spectrum.
Moreover, correcting RL truncation flaws and reproducing empirical results:
$m_{K^\prime} > m_{\pi^\prime}$,
$m_{\rho^\prime} > m_{\pi^\prime}$,
$m_{\rho^\prime} \approx m_{K^{\ast \prime}}$;
the mass splittings $a_1$-$\rho$ and $b_1$-$\rho$ match empirical values because including the ACM in the kernel has markedly increased the masses of the $a_1$ and $b_1$ mesons, whilst $m_\rho$ was deliberately kept unchanged;
$m_{\phi^\prime}-m_\phi$ agrees with experiment to within 2\%;
the $K_1^{+-}$, $K_1^{++}$ level order is correct;
and quark+antiquark scalar mesons are heavy, providing room for the addition of strongly attractive resonant contributions to the bound-state kernels \cite{Holl:2005st, Santowsky:2020pwd}.

\begin{figure*}[!t]
\centerline{%
\includegraphics[clip, width=0.95\textwidth]{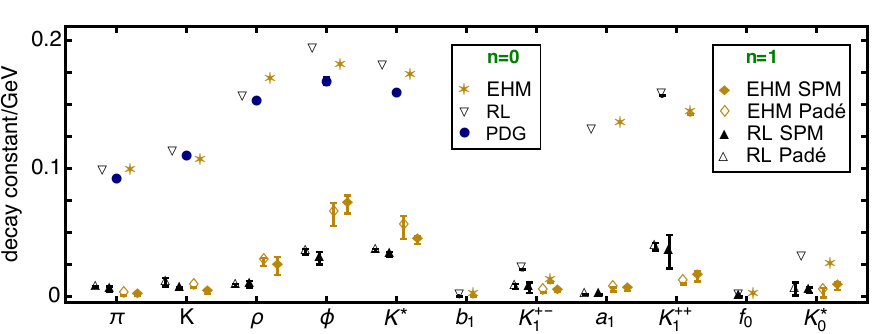}}
\caption{\label{FigLeptonic}
Leptonic decay constants for all states whose masses are reported in Fig.\,\ref{figAllstrange}: ground states, $n=0$; and lowest lying radial excitations, $n=1$.
For the excited states, two extrapolation results are presented for each state, \emph{viz}.\ one obtained with Pad\'e approximants and the other employing the Schlessinger point method (SPM) \cite{Cui:2022fyr, Binosi:2018rht, Yao:2021pyf, Yao:2021pdy} -- the distinct approaches yield consistent outcomes.
Results inferred from data are also plotted, where available \cite[PDG]{Workman:2022ynf}.
%
}
\end{figure*}

Using the Bethe-Salpeter amplitudes obtained in solving for the meson spectrum, canonically normalised in the standard fashion \cite[Sec.\,3]{Nakanishi:1969ph}, Ref.\,\cite{Xu:2022kng} also delivered predictions for the entire array of associated leptonic decay constants, $f_H$, including many which have not yet been measured.  The predicted values are depicted in Fig.\,\ref{FigLeptonic}, which also includes the few available empirical results.
The ground state leptonic decay constants in Fig.\,\ref{FigLeptonic} were calculated directly on-shell, but extrapolation was necessary to obtain on-shell values for those of the excited states.  For these observables, two extrapolation schemes were used and they yielded consistent results in all cases.

Given that Ref.\,\cite{Xu:2022kng} used a simplified interaction, \emph{viz}.\ Eq.\,\eqref{GapKernel}, then the Fig.\,\ref{FigLeptonic} comparison between predicted ground-state decay constants and the few known empirical values is favourable, particularly because decay constants are sensitive to ultraviolet physics, which was omitted.  There are also indications that the EHM-improved kernels deliver better agreement.

The decay constants of radially excited states are especially interesting.  Quantum mechanics models of positronium-like systems produce a single zero in the radial wave function of $n=1$ states.  The decay constant of a first radial excitation is thus $(1/8)$-times that of the ground state.  The predictions in Ref.\,\cite{Xu:2022kng} are generally consistent with this pattern, except for $J^P=0^-$ mesons.  In the pseudoscalar channel, as a corollary of EHM, QCD predicts $f_H^{n=1}\equiv 0$ in the chiral limit \cite{Holl:2004fr, Holl:2005vu, McNeile:2006qy, Ballon-Bayona:2014oma}.  The results in Ref.\,\cite{Xu:2022kng} meet this requirement, whereas such outcomes cannot be achieved in quark models without tuning parameters.  For this reason alone, the Ref.\,\cite{Xu:2022kng} decay constant predictions warrant testing.

Notwithstanding simplifications used in formulating the problem, Ref.\,\cite{Xu:2022kng} delivered
the first Poincar\'e-invariant analysis of the spectrum and decay constants of $u$, $d$, $s$ meson ground- and first-excited states.  The results include predictions for masses of as-yet unseen mesons and many unmeasured decay constants.  One may look forward to extensions of the approach to heavy+light mesons \cite{Binosi:2018rht, Chen:2019otg, Qin:2019oar}, and hybrid/exotic mesons \cite{Burden:2002ps, Qin:2011xq, Hilger:2015hka, Xu:2018cor} and glueballs \cite{Meyers:2012ka, Souza:2019ylx, Kaptari:2020qlt, Huber:2021yfy}.  These directions are especially important owing to worldwide investments in studies of the former and searches for the latter \cite{Denisov:2018unj, BESIII:2020nme, Anderle:2021wcy, AbdulKhalek:2021gbh, Pauli:2021gde, Quintans:2022utc}.

Such progress with meson properties should not obscure the need to calculate the spectrum of baryons.  Indeed, baryons are the most fundamental three-body systems in Nature; and if we do not understand how QCD, a Poincar\'e-invariant quantum field theory, structures the spectrum of baryons, then we don't understand Nature.  Within the context of the truncation scheme introduced in Refs.\,\cite{Munczek:1994zz, Bender:1996bb}, baryon masses and bound-state amplitudes have been calculated using a Poincar\'e-covariant Faddeev equation that describes a six-point Schwinger function for three-quark$\,\to\,$three-quark scattering.  The first solution of this problem for the nucleon ($N$) was presented in Ref.\,\cite{Eichmann:2009qa}; continuing studies are reviewed elsewhere \cite{Eichmann:2016yit, Brodsky:2020vco, Barabanov:2020jvn, Qin:2020rad}; and efforts are now underway to adapt the methods in Refs.\,\cite{Qin:2020jig, Xu:2022kng} to the formulation and solution of baryon Faddeev equations.

\begin{figure}[t]
\centerline{%
\includegraphics[clip, width=0.66\textwidth]{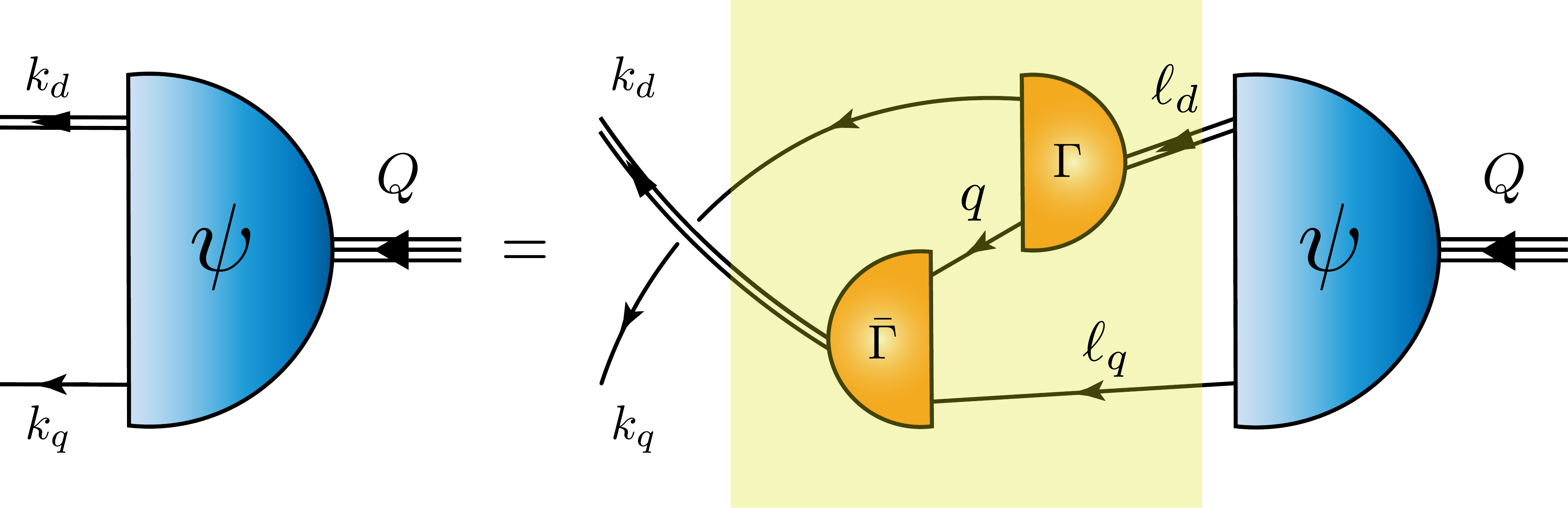}}
\caption{\label{FigFaddeev}
Figurative representation of the integral equation satisfied by the Poincar\'e-covariant matrix-valued function $\Psi$, \emph{viz}.\ the Faddeev amplitude for a baryon with total momentum $Q=\ell_q+\ell_d=k_q+k_d$ built from three valence quarks, two of which are always participants in a nonpointlike, interacting diquark correlation. $\Psi$ describes the sharing of relative momentum between the dressed-quarks and -diquarks.
\emph{Shaded rectangle} -- Faddeev kernel.  Legend: \emph{single line} -- dressed-quark propagator, $S(q)$; $\Gamma^{J^P}(k;K)$ -- diquark correlation amplitude; and \emph{double line} -- diquark propagator, $\Delta^{J^P}(K)$.
}
\end{figure}

Meanwhile, the quark$\,+$\,dynamical-diquark approach to baryon properties, introduced in Refs.\,\cite{Cahill:1988dx, Burden:1988dt, Reinhardt:1989rw, Efimov:1990uz}, is also being pursued vigorously.  This treatment begins with solutions of the equation illustrated in Fig.\,\ref{FigFaddeev}.  As sketched, \emph{e.g}., in Ref.\,\cite[Sec.\,5.1]{Maris:2003vk}, this is an approximation to the three-body Faddeev equation whose kernel is constructed using dressed-quark and nonpointlike diquark degrees-of-freedom.  Binding energy is lodged within the diquark correlation and also produced by the exchange of a dressed-quark, which, as drawn in Fig.\,\ref{FigFaddeev}, emerges in the break-up of one diquark and propagates to be absorbed into formation of another.  In the general case, five distinct diquark correlations are possible: isoscalar-scalar, $(I,J^P=0,0^+)$; isovector-axialvector; isoscalar-pseudoscalar; isoscalar-vector; and isovector-vector.  Channel dynamics within a given baryon determines the relative strengths of these correlations therein.

Given the extensive coverage of the role of diquark correlations in hadron structure presented in Ref.\,\cite{Barabanov:2020jvn}, herein we will only draw some recent highlights from analyses of the baryon spectrum using the Faddeev equation in Fig.\,\ref{FigFaddeev}, drawing largely from Ref.\,\cite{Yin:2021uom}.  That study was built upon a symmetry-preserving treatment of a vector$\times$vector contact interaction (SCI), which was introduced a little over a decade ago \cite{GutierrezGuerrero:2010md} and has since been employed with success in numerous applications, some of which are reviewed in this volume \cite{BashirParticles}.  Amongst the merits of the SCI are its
algebraic simplicity;
limited number of parameters;
simultaneous applicability to many systems and processes;
and potential for generating insights that connect and explain numerous phenomena.

\begin{figure*}[!t]
\begin{tabular}{lc}
{\sf A} & \\[-3ex]
& \hspace*{-2em}\includegraphics[clip, width=0.98\textwidth]{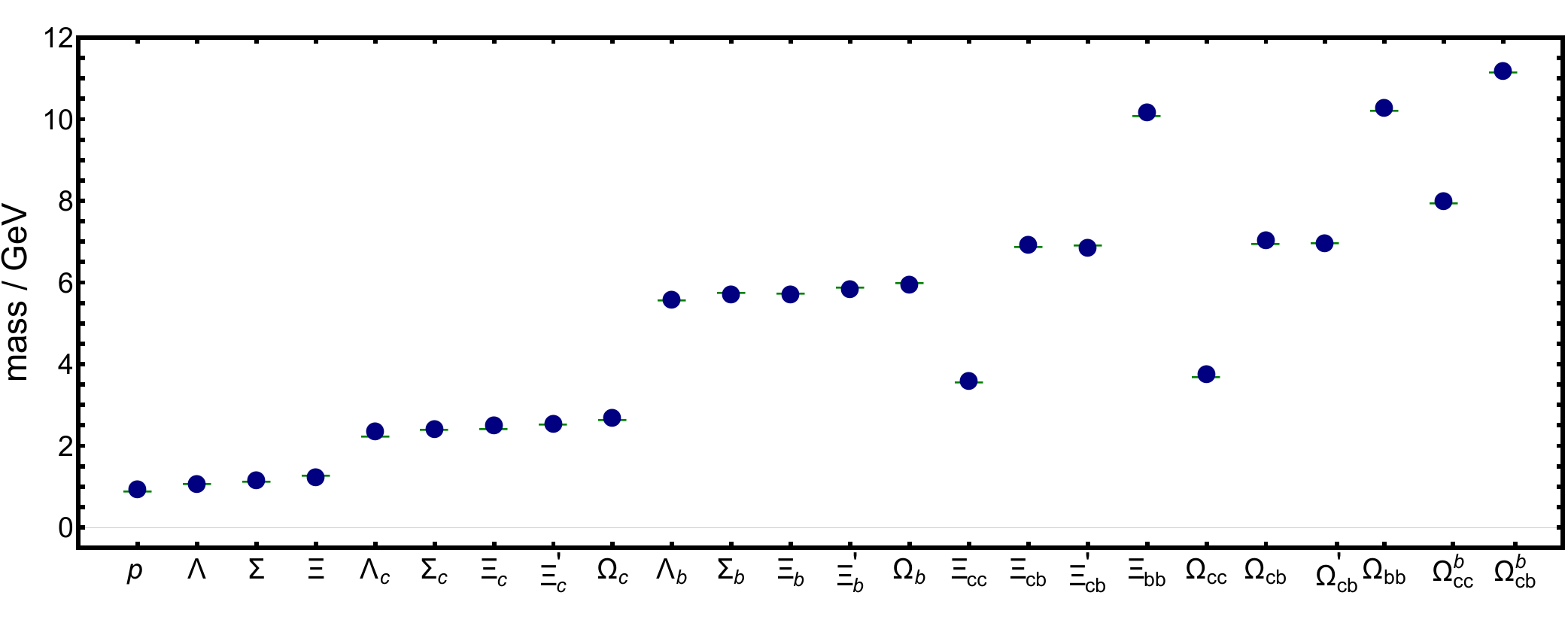} \\
{\sf B} & \\[-3ex]
& \hspace*{-2em}\includegraphics[clip, width=0.98\textwidth]{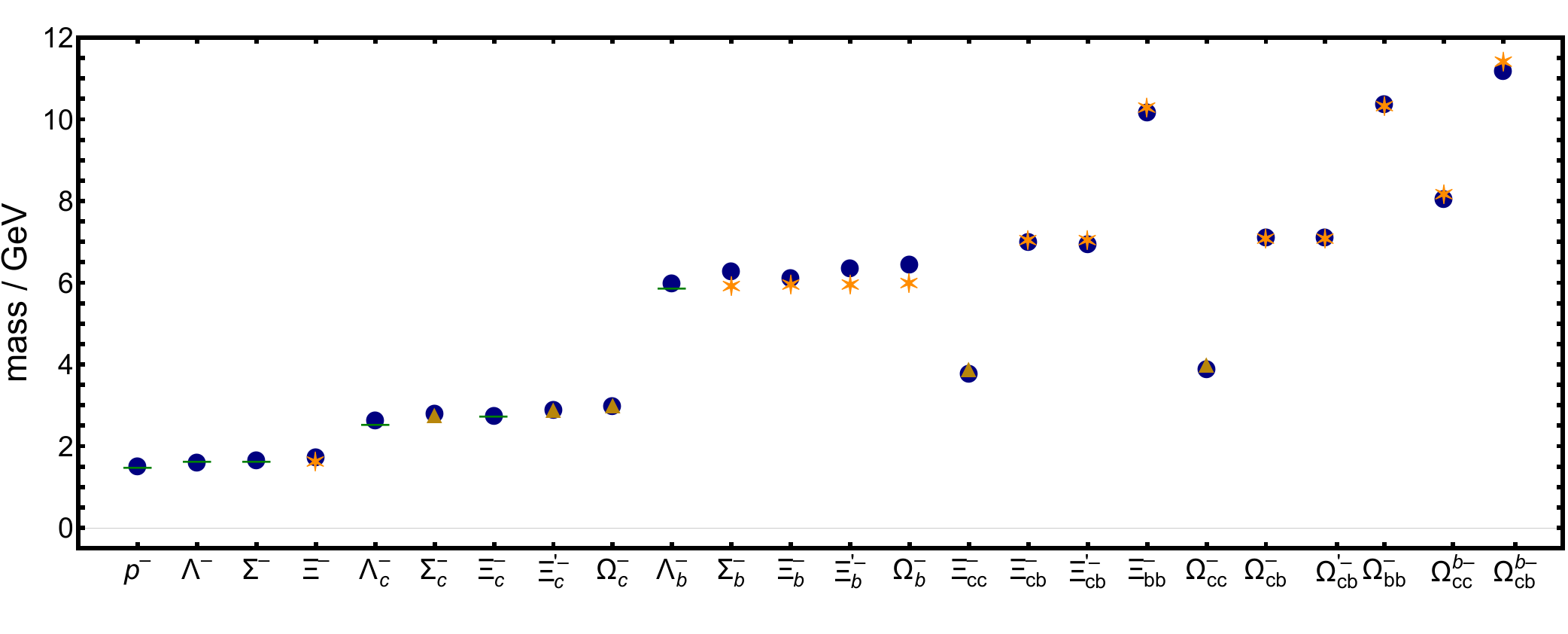}
\end{tabular}
\caption{\label{figResultsHalf}
\emph{Upper panel}\,--\,A.
SCI calculated masses of ground-state flavour-SU$(5)$ \mbox{$J^P=1/2^+$} baryons \cite{Yin:2021uom}
compared with either experiment (first 15) \cite{Workman:2022ynf} or lQCD (last 9) \cite{Brown:2014ena, Mathur:2018epb}.
\emph{Lower panel}\,--\,B.
SCI masses of ground-state flavour-SU$(5)$ $J^P=1/2^-$ baryons in \cite{Yin:2021uom}
compared with experiment \cite{Workman:2022ynf} (green bars), lQCD \cite{Bahtiyar:2020uuj} (gold triangles), or three-body Faddeev equation results \cite{Qin:2019hgk} (orange asterisks).
Analogous plots for $J^P=3/2^\pm$ baryons are presented elsewhere \cite[Figs.\,4B, 5B]{Yin:2021uom}.
}
\end{figure*}

Reference~\cite{Yin:2021uom} used the SCI to calculate ground-state masses of $J^P = 0^\pm, 1^\pm$ $(f\bar g)$ mesons and $J^P=1/2^\pm, 3/2^\pm$ $(fgh)$ baryons, where $f,g,h \in \{u,d,s,c,b\}$.  Using $J^P=\tfrac{1}{2}^\pm$ states as exemplars, Fig.\,\ref{figResultsHalf} highlights the level of quantitative accuracy.

Regarding the 33 mesons, then $\overline{\rm ard}=5(6)$\% when comparing SCI predictions with empirical masses.  In achieving this outcome, it was found that sound expressions of EHM were crucial.
Turning to baryons, the SCI generated 88 distinct bound states; namely, every possible three-quark $1/2^\pm, 3/2^\pm$ ground-state.  In this collection, 34 states have already been identified empirically and lQCD results are available for another 30: for these 64 states, comparing SCI prediction with experiment, where available, or lQCD mass otherwise, $\overline{\rm ard} = 1.4(1.2)$\%.  This level of agreement was only achieved through implementation of EHM-induced effects associated with spin-orbit repulsion in $1/2^-$ baryons.  Notably, the same 88 ground-states are also produced by a three-body Faddeev equation \cite{Qin:2019hgk}: in comparison with those results, $\overline{\rm ard} = 3.4(3.0)$\%.

Overall, Ref.\,\cite{Yin:2021uom} delivered SCI predictions for 164 distinct quantities, 114 of which have either been measured or calculated using lQCD: performing a comparison on this subset yields $\overline{\rm ard} = 4.5(7.1)$\%.
Such quantitative success means that credibility should be given to the qualitative conclusions that follow from the SCI analysis.  We list them here.
%
(\emph{i}) Nonpointlike, dynamical diquarks play a significant role in all baryons.  Usually, the lightest allowed diquark is the most important part of a baryon's Faddeev amplitude.  This remains true, even if the lightest correlation is a (sometimes called bad) axialvector diquark, and also for baryons containing one or more heavy quarks.  In the latter connection, it means one cannot safely assume that singly-heavy baryons may realistically be described as two-body light-diquark+heavy-quark ($qq^\prime+Q$) bound-states or that doubly-heavy baryons ($qQQ^\prime$) can be treated as two-body light-quark+heavy-diquark bound-states, $q + QQ^\prime$.   Corresponding statements apply to the treatment of tetra- and penta-quark problems.
(\emph{ii}) Positive-parity diquarks dominate in positive-parity baryons.  Axialvector diquarks are prominent in all states.
(\emph{iii}) Negative-parity diquarks play a minor role in positive-parity baryons.  On the other hand, owing to EHM, they are significant and sometimes dominant in $J=1/2^-$ baryons.
(\emph{iv}) Curiously, however, $J=3/2^-$ baryons are built (almost) exclusively from $J=1^+$ diquark correlations.
These conclusions are being checked using Faddeev equations with momentum-dependent exchange interactions; hence, a closer link to QCD.  Where results are already available, the SCI conclusions have been confirmed \cite{Chen:2017pse, Chen:2019fzn, Liu:2022ndb, Liu:2022nku}.

Following more than fifty years of hadron spectroscopy based on quark models, we are beginning to see real progress with the use of bound-state equations in quantum field theory.  Poincar\'e-invariant, symmetry-preserving analyses that reveal the expressions of EHM in hadron masses and level orderings are becoming available.  This increases the value of experimental hadron spectra measurements, making them a clearer window onto strong QCD.

\section{Baryon Wave Functions}
\label{SecBaryonWaveFunctions}
%
Concerning baryon structure, as noted when opening Sec.\,\ref{SecSpectroscopy}, quark models are still considered to provide a useful picture \cite{Capstick:2000qj, Crede:2013kia, Giannini:2015zia, Plessas:2015mpa}.  In such models, baryons built from combinations of $u$, $d$, and $s$ valence quarks are grouped into multiplets of SU$(6)\otimes$O$(3)$.
The multiplets are labelled by their flavour content -- SU$(3)$, spin -- SU$(2)$, and orbital angular momentum -- O$(3)$.  However, as has been emphasised, quark potential models do not have an explicit link with QCD, a Poincar\'e-invariant quantum gauge field theory.

For the lightest four $(I,J^P=\tfrac{1}{2},\tfrac{1}{2}^\pm)$ baryons, with $I$ denoting isospin, a comparison between quark model expectations and insights drawn from solutions of the Poincar\'e-covariant Faddeev equation is presented elsewhere \cite{Chen:2017pse}.  Herein, we will illustrate the qualitative character of such comparisons by considering more recent studies of $(\tfrac{3}{2},\tfrac{3}{2}^\pm)$, $(\tfrac{1}{2},\tfrac{3}{2}^\mp)$ baryons \cite{Liu:2022ndb, Liu:2022nku}.

These systems were studied in Ref.\,\cite{Eichmann:2016hgl} using RL truncation and direct calculations of all primary Schwinger functions.  With current algorithms, owing to singularities which enter the integration domains sampled by the Faddeev equations \cite{Maris:1997tm}, this approach limits the ability to compute wave functions because the on-shell point for many systems is inaccessible.  (Similar issues are encountered with meson structure studies -- Sec.\,\ref{SecSpectroscopy}.)
To circumvent this issue, Refs.\,\cite{Liu:2022ndb, Liu:2022nku} employed the QCD-kindred framework introduced elsewhere \cite{Alkofer:2004yf}, in which, instead of calculating all primary Schwinger functions, one uses physics-constrained algebraic representations of the Faddeev kernel elements.  This weakens the connection with QCD, but that loss is well compensated because, with reliably informed choices for the representation functions, the expedient enables access to on-shell baryon wave functions.  The QCD-kindred framework has widely been used with success -- see, \emph{e.g}., Refs.\,\cite{Segovia:2014aza, Chen:2018nsg, Lu:2019bjs, Chen:2020wuq, Chen:2021guo, ChenChen:2022qpy}.

Within the quark model framework and using standard spectroscopic notation, $n \,^{2S+1}\!L_J$, where $n$ is the radial quantum number with ``0'' labelling the ground state, the lightest four $(\tfrac{3}{2},\tfrac{3}{2}^\pm)$ $\Delta$-baryons, constructed from isospin $I=\tfrac{3}{2}$ combinations of three $u$ and/or $d$ quarks, are understood as follows \cite{Klempt:2012fy}:
\begin{enumerate}
\item $\Delta(1232)\tfrac{3}{2}^+$ \ldots\ $0 \,^{4} S_{\tfrac{3}{2}}={\mathsf S}$-wave ground-state;
\item $\Delta(1600)\tfrac{3}{2}^+$ \ldots\ $1 \,^{4} S_{\tfrac{3}{2}}= {\mathsf S}$-wave radial excitation of $\Delta(1232)\tfrac{3}{2}^+$;
%
\item $\Delta(1700)\tfrac{3}{2}^-$ \ldots\ $0 \,^{2} P_{\tfrac{3}{2}}= {\mathsf P}$-wave orbital angular momentum excitation of $\Delta(1232)\tfrac{3}{2}^+$;
%
\item $\Delta(1940)\tfrac{3}{2}^-$ \ldots\ $1 \,^{4} P_{\tfrac{3}{2}}= {\mathsf P}$-wave excitation of $\Delta(1600)\tfrac{3}{2}^+$.  \label{Delta1940}
\end{enumerate}
Analogously, the $(\tfrac{1}{2},\tfrac{3}{2}^\mp)$ states are interpreted thus \cite{Klempt:2012fy}:
\begin{enumerate}
\item $N(1520)\tfrac{3}{2}^-$ \ldots\ $0 \,^{2} P_{\tfrac{1}{2}}={\mathsf P}$-wave ground-state in this channel and an angular-momentum coupling partner of $N(1535)\tfrac{1}{2}^-$;
\item $N(1700)\tfrac{3}{2}^-$ \ldots\ $0\,^{4} P_{\tfrac{3}{2}}= {\mathsf P}$-wave angular-momentum coupling partner of $N(1520)\tfrac{3}{2}^-$;
%
\item $N(1720)\tfrac{3}{2}^+$ \ldots\ $0 \,^{2} D_{\tfrac{3}{2}}= {\mathsf D}$-wave orbital angular momentum excitation of $N(1520)\tfrac{3}{2}^-$;
%
\item $N(1900)\tfrac{3}{2}^+$ \ldots\ $0 \,^{4} D_{\tfrac{3}{2}}= {\mathsf D}$-wave orbital angular momentum excitation of $N(1700)\tfrac{3}{2}^-$.
\end{enumerate}

On the other hand, Poincar\'e-invariant quantum field theory does not readily admit such assignments.  Instead, the states appear as poles in the six-point Schwinger functions associated with the given $(I,J^P)$ channels.  Here, ``$(1\leftrightarrow 3)$'' and ``$(2\leftrightarrow 4)'$' in each block above are related as parity partners.  All differences between positive- and negative-parity states can be attributed to chiral symmetry breaking in quantum field theory.  This is highlighted by the $\rho$-$a_1$ meson complex \cite{Weinberg:1967kj, Chang:2011ei, Williams:2015cvx, Qin:2020jig, Xu:2022kng}.  Regarding light-quark hadrons, such symmetry breaking is almost exclusively dynamical \cite{Lane:1974he, Politzer:1976tv, Pagels:1978ba, Higashijima:1983gx, Pagels:1979hd, Cahill:1985mh, Bashir:2012fs}.  As noted above, DCSB is a corollary of EHM \cite{Roberts:2020udq, Roberts:2020hiw, Roberts:2021xnz, Roberts:2021nhw, Binosi:2022djx, Papavassiliou:2022wrb}; hence, quite probably linked tightly with confinement -- Sec.\,\ref{SecConfinement}.  These features imbue quantum field theory analyses of $(\tfrac{3}{2},\tfrac{3}{2}^\pm)$, $(\tfrac{1}{2},\tfrac{3}{2}^\mp)$  baryons with particular interest; consequently, experiments that test predictions made for structural differences between parity partners are highly desirable.

Working with the Faddeev equation sketched in Fig.\,\ref{FigFaddeev}, then, \emph{a priori}, the $(\tfrac{3}{2},\tfrac{3}{2}^\pm)$ baryons are the simpler systems because they can only contain isovector-axialvector and isovector-vector diquarks, whereas the $(\tfrac{1}{2},\tfrac{3}{2}^\mp)$ systems may involve all five distinct types of diquarks: $(0,0^+)$, $(1,1^+)$, $(0,0^-)$, $(1,1^-)$, $(0,1^-)$.
Nonetheless, the formulation of the bound-state problems in both channels is practically identical, using the same dressed-quark and -diquark propagators; diquark correlation amplitudes; \emph{etc}.  This way, one guarantees a unified description of all states in the spectrum.  The propagators are parametrised using entire functions \cite{Munczek:1986sa, Efimov:1988yd, Burden:1991gd}; hence, satisfy the confinement constraints described in Sec.\,\ref{SecConfinement}.  It is this feature which enables on-shell calculations for all baryons.

\begin{figure*}[!t]
\centerline{%
\includegraphics[clip, width=0.98\textwidth]{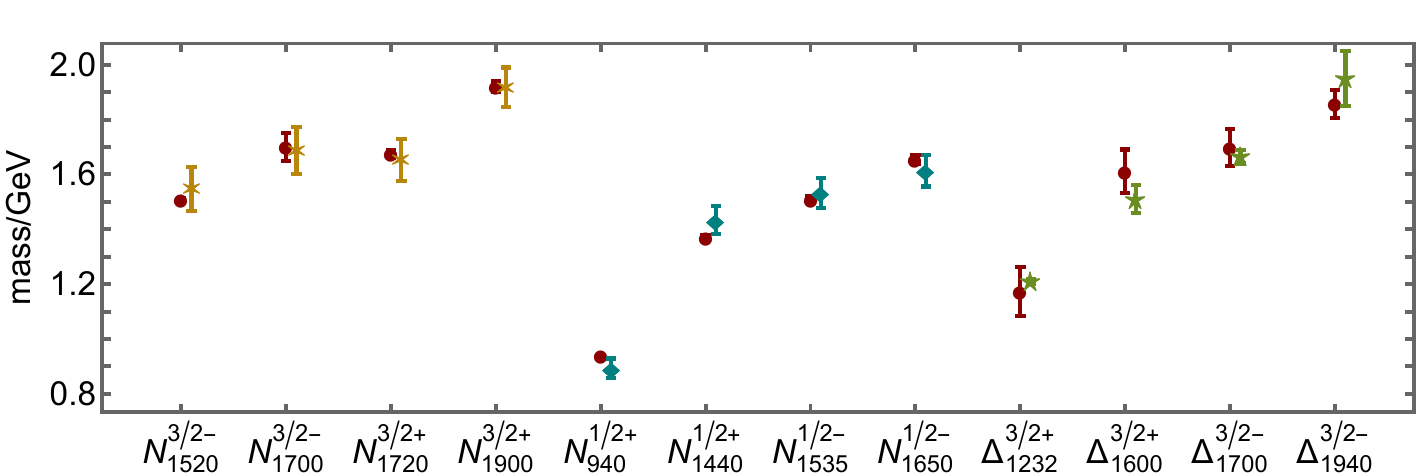}}
\caption{\label{MassCompare}
Real part of empirical pole position for each designated baryon \cite{Workman:2022ynf} (red circle) compared with:
calculated masses in Ref.\,\cite{Liu:2022nku} (gold asterisks) after subtracting $\delta^{N_{3/2}}_{\rm MB}=0.13\,$GeV from each;
Ref.\,\cite{Chen:2017pse} (teal diamonds) after subtracting $\delta^{N_{1/2}}_{\rm MB}=0.30\,$GeV;
and Ref.\,\cite{Liu:2022ndb} (green five-pointed stars) after subtracting $\delta^{\Delta_{3/2}}_{\rm MB}=0.17\,$GeV.
All Faddeev equation predictions are drawn with an uncertainty that reflects a $\pm 5$\% change in diquark masses.
}
\end{figure*}

\begin{figure*}[!t]
\hspace*{-1ex}\begin{tabular}{lcl}
\large{\textsf{A}} & & \large{\textsf{B}}\\[-0.7ex]
%
\includegraphics[clip, width=0.44\textwidth]{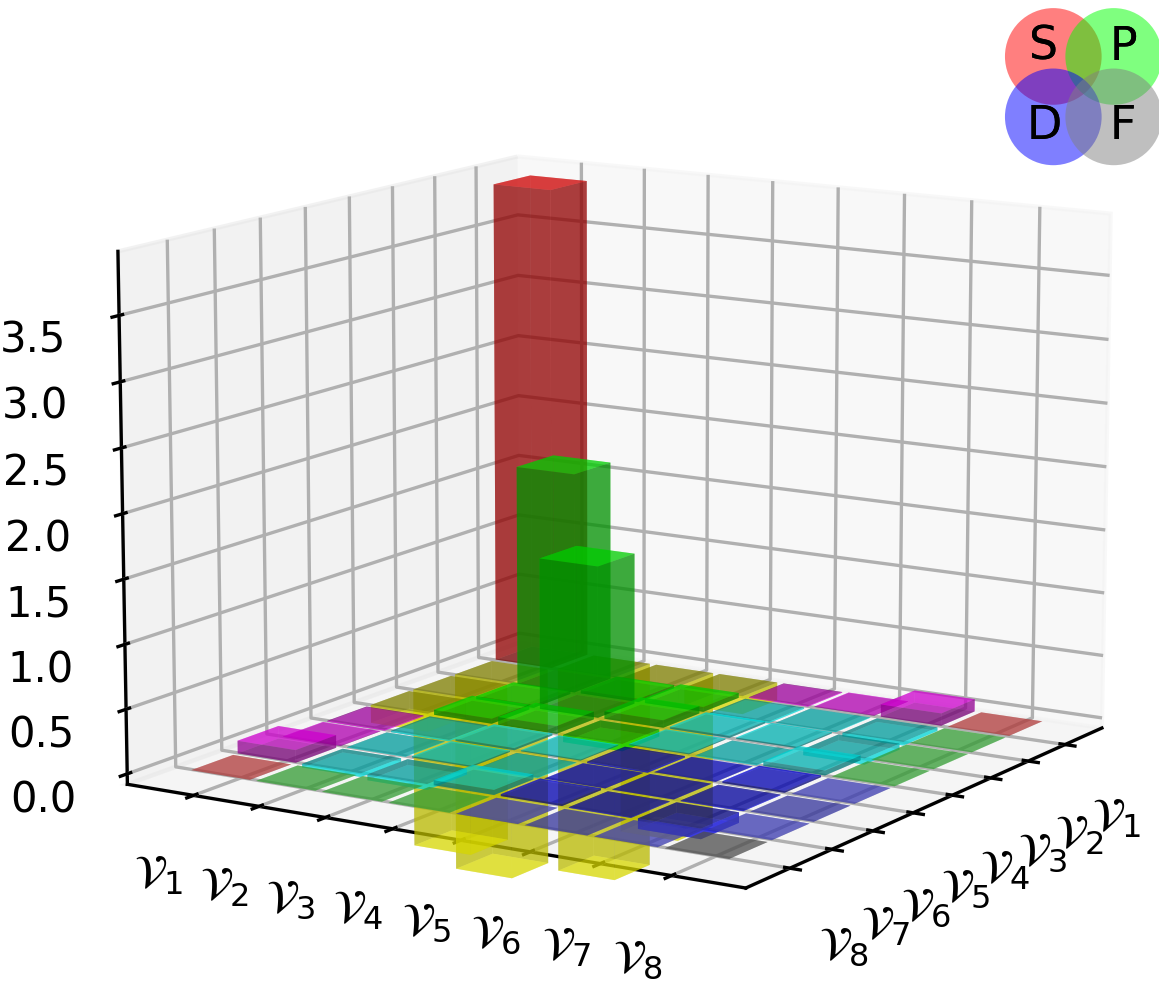} & \hspace*{1.8em} &
\includegraphics[clip, width=0.44\textwidth]{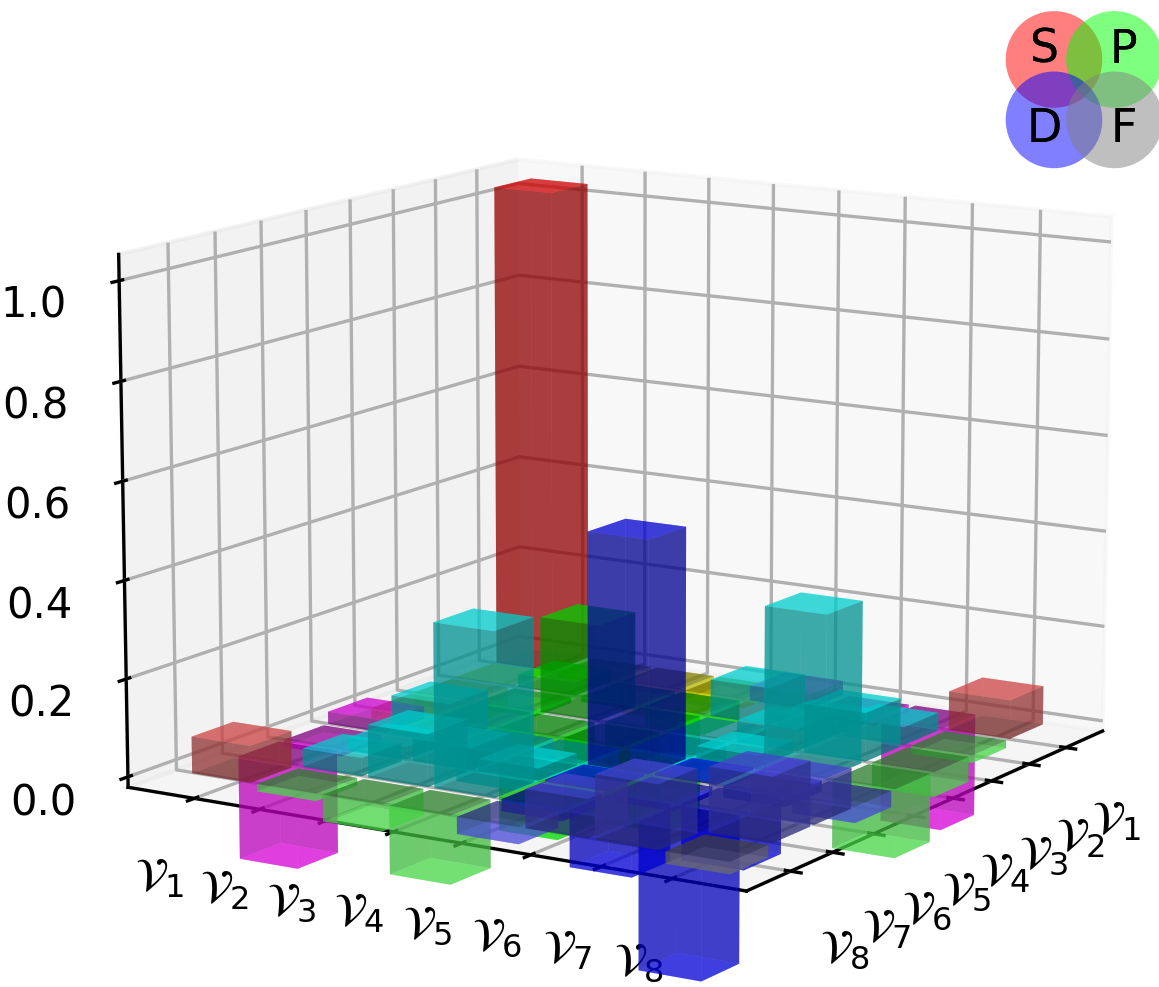} \\[1ex]
& \hspace*{-18ex} \includegraphics[clip, width=0.395\textwidth]{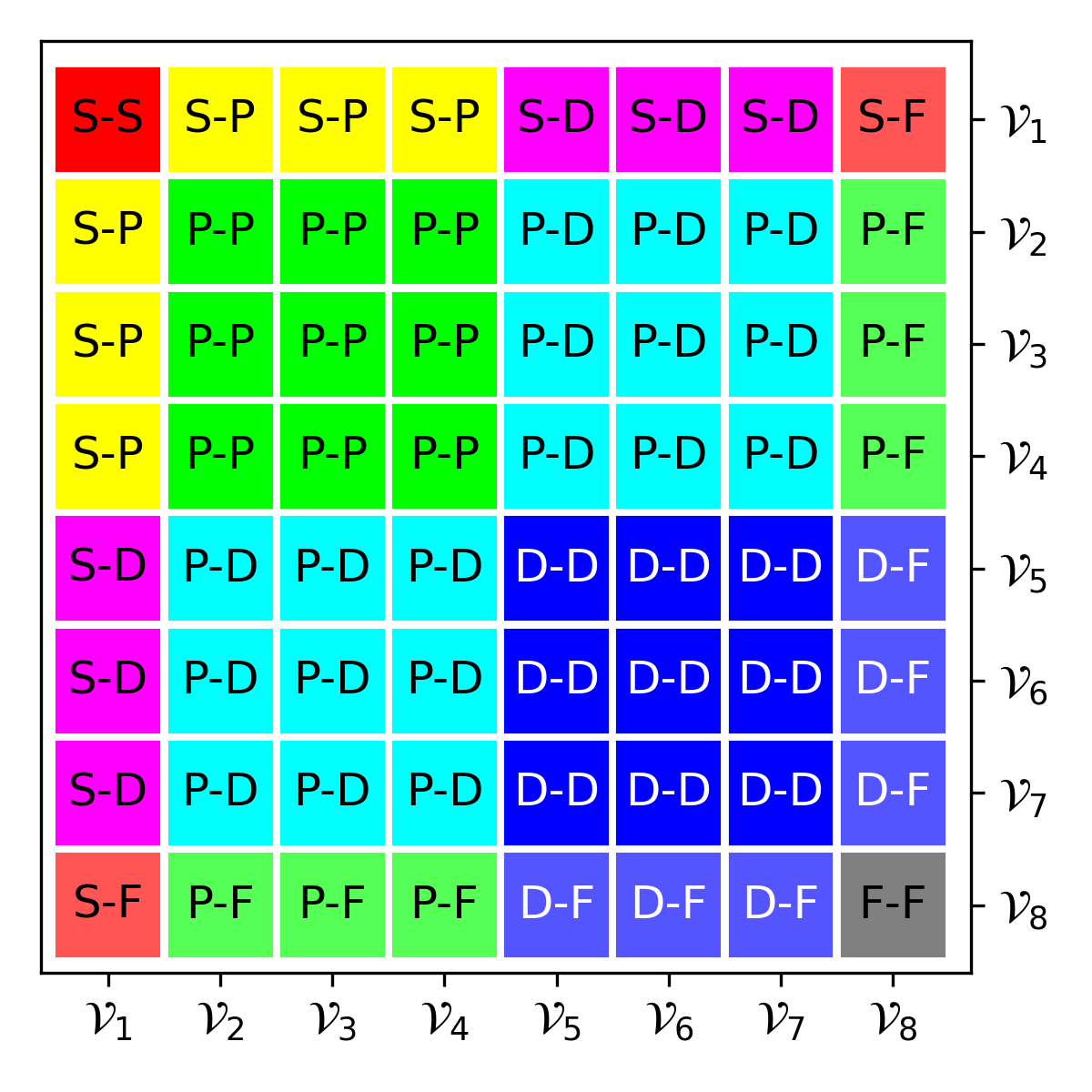} \hspace*{-18ex}  & \\[1ex]
\large{\textsf{C}} & & \large{\textsf{D}}\\[-0.7ex]
\includegraphics[clip, width=0.44\textwidth]{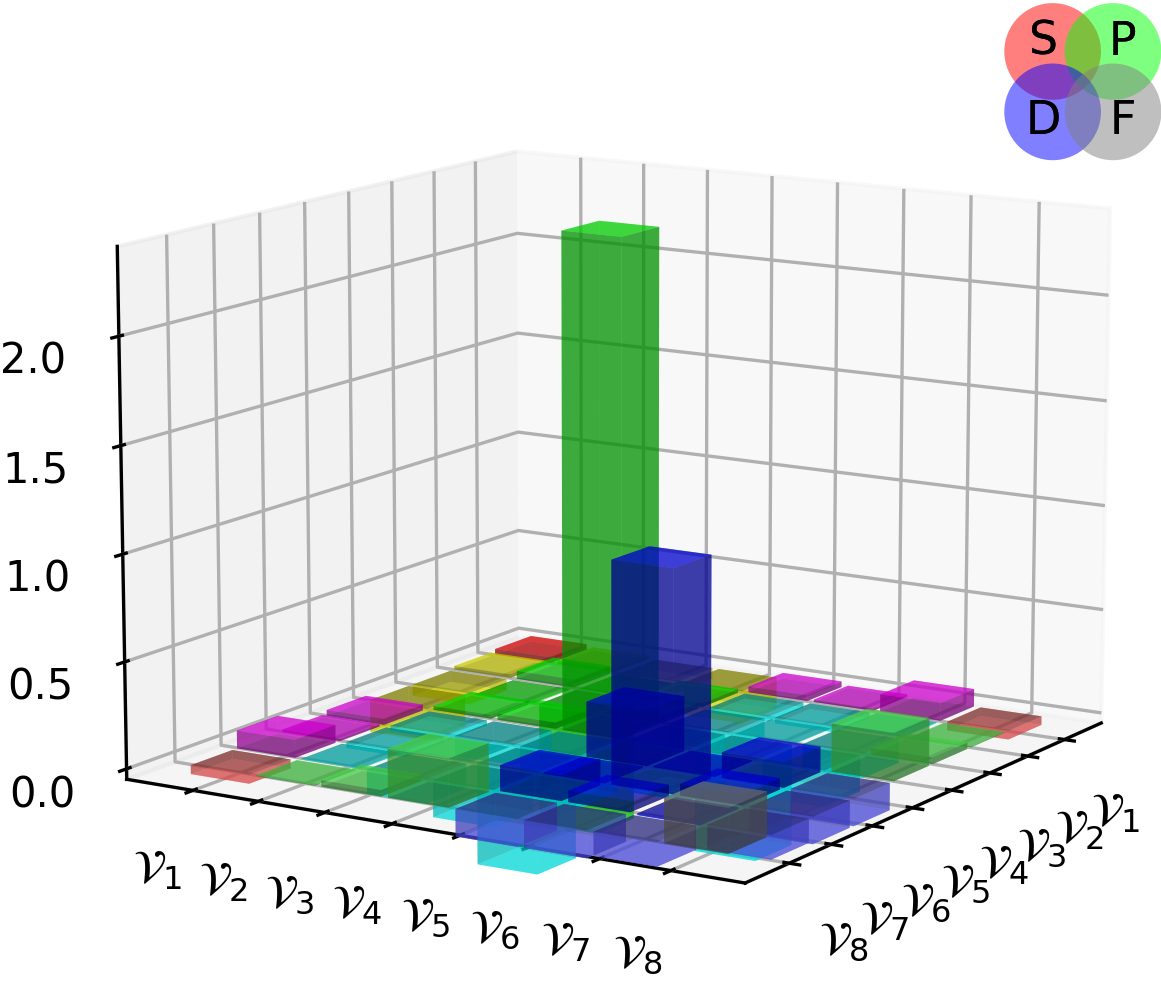} & \hspace*{1.8em} &
\includegraphics[clip, width=0.44\textwidth]{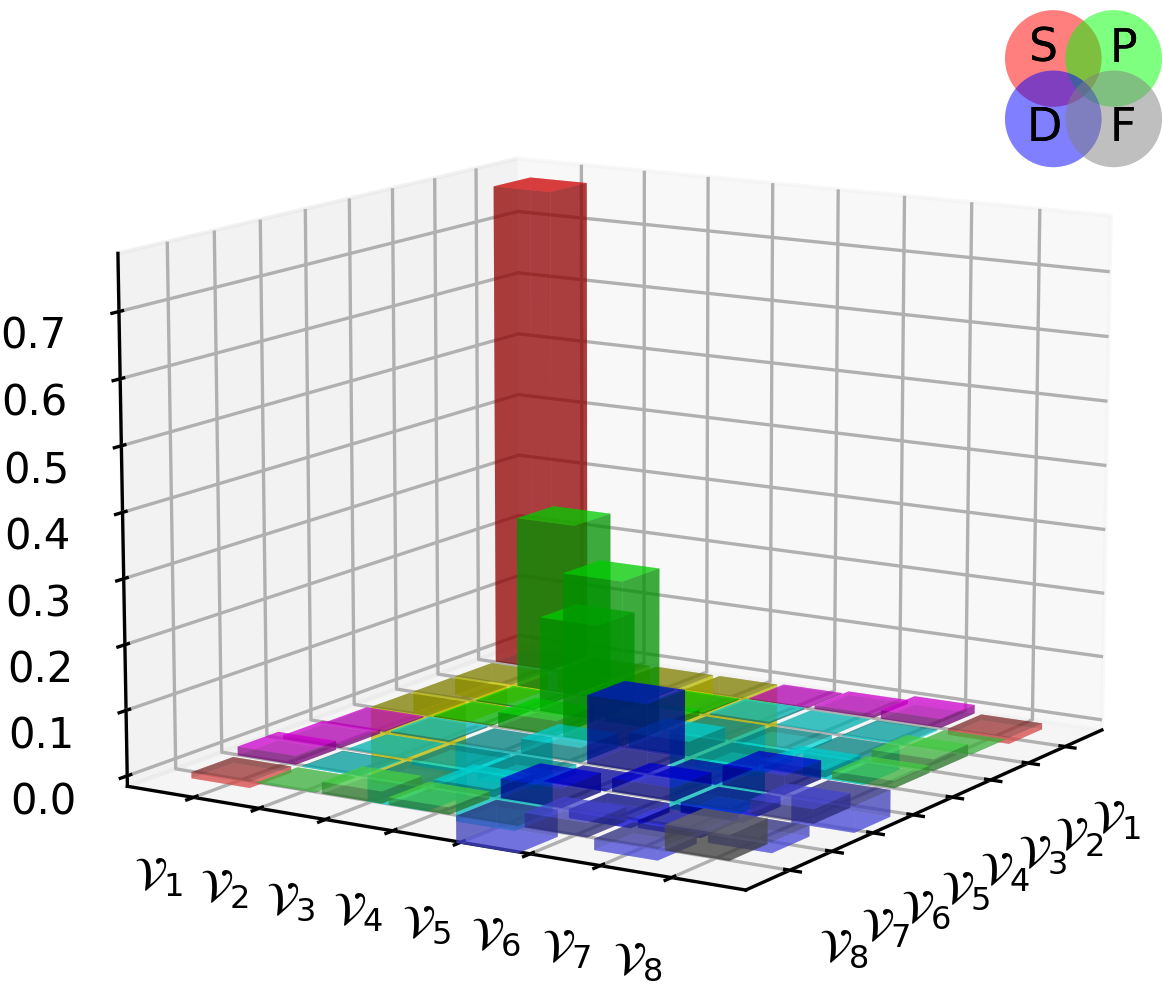} \\
\end{tabular}
\caption{\label{LFigures}
Rest frame quark+axialvector-diquark orbital angular momentum content of $(\tfrac{3}{2},\tfrac{3}{2}^\pm)$ baryons considered in Ref.\,\cite{Liu:2022ndb}, as measured by the contribution from the various components to the associated canonical normalisation constant:
{A} -- $ \Delta(1232)\tfrac{3}{2}^+$;
{B} -- $ \Delta(1600)\tfrac{3}{2}^+$;
{C} -- $ \Delta(1700)\tfrac{3}{2}^-$; and
{D} -- $ \Delta(1940)\tfrac{3}{2}^-$.
The overall-positive normalisations receive both positive (above plane) and negative (below plane) contributions.
The central image provides the legend for interpretation of the other panels, identifying interference between the various identified orbital angular momentum basis components in the baryon rest frame.
}
\end{figure*}

The calculated spectrum of states is displayed in Fig.\,\ref{MassCompare}.  As highlighted elsewhere \cite{Hecht:2002ej, Sanchis-Alepuz:2014wea, Chen:2017pse, Burkert:2019bhp, Liu:2022ndb, Liu:2022nku}, the kernel in Fig.\,\ref{FigFaddeev} does not include contributions that may be understood as meson-baryon final-state interactions.  These are the interactions which transform a bare-baryon into the observed state, \emph{e.g}., via dynamical coupled channels calculations \cite{JuliaDiaz:2007kz, Suzuki:2009nj, Ronchen:2012eg, Kamano:2013iva, Garcia-Tecocoatzi:2016rcj}.  The Faddeev amplitudes and masses calculated in Refs.\,\cite{Chen:2017pse, Liu:2022ndb, Liu:2022nku} should therefore be seen as describing the \emph{dressed-quark core} of the bound-state, not the fully-dressed, observable object \cite{Eichmann:2008ae, Eichmann:2008ef, Roberts:2011cf}.  That explains why the masses are uniformly too large.  Evidently and importantly, in each sector, a single subtraction constant is sufficient to realign the masses and produce a good description of the spectrum.

Regarding $(\tfrac{3}{2},\tfrac{3}{2}^\pm)$ baryons, Ref.\,\cite{Liu:2022ndb} found that although these states may contain both $(1,1^+)$ and $(1,1^-)$ quark+quark correlations, one can neglect the $(1,1^-)$ diquarks because they have practically no impact on the masses or wave functions.  After this simplification, the Poincar\'e-covariant wave functions involve eight independent matrix-valued terms, each multiplied by a scalar function of two variables: $(k^2,k\cdot Q)$, with $k$ the quark+diquark relative momentum.  Studying the properties of these functions, one may conclude that the $\Delta(1600)\tfrac{3}{2}^+$ is fairly interpreted as a radial excitation of the $\Delta(1232)\tfrac{3}{2}^+$, as suggested by the quark model.  However, the wave functions of the $\Delta(1700)\tfrac{3}{2}^-$, $\Delta(1940)\tfrac{3}{2}^-$ states are complicated and do not readily admit direct analogies with quark model pictures.

Projecting the Poincar\'e-covariant Faddeev wave functions of $(\tfrac{3}{2},\tfrac{3}{2}^\pm)$ baryons into their respective rest frames, one arrives at a $J=L+S$ separation which is comparable to that associated with quark models.  (Here $L$ is quark-diquark orbital angular momentum.)  Following this procedure, Ref.\,\cite{Liu:2022ndb} found that the angular momentum structure of all these states is much more complicated than is typically generated in quark models -- see Fig.\,\ref{LFigures}.

Evidently, making a link to quark models, the $\Delta(1232)\tfrac{3}{2}^+$ and $\Delta(1600)\tfrac{3}{2}^+$ are characterised by a dominant $\mathsf S$-wave component, and the $\Delta(1700)\tfrac{3}{2}^-$ by a prominent $\mathsf P$-wave.  The $\Delta(1940)\tfrac{3}{2}^-$, however, does not fit this picture: contrary to quark model expectations, indicated on page~\pageref{Delta1940}, this state is $\mathsf S$-wave dominated.  Moreover, each state contains every admissible partial wave.

Combining all gathered information, Ref.\,\cite{Liu:2022ndb} furthermore concluded that the negative parity $\Delta$-baryons are not merely orbital angular momentum excitations of positive parity ground states.  In this observation, the results match those obtained earlier for $(\tfrac{1}{2},\tfrac{1}{2}^\pm)$ baryons \cite{Chen:2017pse}.

Recalling now that the interpolating fields for positive and negative parity hadrons are related by chiral rotation of the quark spinors used in their construction, then the highlighted structural differences are largely generated by DCSB.
Regarding the $\Delta(1940)\tfrac{3}{2}^-$ in particular, these novel structural predictions may be expected to encourage new experimental efforts aimed at extracting reliable information about this little understood state from exclusive $\pi^+\pi^- p$ electroproduction data \cite{CLAS:2017fja, Trivedi:2018rgo}.

Observations upon similar features were made about $(\tfrac{1}{2},\tfrac{3}{2}^\pm)$ baryons in Ref.\,\cite{Liu:2022nku}.  To begin, despite the fact that such states may contain all possible diquark correlations, the analyses showed that a good approximation is obtained by keeping only $(0,0^+)$, $(1,1^+)$ correlations.  This runs counter to the nature of $(\tfrac{1}{2},\tfrac{1}{2}^-)$ systems, in which $(0,0^-)$, $(0,1^-)$ diquarks are also important \cite{Chen:2017pse, Raya:2021pyr}.
Projecting the Poincar\'e-covariant Faddeev wave functions into the baryon rest-frames and considering the baryon mass-fraction contributed by each partial wave, this collection of states form a set related via orbital angular momentum excitation: as in quark models, the parity-negative states are primarily $\mathsf P$-wave in nature whereas the parity-positive states are $\mathsf D$ wave -- see Ref.\,\cite[Fig.\,4]{Liu:2022nku}.  However, looking with finer resolution, using charts of the canonical normalisation constant contributions from the various partial waves in the Poincar\'e-covariant wave functions, alike with Fig.\,\ref{LFigures} herein, far greater $L$-complexity was observed than is usually found in quark models -- see Ref.\,\cite[Fig.\,7]{Liu:2022nku}.  Here, too, one may anticipate that these structural predictions can be tested using data from measurements of resonance electroexcitation at large momentum transfers.  For the $N(1520)\tfrac{3}{2}^-$, data are already available \cite{CLAS:2009ces, Mokeev:2015lda, Mokeev:2020vab, Carman:2020qmb, Mokeev:2021dab, Mokeev:2022xfo} and calculations of the electroproduction form factors are underway.  Large-$Q^2$ data on the other states is not available; so, the predictions in Ref.\,\cite{{Liu:2022nku}} will also encourage new experimental efforts in this area.

Parity partner channels are identical when chiral symmetry is restored \cite{Roberts:2000aa, Fischer:2018sdj}.  It is therefore interesting to note that the mass splitting between partner states does not exhibit a simple pattern, \emph{viz}.\ empirically \cite{Workman:2022ynf}:
{\allowdisplaybreaks
\begin{equation}
\begin{array}{lc}
{\rm states} & {\rm mass\;splitting/GeV}\\
N(1535)\tfrac{1}{2}^- - N(940)\tfrac{1}{2}^+ & 0.57\,, \\
N(1650)\tfrac{1}{2}^- - N(1440)\tfrac{1}{2}^+& 0.29\,,\\
\Delta(1700)\tfrac{3}{2}^- - \Delta(1232)\tfrac{3}{2}^+ & 0.46 \,,\\
\Delta(1940)\tfrac{3}{2}^- - \Delta(1600)\tfrac{3}{2}^+ & 0.44 \,, \\
N(1720)\tfrac{3}{2}^+ - N(1520)\tfrac{3}{2}^- & 0.17 \,, \\
N(1900)\tfrac{3}{2}^+ - N(1700)\tfrac{3}{2}^- & 0.22 \,. \\
\end{array}
\end{equation}
This system dependence of the mass splitting is also linked to the deeper structural differences between these states that are expressed in their complex wave functions.

Using a familiar quantum mechanics framework, quark models produce baryon wave functions that have an appealing simplicity.  However, far richer structures are found when quantum field theory is used to solve baryon bound-state problems.  The growing body of quantum field theory predictions can be tested, \emph{e.g}., in modern and future large $Q^2$ measurements of baryon elastic and transition form factors.  In fact, the large $Q^2$ character of such experiments is alone sufficient to demand the sort of Poincar\'e-invariant, symmetry-preserving treatment that only analyses in quantum field theory can deliver.  One may therefore expect studies using the Faddeev equation approach outlined in this section to become steadily more widespread.

\section{Meson Form Factors}
\label{SecFormFactors}
The truncation scheme explained in Refs.\,\cite{Munczek:1994zz, Bender:1996bb} has been used to calculate many meson elastic and transition form factors \cite{Maris:2003vk}; and modern algorithms have enabled predictions to be delivered on the entire domain of spacelike $Q^2$ \cite{Chang:2013nia, Raya:2015gva, Raya:2016yuj, Gao:2017mmp, Chen:2018rwz, Ding:2018xwy}, making it possible to draw connections with hard scattering formulae derived using QCD perturbation theory \cite{Lepage:1979zb, Efremov:1979qk, Lepage:1980fj}.  These new predictions, which unify the infrared and ultraviolet $Q^2$ domains, are providing the impetus for measurements at new generation high-energy and high-luminosity facilities \cite{Aguilar:2019teb, Chen:2020ijn, Roberts:2021nhw, Arrington:2021biu}.  Recalling that QCD is not found in form factor scaling but in scaling violations, then the goal of these new experiments is, of course, to discover the breakaway from scaling in a hard exclusive process and thus reveal the hand of QCD.

In connection with these new facilities, it has been argued that the interaction of a heavy vector-meson, $V=J/\psi, \Upsilon$, with a proton, $p$, may provide access to a QCD van der Waals interaction, produced by multiple gluon exchange \cite{Brodsky:1989jd, TarrusCastella:2018php}, and/or the QCD trace anomaly \cite{Luke:1992tm, Kharzeev:1995ij, Roberts:2016vyn, Krein:2020yor}.  The van der Waals interaction is of interest because it might relate to, amongst other things, the observation of hidden-charm pentaquark states \cite{Aaij:2015tga}; and the trace anomaly is topical because of its connection with EHM.

\begin{figure}[t] 
\centerline{\includegraphics[clip,width=0.45\textwidth]{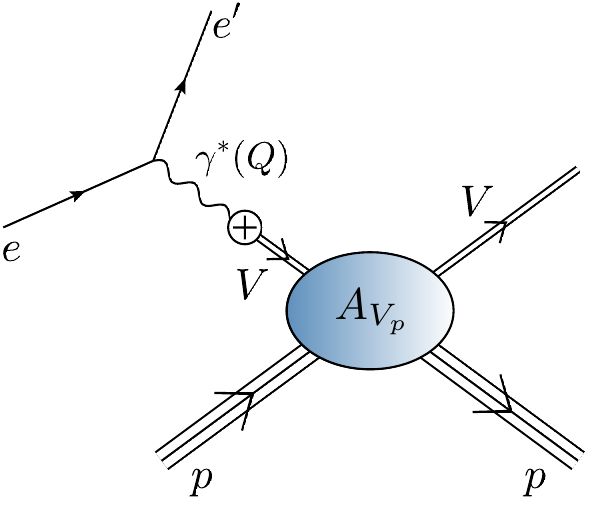}}
\caption{\label{FigVMD} Electroproduction of a vector-meson from the proton: $e + p \to e^\prime + V + p$, often interpreted as providing access to $V + p \to V+p$ using a vector meson dominance model.  The $\gamma^{(\ast)}(Q)\to V$ VMD transition is indicated by the crossed-circle.  It is usually assumed to occur with a momentum-independent strength $\gamma_{\gamma V}$ \cite{Sakurai:1960ju, sakurai:1969, Fraas:1970vj}.
However, regarding heavy mesons, at least, the VMD hypothesis is unsound, as discussed below and in, \emph{e.g}.,  Refs.\,\cite{Xu:2021mju, Sun:2021pyw}.
%
}
\end{figure}

Lacking vector-meson beams, ongoing and anticipated experiments at electron ($e$) accelerators are based on an expectation that the desired $V+p$ interactions can be accessed through the electromagnetic production of vector-mesons from the proton, in reactions like $e + p \to e^\prime + V + p$ \cite{Ali:2019lzf, Anderle:2021wcy, AbdulKhalek:2021gbh}.  This is because some practitioners imagine that single-pole vector meson dominance (VMD) \cite{Sakurai:1960ju, sakurai:1969, Fraas:1970vj} can reliably be employed to draw a clean link between $e + p \to e^\prime + V + p$ and the desired $V p\to V p$ cross-sections.  In this picture -- see Fig.\,\ref{FigVMD}, the interaction is supposed to proceed via the following sequence of steps:
(\emph{i}) $e \to e^\prime + \gamma^{(\ast)}(Q)$;
(\emph{ii}) $\gamma^{(\ast)}(Q)\to V$;
and (\emph{iii}) $V+ p \to V+p$.  $\gamma^{(\ast)}(Q)$ is a virtual photon and step (\emph{ii}) expresses the VMD hypothesis.
As commonly used, VMD assumes: (\emph{a}) that a photon, which is generally spacelike, so that $Q^2 > 0$, transforms into an on-shell vector-meson, with timelike momentum $Q^2=-m_V^2$;
and (\emph{b}) that the $Q^2 > 0$ strength and form of the transition in (\emph{ii}) is the same as that measured in the real vector meson decay process, $V \to \gamma^\ast(Q^2=-m_V^2) \to e^+ + e^- $.
Property (\emph{b}) means that $\gamma_{\gamma V}$, the associated decay constant, is fixed at its meson on-shell value and acquires no momentum dependence:
\begin{equation}
\label{EqCFI}
\gamma_{\gamma V}^2 = 4\pi \alpha_{\rm em} m_V^2 f_V^2 \,,
\end{equation}
where $\alpha_{\rm em}$ is the QED fine-structure constant, $m_V$ is the vector meson mass, and $f_V$ measures the strength of the meson's Bethe-Salpeter wave function at the origin in configuration space \cite[Sec.\,IIB]{Ivanov:1998ms}.

The fidelity of these VMD assumptions was recently subjected to scrutiny via analyses of the photon vacuum polarisation and photon-quark vertex \cite{Xu:2021mju}.
Regarding the photon vacuum polarisation, it was shown that there is no vector-meson contribution to this polarisation at the photoproduction point, $Q^2=0$.  Consequently, massless real photons cannot readily be linked with massive vector-bosons and the current-field identity, Eq.\,\eqref{EqCFI}, typical of VMD implementations, should not be used literally because it entails violations of Ward-Green-Takahashi identities (breaking of symmetries) in QED.

\begin{figure}[t] 
\hspace*{-1ex}\begin{tabular}{lcl}
\large{\textsf{A}} & & \large{\textsf{B}}\\[-0.7ex]
%
\includegraphics[clip, width=0.47\textwidth]{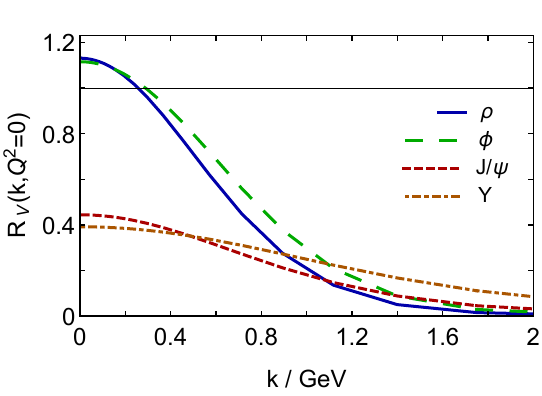} & \hspace*{0.em} &
\includegraphics[clip, width=0.47\textwidth]{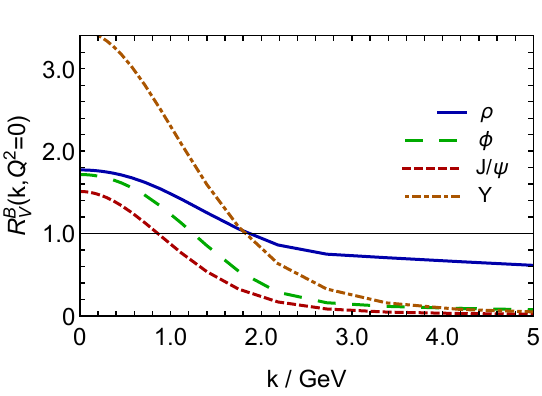}
\end{tabular}
\vspace*{2ex}
\caption{\label{FigRVk2Q20}
\emph{Left panel}\,--\,{A}: First ratio in Eq.\,\eqref{RatioVMD} computed using matched solutions of the gap and Bethe-Salpeter equations for $V=\rho, \phi, J/\psi, \Upsilon$.
\emph{Right panel}\,--\,{B}: Second ratio in Eq.\,\eqref{RatioVMD}, computed similarly.
In cases where the VMD hypothesis were sound, all these curves would lie close to the thin horizontal line drawn at unity.
}
\end{figure}

Turning to the dressed photon-quark vertex, $\Gamma_\nu^\gamma(k;Q)$, this three-point Schwinger function exhibits a pole at the mass of any vector-meson bound-state.  A physical property, it expresses the fact that the decay $V\to e^+ e^-$ proceeds via a timelike virtual photon.  The VMD hypothesis may thus be seen as a claim that $\left.\Gamma_\nu^\gamma(k;Q)\right|_{Q^2\simeq 0}$ maintains a rigorous link, in both magnitude and momentum-dependence, with the Bethe-Salpeter amplitude of an on-shell vector-meson.  However, as shown in Fig.\,\ref{FigRVk2Q20}, that is not the case.  The panels in Fig.\,\ref{FigRVk2Q20} depict the ratios
%
\begin{equation}
\label{RatioVMD}
R_V(k^2;Q^2=0) = \frac{2 f_V}{m_V} \frac{F_1^0(k^2;-m_V^2)}{G_1^0(k^2; Q^2=0)}\,,
\quad
R_V^B(k^2;Q^2=0)  = \frac{f_V}{m_V} \frac{F_8^0(k^2;-m_V^2)}{G_3^0(k^2; Q^2=0)}\,,
\end{equation}
where
$G_1^0$ is the zeroth Chebyshev moment\footnote{The Chebyshev (or hyperspherical) expansion of Poincar\'e-invariant functions of two scalar variables is discussed, \emph{e.g}., in Ref.\,\cite[IV.B]{Maris:1997tm}.} of the dominant amplitude in the photon-quark vertex, associated with the matrix structure $\gamma\cdot \epsilon(Q)$, where $\epsilon(Q)$ is the photon polarisation vector -- in fact, using the vector Ward-Green-Takahashi identity and Eq.\,\eqref{gendseN}, $G_1^0(k^2; Q^2=0) = A_g(k^2)$, where $g$ is the flavour of the meson's valence quark;
$F_1^0$ is its analogue in the vector meson bound-state amplitude;
$G_3^0$ is the zeroth moment of that term in the photon-quark vertex which is directly linked to the scalar piece of the dressed-quark self-energy via the vector Ward-Green-Takahashi identity, \emph{i.e}., $G_3^0(k^2; Q^2=0)=-2 B_g^\prime(k^2)$;
and $F_8^0$ is its analogue in the vector meson bound-state amplitude.
Were VMD to be a sound assumption, then all these curves would lie near to the thin horizontal line drawn at unity in both panels of Fig.\,\ref{FigRVk2Q20}.  However, whilst one might discuss the case for lighter vector mesons, the VMD hypothesis is plainly false for heavy vector mesons: the momentum-dependence of the $Q^2=0$ photon-quark vertex is completely different from that of the vector-meson Bethe-Salpeter amplitude.

One is consequently led to conclude that insofar as heavy mesons are concerned, no extant attempt to link $e + p \to e^\prime + V + p$ reactions with $V + p \to V+p$ via VMD is reliable.  A similar conclusion is drawn in Ref.\,\cite{Sun:2021pyw} using arguments within perturbative QCD.  It is strengthened by Ref.\,\cite{Du:2020bqj}, which demonstrates that even if VMD were valid, then contributing relevant coupled-channels processes would obscure connections between $e + p \to e^\prime + V + p$ and $V + p \to V+p$ reactions.
These analyses make tenuous any interpretation of $e + p \to e^\prime + V + p$ reactions as a path to hidden-charm pentaquark production or as a route to understanding the origin of the proton mass.
In fact, the crudity of existing models prevents sound conclusions being drawn about the capacity of heavy meson production to reveal something about the SM \cite{Lee:2022ymp}.

On the other hand, numerous applications \cite{Eichmann:2016yit, Roberts:2020udq, Eichmann:2020oqt, Roberts:2020hiw, Qin:2020rad}, including $\gamma^\ast \gamma \to \pi^0$, $\eta$, $\eta^\prime$, $\eta_c$, $\eta_b$ \cite{Raya:2016yuj, Ding:2018xwy}, show that a viable alternative to the VMD hypothesis exists in adapting the CSMs discussed herein to a direct analysis of processes like $\gamma^{(\ast)} + p \to V + p$.  Considering vector-meson photo/electroproduction from the proton, Refs.\,\cite{Pichowsky:1996tn, Shi:2021taf, Lee:2022ymp} illustrate how one might proceed; and given developments in the past vicennium, improvements of such studies are now possible.

Following the first studies almost thirty years ago \cite{Roberts:1994hh}, the CSM treatment of meson elastic and transition form factors has now reached a mature level.  Today, as sketched above, sound predictions with a traceable connection to QCD are being delivered.  This enables new opportunities to be exploited, such as
informative comparison with results from lQCD \cite{Chen:2018rwz},
weak transitions of heavy mesons -- see Refs.\,\cite{Yao:2020vef, Yao:2021pyf, Yao:2021pdy} and Sec.\,\ref{SecTransition},
calculations of form factors describing light-by-light scattering contributions to the muon anomalous magnetic moment \cite{Miramontes:2021exi},
and a beginning to the analysis of meson gravitational form factors \cite{PepePresentation}.

\section{Baryon Form Factors}
\label{SecBaryonFormFactors}
Advances in the calculation of meson form factors are complemented by progress with the parameter-free prediction of baryon elastic and transition form factors.  The first direct RL truncation study of nucleon elastic form factors was presented in Ref.\,\cite{Eichmann:2011vu}.  It was extended to nucleon axial and pseudoscalar form factors in Ref.\,\cite{Eichmann:2011pv}, the elastic form factors of $\Delta$- and $\Omega$-baryons in Ref.\,\cite{Sanchis-Alepuz:2013iia}, and nucleon tensor charges in Ref.\,\cite{Qin:2020rad}.

\begin{figure}[t]
\centerline{\includegraphics[clip, height=0.66\textwidth, width=0.8\textwidth]{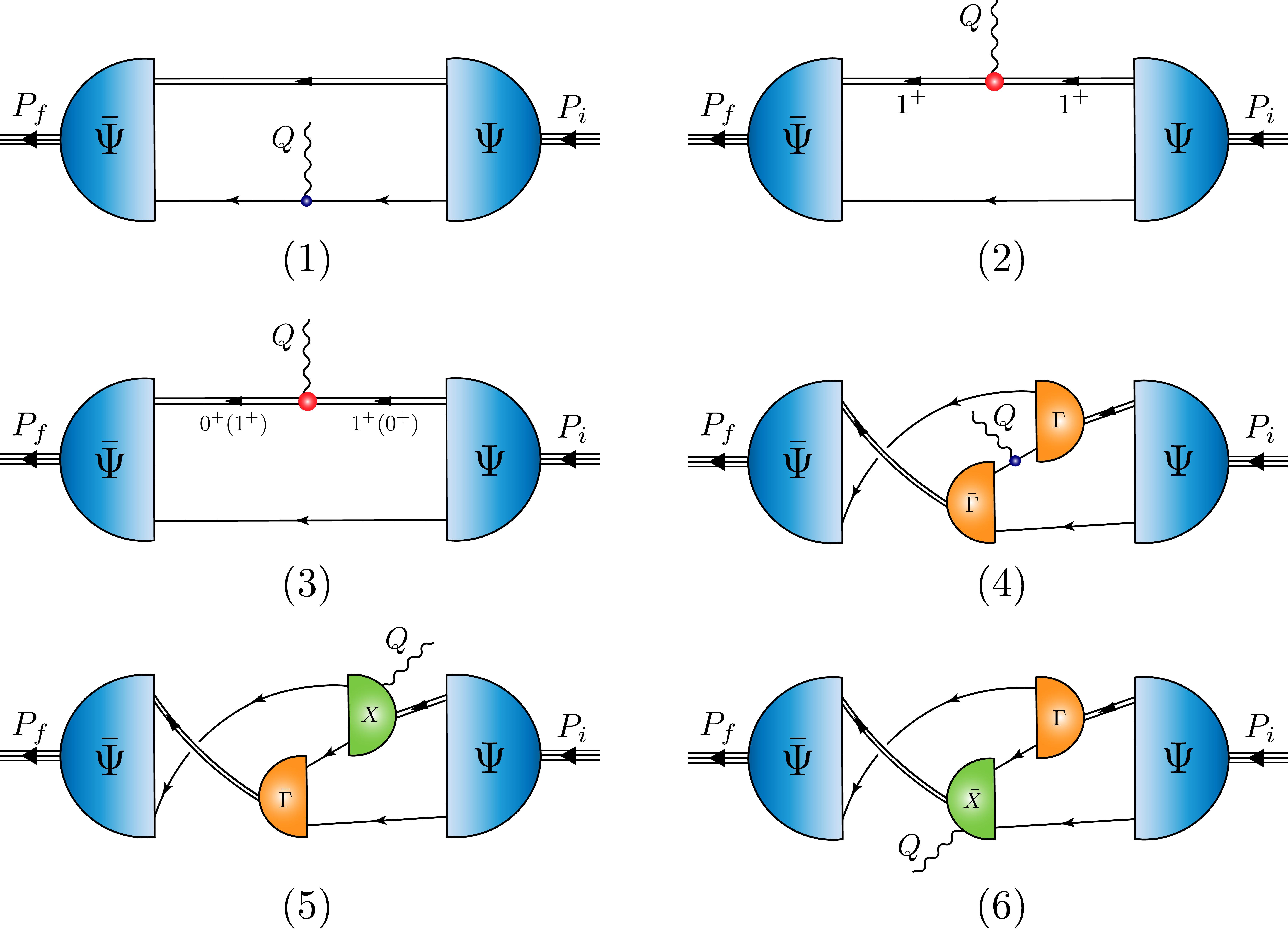}}
\caption{\label{figcurrent} Axial/pseudoscalar current that ensures a symmetry-preserving interaction with an on-shell baryon described by a Faddeev amplitude obtained from the equation depicted in Fig.\,\ref{FigFaddeev}: \emph{single line}, dressed-quark propagator; \emph{undulating line}, axial/pseudoscalar current; $\Gamma$,  diquark correlation amplitude; \emph{double line}, diquark propagator; and $\chi$, seagull terms.  A detailed legend is provided in Table~\ref{DiagramLegend}.}
\end{figure}

Many other direct RL truncation analyses are in train.  However, in all such studies, the challenge of quark propagator singularities moving into the complex integration domain must also be overcome \cite{Maris:1997tm}.  Using existing algorithms, the singularities limit the $Q^2$-reach of baryon form factor calculations.  Here, again, the QCD-kindred quark plus fully-interacting, nonpointlike diquark picture of baryon structure -- outlined in Sec.\,\ref{SecBaryonWaveFunctions} -- can profitably be exploited.
Some of the successes are summarised elsewhere \cite{Brodsky:2020vco, Barabanov:2020jvn}.  Recently, the predictions made for $\gamma+p \to \Delta(1600)$ transition form factors \cite{Lu:2019bjs} have been tested in analyses of $\pi^+ \pi^- p$ electroproduction data collected at Jefferson Lab (JLab) \cite{MokeevPrivate2022}, with preliminary results supporting the quark+diquark picture \cite{VictorMokeevPresentation}.
%

Scanning the sources indicated above, it becomes apparent that, hitherto, nucleon properties have largely been probed in $e+N$ scattering.  Aspects of this field are reviewed elsewhere \cite{Perdrisat:2006hj}.  An entirely new window onto baryon structure is opened when one uses neutrino scattering.  In fact, reliable predictions of nucleon and $N \to \Delta(1232)$ electroweak form factors are crucial for understanding new-generation long-baseline neutrino oscillation experiments \cite{Mosel:2016cwa, Alvarez-Ruso:2017oui, Hill:2017wgb, Gysbers:2019uyb, Lovato:2020kba, King:2020wmp, Simons:2022ltq}.  Importantly in this connection, recent developments within the framework of CSMs have enabled practitioners to deliver the first Poincar\'e-invariant parameter-free predictions for such form factors on a momentum transfer domain that extends to $Q^2=10\,$GeV$^2$ \cite{Chen:2020wuq, Chen:2021guo, ChenChen:2022qpy}.  Extensions to even larger $Q^2$ are feasible.  Where data are available, the predictions confirm the measurements.  More significantly, the results are serving as motivation for new experiments at high-luminosity facilities.

\begin{table}[t]
\caption{\label{DiagramLegend}
Enumeration of terms in the baryon axialvector/pseudoscalar current, drawn in Fig.\,\ref{figcurrent}.}
\begin{enumerate}
\item Diagram~1, two terms: $\langle J \rangle^{S}_{\rm q}$ -- probe strikes dressed-quark with scalar diquark spectator; and $\langle J \rangle^{A}_{\rm q}$ -- probe strikes dressed-quark with axialvector diquark spectator.
\item Diagram~2: $\langle J \rangle^{AA}_{\rm qq}$ -- probe strikes axialvector diquark with dressed-quark spectator.
\item Diagram~3: $\langle J \rangle^{\{SA\}}_{\rm qq}$ -- probe mediates transition between scalar and axialvector diquarks, with dressed-quark spectator.
\item Diagram~4, three terms:
    $\langle J \rangle_{\rm ex}^{SS}$ -- probe strikes dressed-quark ``in-flight'' between one scalar diquark correlation and another;
    $\langle J \rangle_{\rm ex}^{\{SA\}}$ -- dressed-quark ``in-flight'' between a scalar diquark correlation and an axialvector correlation;
    and $\langle J \rangle_{\rm ex}^{AA}$ -- ``in-flight'' between one axialvector correlation and another.
\item Diagrams~5 and 6 -- seagull diagrams describing the probe coupling into the diquark correlation amplitudes: $\langle J\rangle_{\rm sg}$.  There is one contribution from each diagram to match every term in Diagram (4).
\end{enumerate}
\end{table}

The key step in the use of CSMs was construction of a symmetry-preserving current that describes the coupling between axialvector and pseudoscalar probes and baryons whose structure is determined by the Faddeev equation in Fig.\,\ref{FigFaddeev}.  This current is illustrated in Fig.\,\ref{figcurrent} and explained by the legend in Table~\ref{DiagramLegend}.
The origins and characters of Diagrams~1-3 are obvious: the probe must interact with every ``constituent'' that carries a weak charge.
Diagram~4 is a two-loop diagram, made necessary by the quark-exchange nature of the kernel in Fig.\,\ref{FigFaddeev}: the object exchanged in binding also carries a weak charge.
Given the presence of Diagram~4, so-called seagull diagrams -- Diagrams 5 and 6 -- are necessary to ensure symmetry preservation.  The analogous contributions for baryon electromagnetic currents were derived in Ref.\,\cite{Oettel:1999gc}, but it took more than twenty years before the seagull terms were derived for axialvector and pseudoscalar currents \cite{Chen:2021guo}.  These contributions are both two-loop diagrams.

Axialvector interactions of a nucleon are described by two form factors -- $G_A(Q^2)$ (axial) and $G_P(Q^2)$ (induced-pseudoscalar), associated with the following matrix element:
\begin{subequations}
\label{jaxdq0}
\begin{align}
\label{jaxdq}
\hat J^j_{5\mu}(K,Q) &
:= \langle N(P_f)|{\mathpzc A}^j_{5\mu}(0)|N(P_i)\rangle \\
\label{jaxdqb}
=&\bar{u}(P_f)\frac{\tau^j}{2}\gamma_5
\bigg[ \gamma_\mu G_A(Q^2) +i\frac{Q_\mu}{2m_N}G_P(Q^2) \bigg]\,u(P_i)\,,
\end{align}
\end{subequations}
where $P_{i,f}$ are, respectively, the initial and final nucleon momenta, with $P_{i,f}^2=-m_N^2$, $m_N$ is the nucleon mass,
and $u(P)$ is the associated Euclidean spinor.  (Associated conventions are specified elsewhere, \emph{e.g}., Ref.\,\cite[Appendix\,B]{Segovia:2014aza}.)
The average momentum of the system is $K=(P_i+P_f)/2$ and $Q=P_f-P_i$ is the momentum transferred between initial and final states.  It is usual to consider the SU$(2)_{\rm F}$ isospin limit $m_u=m_d=:m_q$, with the flavour structure described using Pauli matrices $\{\tau^j|j=1,2,3\}$: $\tau^{1\pm i2}:=(\tau^1\pm i\tau^2)/2$ correspond to the weak charged currents and $\tau^3$ is the neutral current. Moreover, the isovector axial current operator is
\begin{equation}
\label{jaxx}
{\mathpzc A}_{5\mu}^j(x) = \bar{q}(x) \frac{\tau^j}{2} \gamma_5 \gamma_\mu q(x), \,\,\,\,\, q=\left(\begin{array}{c} u \\ d \end{array}\right) \,.
\end{equation}
A third form factor is defined via the kindred pseudoscalar current, a matrix element of ${\mathpzc P}^j_5(x) = \bar{q}(x)\frac{\tau^j}{2}\gamma_5q(x)$:
\begin{equation}
\label{jpsdq0}
\hat J^j_{5}(K,Q) :=
\langle N(P_f)|{\mathpzc P}^j_5(0)|N(P_i)\rangle
=\bar{u}(P_f)\frac{\tau^j}{2}\gamma_5\,G_5(Q^2)\,u(P_i)\,.
\end{equation}

Using the framework deployed to calculate the baryon wave functions discussed in Sec.\,\ref{SecBaryonWaveFunctions}, Refs.\,\cite{Chen:2020wuq, Chen:2021guo} delivered the parameter-free prediction for $G_A(Q^2)$ displayed in Fig.\,\ref{FigGAx}: the lighter blue band expresses the impact of $\pm 5$\% variations in the diquark masses: $m_{0^+} = 0.80 (1\pm 0.05)$; $m_{1^+} = 0.89 (1\pm 0.05)$.  The calculated values for the nucleon axial charge $g_A=G_A(Q^2=0)$, associated axial charge radius, and axial mass are, respectively,
\begin{equation}
g_A = 1.25(3)\,, \quad
\langle r^2_A\rangle^{1/2} m_N = 3.25(4)\,, \quad
m_A/m_N = 1.23(3)\,.
\end{equation}
For comparison, empirically $g_A=1.2756(13)$ \cite[PDG]{Workman:2022ynf} and \cite{Meyer:2016oeg} $\langle r^2_A\rangle^{1/2} m_N = 3.23(72)$, $m_A/m_N = 1.15(08)$.  Evidently, the CSM predictions agree with extant data.
Regarding the axial mass, we note that it is sometimes convenient, when comparing with other analyses, to use a dipole \emph{Ansatz} as an approximation for the axial form factor:
\begin{equation}
\label{ma}
G_A(Q^2) = \frac{g_A}{\big(1+Q^2/m_A^2\big)^2}\,;
\end{equation}
so Refs.\,\cite{Chen:2020wuq, Chen:2021guo} extracted $m_A$ using Eq.\,\eqref{ma} to interpolate the \emph{global} $Q^2$-behaviour of $G_A$ on $x\in[0,1.6]$, in which case $m_A$ is not simply related to the axial radius.
It is worth remarking that scalar and axialvector diquark mass variations interfere destructively, \emph{e.g}., reducing $m_{0^+}$ increases $g_A$, whereas $g_A$ decreases with the same change in the axialvector mass.

\begin{figure}[t]
\centering
\includegraphics[clip, width=0.66\textwidth]{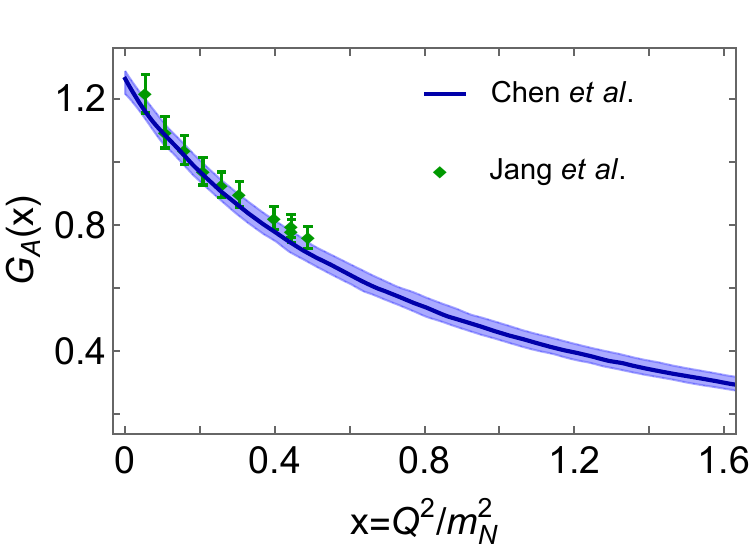}
\caption{\label{FigGAx}
Predicted result for $G_A(x)$ in Ref.\,\cite[Chen \emph{et al}.]{Chen:2021guo} -- blue curve within lighter blue uncertainty band, compared with lQCD results from Ref.\,\cite[Jang \emph{et al}.]{Jang:2019vkm} -- green diamonds. With respect to the CSM central results, this comparison may be quantified by reporting the mean-$\chi^2$ value, which is 0.27.
}
\end{figure}

Turning to the induced pseudoscalar form factor, muon capture experiments ($\mu\,+\,p\,\to\,\nu_\mu\,+\,n$) may be used to determine the induced pseudoscalar charge:
\begin{equation}
g_p^\ast = \frac{m_\mu}{2m_N}
G_P(Q^2 = 0.88\,m_\mu^2)\,.
\end{equation}
The CSM prediction is $g_p^\ast=8.80(23)$.  Compared with the recent MuCap Collaboration result, $g_p^\ast=8.06(55)$\,\cite{Andreev:2012fj, Andreev:2015evt}, it agrees within uncertainties but is slightly larger.  The CSM value is nicely aligned with the world average \cite{Bernard:2001rs}: $g_p^\ast=8.79(1.92)$.

\begin{figure}[t]
\centering
\includegraphics[clip, width=0.66\textwidth]{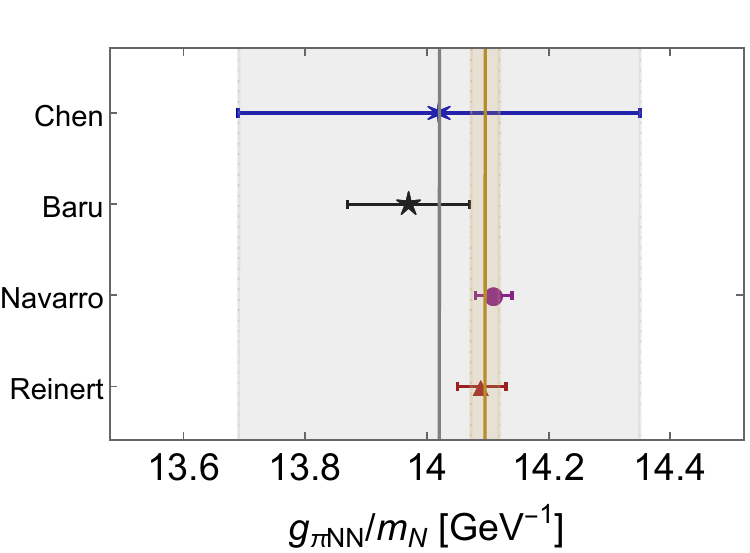}
\caption{\label{Figgpinn}
Comparison of the CSM prediction for $g_{\pi NN}/m_N$ (blue asterisk) \cite{Chen:2021guo} with
values extracted from pion-nucleon scattering \cite[Baru]{Baru:2011bw} (black star),
Granada 2013 $np$ and $pp$ scattering database \cite[Navarro]{NavarroPerez:2016eli} (purple circle),
and nucleon-nucleon scattering \cite[Reinert]{Reinert:2020mcu} (red triangle).
The vertical grey band marks the estimated uncertainty in the CSM prediction.
The error-weighted average of the depicted results, Eq.\,\eqref{ErrorWeighted}, is drawn as the gold line within like-coloured band.
}
\end{figure}

The pseudoscalar form factor, $G_5(Q^2)$, is of interest because, \emph{inter alia}, it is used to define the pion-nucleon form factor $G_{\pi NN}(Q^2)$ via
\begin{align}
\label{gpinn}
G_5(Q^2) =: \frac{m_\pi^2}{Q^2+m_\pi^2}\frac{f_\pi}{m_q}G_{\pi NN}(Q^2)\,,
\end{align}
where $f_\pi$ is the pion leptonic decay constant (appearing in Fig.\,\ref{FigLeptonic}) and $G_{\pi NN}(-m_\pi^2)=g_{\pi NN}$ is the $\pi N N$ coupling, a key input to nucleon+nucleon potentials.  The CSM prediction is \cite{Chen:2020wuq, Chen:2021guo}:
$g_{\pi NN}/m_N= 14.02(33)/{\rm GeV}$.
This value overlaps with that inferred from pion-nucleon scattering \cite{Baru:2011bw} ($g_{\pi NN}/m_N = 13.97(10)/{\rm GeV}$);
and compares favourably with a determination based on the Granada 2013 $np$ and $pp$ scattering database \cite{NavarroPerez:2016eli} ($g_{\pi NN}/m_N = 14.11(3)/{\rm GeV}$) and a recent analysis of nucleon-nucleon scattering using effective field theory and related tools \cite{Reinert:2020mcu} ($g_{\pi NN}/m_N = 14.09(4)/{\rm GeV}$).  All these results are compared in Fig.\,\ref{Figgpinn}, which also highlights their error-weighted average:
\begin{equation}
\label{ErrorWeighted}
g_{\pi NN}/m_N = 14.10(2)/{\rm GeV}\,.
\end{equation}

Continuing with an effort to inform nuclear physics potentials using hadron physics results, it is worth noting that on $-m_\pi^2 <Q^2 < 2\,m_N^2$, a fair approximation to the CSM prediction for the pion-nucleon form factor is provided by $(x=Q^2/m_N^2)$:
\begin{equation}
\label{DipolepiNN}
G_{\pi NN}^d(x) = \frac{13.47 m_N}{ ( 1 + x /0.845^2 )^2 }\,.
\end{equation}
This corresponds to a $\pi NN$ dipole mass $\Lambda_{\pi N N} = 0.845 m_N = 0.79\,$GeV, \emph{viz}.\ a soft form factor.
(A commensurate value was obtained previously in a simpler quark+scalar-diquark model \cite{Bloch:1999rm}.)
Being just $\sim 20$\% greater, the CSM prediction is qualitatively equivalent to the $\pi N N$ dipole mass inferred in a dynamical coupled-channels analysis of $\pi N$, $\gamma N$ interactions \cite{Kamano:2013iva}.  Future such coupled channels studies may profit by implementing couplings and range parameters determined in CSM analyses.

An extension of this effort to the more general case of hyperon+nucleon potentials can be found in Ref.\,\cite{Cheng:2022jxe}.  Using a SCI, predictions were made therein for an array of meson+octet-baryon couplings.  Comparing the results with extant phenomenological interactions \cite{Haidenbauer:2005zh, Rijken:2010zzb, Kamano:2013iva}, one finds a mean absolute relative deviation ($P$ is the pseudoscalar meson absorbed in the baryon transition $B\to B^\prime$ and $g_{P B^\prime B}$ is the associated coupling)
\begin{equation}
\delta_r^g:=\{|g_{P B^\prime B}^{\rm SCI}/g_{P B^\prime B}^{\mbox{\rm\footnotesize phen.}}-1|\} = 0.18(14)\,.
\end{equation}
This result strengthens the case in favour of using CSM predictions for the couplings as new constraints in the development of baryon+baryon potentials.

A little algebra reveals that the proton axial charge can be interpreted as a measure of the valence-quark contributions to the proton light-front helicity, \emph{e.g}., Ref.\,\cite[Eqs.\,(6)--(8)]{Chang:2012cc}:
\begin{equation}
\label{gAhelicity}
g_A = \int_0^1 dx\, [\Delta {\mathpzc u}^p(x;\zeta_{\cal H}) - \Delta {\mathpzc d}^p(x;\zeta_{\cal H})]
=: g_A^u - g_A^d\,,
\end{equation}
where $\Delta {\mathpzc q}^p(x;\zeta_{\cal H})= {\mathpzc q}^p_\uparrow(x;\zeta_{\cal H})-{\mathpzc q}^p_\downarrow(x;\zeta_{\cal H})$ is the light-front helicity DF for a quark $q$.  Plainly, $\Delta {\mathpzc q}^p$ is the difference between the light-front number-density of quarks with helicity parallel and antiparallel to that of the proton.  It is scale dependent.

Eq.\,\eqref{gAhelicity} conveys additional significance to a flavour separation of the axial charge form factor:
\begin{equation}
G_A(Q^2) = G_A^u(Q^2) - G_A^d(Q^2)\,.
\end{equation}
A detailed analysis is presented in Ref.\,\cite[Sec.\,4]{ChenChen:2022qpy}, which reveals the following diagram contributions to the separate $u$, $d$ axial form factors:
{\allowdisplaybreaks\begin{subequations}
\label{FlavourSep}
\begin{align}
G_A^u & = \langle J \rangle^{S}_{\rm q} - \mbox{\underline{$\langle J \rangle^{A}_{\rm q}$}} + \langle J \rangle^{AA}_{\rm qq} + \tfrac{1}{2} \langle J \rangle^{\{SA\}}_{\rm qq}  + 2 \langle J \rangle^{\{SA\}}_{\rm ex} + \tfrac{4}{5} \langle J \rangle^{AA}_{\rm ex}, \label{T1}\\
-G_A^d & = \phantom{\langle J \rangle^{S}_{\rm q} - } \mbox{\underline{$2 \langle J \rangle^{A}_{\rm q}$}} + \tfrac{1}{2}\langle J \rangle^{\{SA\}}_{\rm qq}  + \langle J \rangle^{SS}_{\rm ex} - \langle J \rangle^{\{SA\}}_{\rm ex} + \tfrac{1}{5} \langle J \rangle^{AA}_{\rm ex}, \label{T2}
\end{align}
\end{subequations}
where the nomenclature of Table~\ref{DiagramLegend} is used.
Identified according to Eqs.\,\eqref{FlavourSep}, the calculated $Q^2=0$ contributions are listed in Table~\ref{isovectorcharge}.
It is worth stressing that Eqs.\,\eqref{FlavourSep} express the fact that since a $0^+$ diquark cannot couple to an axialvector current, then Diagram\,1 in Fig.\,\ref{figcurrent} only supplies a $u$-quark contribution to the proton $G_A(Q^2)$, \emph{viz}.\ $\langle J \rangle^{S}_{\rm q}$.  It follows that in a scalar-diquark-only proton, a $d$-quark contribution to $G_A(Q^2)$ can only arise from Fig.\,\ref{figcurrent}\,-\,Diagram\,4, \emph{i.e}., $\langle J \rangle^{SS}_{\rm ex}$; and $|\langle J \rangle^{SS}_{\rm ex}/\langle J \rangle^{S}_{\rm q}| \approx 0.06$.  Notably, many scalar-diquark-only models omit Diagram~4, in which cases $G_A^d(Q^2) \equiv 0$.
An extension of these observations to the complete array of octet baryons is described elsewhere \cite[Sec.\,IV]{Cheng:2022jxe}.

\begin{table}[t]
\caption{\label{isovectorcharge}
Flavour and diagram -- Fig.\,\ref{figcurrent} -- separation of the proton axial charge: $g_A^u=G_A^u(0)$, $g_A^d=G_A^d(0)$; $g_A^u - g_A^d = 1.25(3)$.
The listed uncertainties express the effect of $\pm 5$\% variations in the diquark masses, \emph{e.g}.\ $0.88_{6_\mp} \Rightarrow 0.88 \mp 0.06$. }
\begin{center}
\begin{tabular*}
{\hsize}
{
r@{\extracolsep{0ptplus1fil}}
|l@{\extracolsep{0ptplus1fil}}
l@{\extracolsep{0ptplus1fil}}
l@{\extracolsep{0ptplus1fil}}
l@{\extracolsep{0ptplus1fil}}
l@{\extracolsep{0ptplus1fil}}
l@{\extracolsep{0ptplus1fil}}
l@{\extracolsep{0ptplus1fil}}}\hline
 & $\langle J \rangle^{S}_{\rm q}$  & $\langle J \rangle^{A}_{\rm q}$ &$\langle J \rangle^{AA}_{\rm qq}$ & $\langle J \rangle^{\{SA\}}_{\rm qq}$
 & $\langle J \rangle_{\rm ex}^{SS}$
 & $\langle J \rangle_{\rm ex}^{\{SA\}}$
 & $\langle J \rangle_{\rm ex}^{AA}$  \\\hline
 $g_A^u$ & $0.88_{6_\mp}$ & $-0.08_{0_\pm} $ & $0.03_{0_\pm}$ & $0.08_{0_\mp}$ & $0$ & $\approx 0$ & $0.03_{1_\pm}$ \\
 $-g_A^d$ & $0$ & $\phantom{-}0.16_{0_\pm} $ & $0$ & $0.08_{0_\mp}$ & $0.05_{1_\pm}$ & $\approx 0$ & $0.01_{0_\pm }$\\[0.4ex]  \hline
\end{tabular*}
\end{center}
\end{table}

Using the solution of the Faddeev equation -- Fig.\,\ref{FigFaddeev}, Ref.\,\cite{ChenChen:2022qpy} reports
\begin{equation}
\label{Separations}
g_A^u/g_A  = \phantom{-} 0.76 \pm 0.01\,, \quad
g_A^d/g_A  =- 0.24 \pm 0.01 \,, \quad
g_A^d/g_A^u  = - 0.32 \pm 0.02 \,.
\end{equation}
In nonrelativistic quark models with uncorrelated wave functions, $g_A^d/g_A^u = -1/4$.  Hence, the relevant comparison reveals that the highly-correlated wave function obtained by solving the Faddeev equation gives the valence $d$ quark a markedly larger fraction of the proton's light-front helicity than is found in simple quark models.  Reviewing the discussion after Eq.\,\eqref{FlavourSep}, it becomes apparent that this feature owes to the presence of axialvector diquarks in the proton: the current contribution arising from the $\{uu\}$ correlation -- underlined term in Eq.\,\eqref{T2}, measuring the probe striking the valence $d$ quark, is twice as strong as that from the $\{ud\}$ correlation -- underlined term in Eq.\,\eqref{T1}, in which the probe strikes the valence $u$ quark.

It is natural to enquire after the robustness of the results in Eq.\,\eqref{Separations}.  Consider, therefore, that assuming SU$(3)$-flavour symmetry in analyses of octet baryon axial charges, these charges are expressed in terms of two low-energy constants, \cite[Table~1]{Cabibbo:2003cu}: $D$, $F$, with $g_A^u = 2 F$, $g_A^d=F-D$.
(This assumption is accurate to roughly 4\% -- see, \emph{e.g}., Ref.\,\cite{Cheng:2022jxe}.)
In this case, the values in Eq.\,\eqref{Separations} correspond to
\begin{equation}
\label{TDF}
D= 0.78(2)\,, \quad F=0.48(1)\,, \quad F/D=0.61(2)\,.
\end{equation}
On the other hand, using available empirical information \cite{Workman:2022ynf}, one obtains $D= 0.774(26)$, $F=0.503(27)$, and $g_A^u/g_A = 0.79(4)$, $g_A^d/g_A= -0.21(3)$, $g_A^d/g_A^u= -0.27(4)$, values which are consistent with the results in Eqs.\,\eqref{Separations}, \eqref{TDF}.%
\footnote{If one eliminates axialvector diquarks from the proton wave function, then $g_A^d/g_A^u=-0.054(13)$, a result disfavoured by experiment at the level of $5.1\sigma$, \emph{i.e}., the probability that the scalar-diquark-only proton result could be consistent with data is $1/7,100,000$.}
In addition, the SCI predicts \cite{Cheng:2022jxe}
\begin{equation}
\label{DFvalues}
D = 0.78\,, \; F = 0.43\,,\; F/D = 0.56\,,
\end{equation}
and a covariant baryon chiral perturbation theory analysis yields $D=0.80(1)$, $F=0.47(1)$, $F/D=0.59(1)$ \cite{Ledwig:2014rfa}.

Given the favourable realistic proton wave function comparisons presented above, the values in Eq.\,\eqref{Separations} can be viewed as reliable.  This is important because of the connection between flavour-separated axial charges and the so-called proton ``spin crisis'' \cite{Aidala:2012mv, Deur:2018roz}.  At any given resolving scale, the singlet, triplet, and octet axial charges of the proton are, respectively:
\begin{equation}
a_0 = g_A^u + g_A^d + g_A^s \,, \quad
a_3 = g_A =  g_A^u - g_A^d \,, \quad
a_8 = g_A^u + g_{A}^{d} - 2 g_{A}^{s} \,.
\end{equation}
If working at the hadron scale, $\zeta_{\cal H}$, where dressed valence quasiparticles carry all proton properties \cite{Cui:2020dlm, Cui:2020tdf, Cui:2021mom, Cui:2022bxn, Chang:2022jri, Lu:2022cjx}, then $g_A^s\equiv 0$ $a_0=a_8$; hence \cite{ChenChen:2022qpy}
\begin{equation}
\label{Eqa0proton}
a_0 = 0.65(2)\,.
\end{equation}
In general, $a_{3,8}$ are conserved charges, \emph{viz}.\ they are the same at all resolving scales, $\zeta$.  However, that is not true of the individual terms in their definitions: the separate valence quark charges $g_{A}^u$, $g_{A}^d$, $g_{A}^s$ evolve with $\zeta$ \cite{Deur:2018roz}.
Consequently, the value of $a_0$, which is the fraction of the proton's total $J=1/2$ carried by valence degrees-of-freedom, changes with scale: the result in Eq.\,\eqref{Eqa0proton} is the maximum value of $a_0$ and the fraction falls slowly with increasing $\zeta$.

Returning to expectations based on simple, nonrelativistic quark models, textbook-level algebra yields $a_0=1$.  So, in such pictures, all the proton spin derives from that of the constituent quarks.  On the other hand, the CSM analysis in Ref.\,\cite{ChenChen:2022qpy} predicts that proton dressed-valence degrees-of-freedom carry only two-thirds of the spin.  Since there are no other degrees-of-freedom at $\zeta_{\cal H}$ and the Poincar\'e-covariant proton wave function properly describes a $J=1/2$ system, then the ``missing'' part of the total-$J$ must be associated with quark+diquark orbital angular momentum.  Similar conclusions apply for all ground-state octet baryons \cite{Cheng:2022jxe}.

The study in Ref.\,\cite{ChenChen:2022qpy} delivered Poincar\'e-invariant parameter-free predictions for the proton axial form factor and its flavour separation out to $Q^2 \approx 10\,m_N^2$.  The axial form factor itself agrees with available data \cite{DelGuerra:1976uj, CLAS:2012ich}, which extends to $Q^2 \approx 5\,m_N^2$ -- see Ref.\,\cite[Fig.\,3]{ChenChen:2022qpy}.  More importantly, the results will likely serve as motivation for new experiments aimed at exploring nucleon structure with axialvector probes instead of the photon, opening a new window onto hadron structure.
It is worth highlighting here that a dipole fit to data is only a good approximation on the fitted domain.  With increasing $Q^2$, the dipole increasingly overestimates the actual result, being $56(5)$\% too large at $Q^2=10\,m_N^2$ -- see Ref.\,\cite[Fig.\,4B]{ChenChen:2022qpy}.  It therefore becomes an unsound tool for developing qualitative insights and quantitative cross-section estimates.

\begin{figure*}[!t]
\hspace*{-1ex}\begin{tabular}{lcl}
\large{\textsf{A} \hspace*{1em}$0^+\!, d $} & & \large{\textsf{B} \hspace*{1em}$0^+\!, u $}\\[1.0ex]
\includegraphics[clip, width=0.4\textwidth]{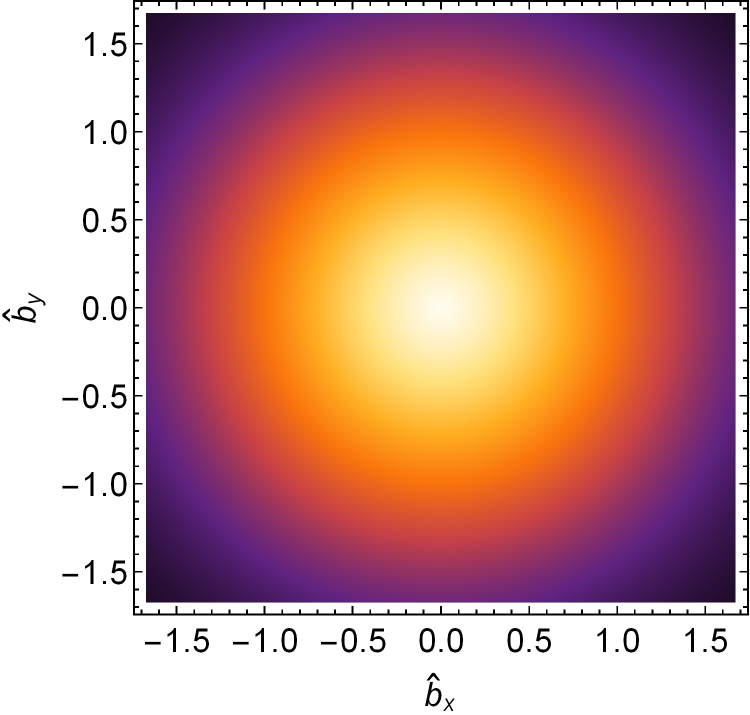} &
\hspace*{2em} \includegraphics[clip, width=0.054\textwidth]{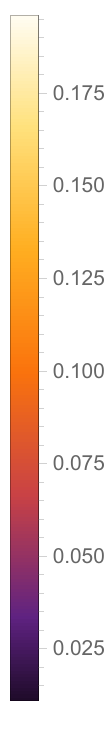} \hspace*{0.4em} &
\includegraphics[clip, width=0.4\textwidth]{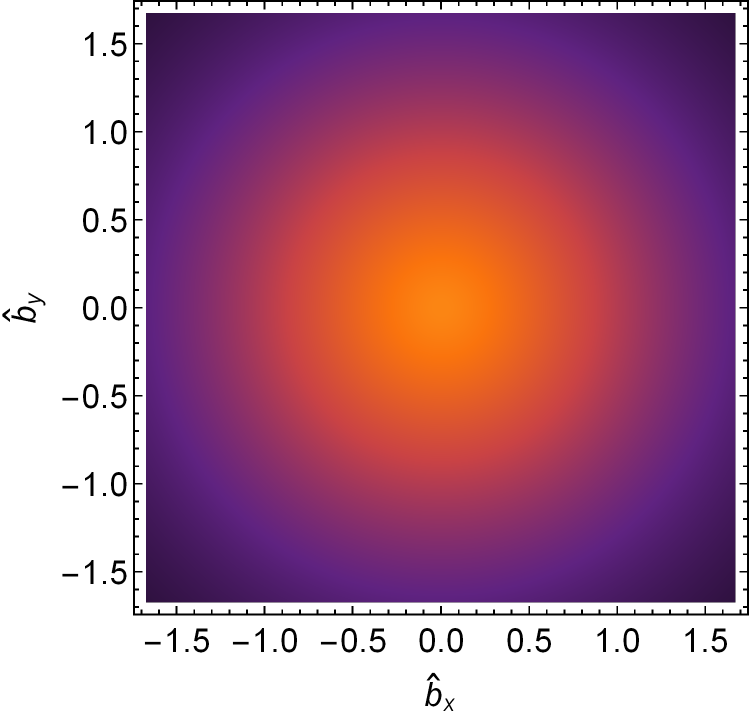} \\[2ex]
\large{\textsf{C} \hspace*{1em}$0^+$\&$1^+\!, d $} & & \large{\textsf{D} \hspace*{1em}$0^+$\&$1^+\!, u $}\\[-3.0ex]
\includegraphics[clip, width=0.4\textwidth]{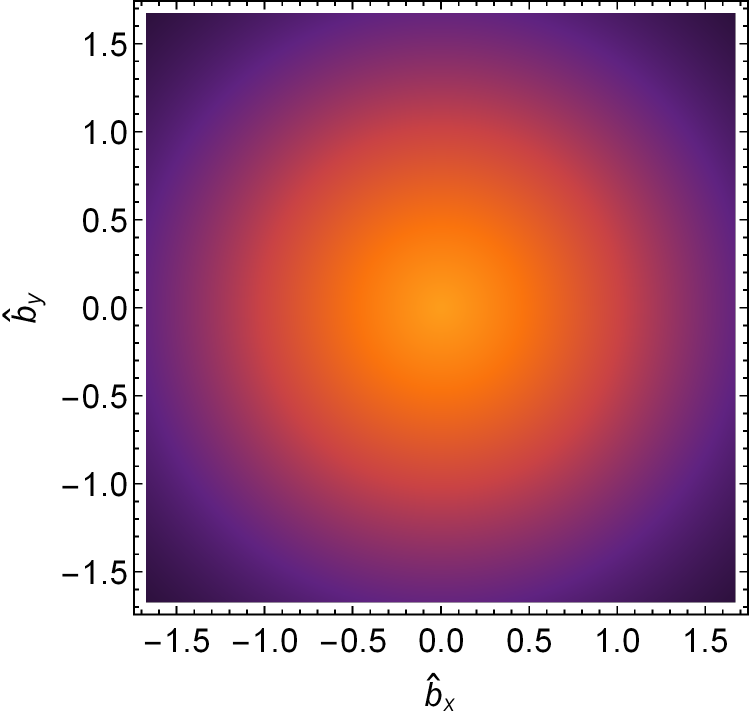} &
\hspace*{2em} \includegraphics[clip, width=0.054\textwidth]{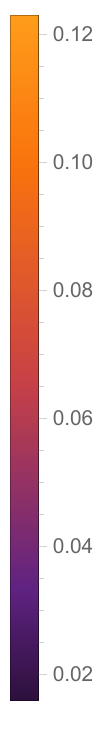} \hspace*{0.4em} &
\includegraphics[clip, width=0.4\textwidth]{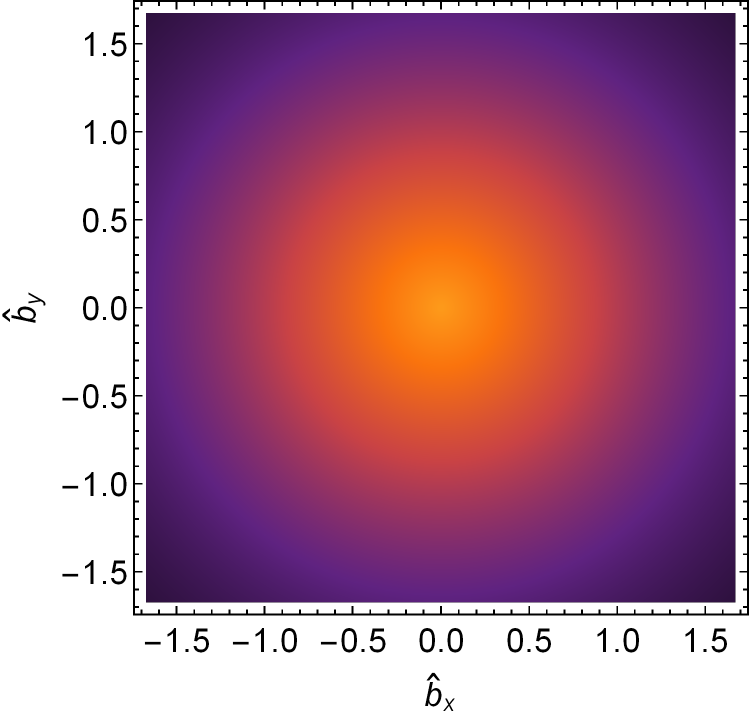}
\end{tabular}
\caption{\label{FigProfiles}
Transverse density profiles,  Eq.\,\eqref{density}, calculated in Ref.\,\cite{ChenChen:2022qpy} from flavour separated proton axial form factors.
Top row: scalar-diquark-only proton.
{Panel A} -- two-dimensional plot of $\hat\rho_A^d(|\hat b|)/g_A^d$;
{Panel B} -- similar plot of $\hat\rho_A^u(|\hat b|)/g_A^u$.
Removing the $1/g_A^f$ normalisation, the $b=0$ values are $\hat \rho_A^d(0) = -0.009$, $\rho_A^u(0)= 0.097$.
Bottom row: realistic proton Faddeev amplitude, including axialvector diquarks as predicted by the Faddeev equation in Fig.\,\ref{FigFaddeev}.
{Panel C} -- $\hat\rho_A^d(|\hat b|)/g_A^d$;
{Panel D} -- $\hat\rho_A^u(|\hat b|)/g_A^u$.
Removing the $1/g_A^f$ normalisation, the $b=0$ values of these profiles are $\hat \rho_A^d(0) = -0.038$, $\hat\rho_A^u(0)= 0.12$.
\emph{N.B}.\ $\int d^2 \hat b \,\hat\rho_A^f(|\hat b|)/g_A^f = 1$, $f=u,d$.
}
\end{figure*}

Furthermore, with flavour-separated form factors in-hand on such a large $Q^2$-domain, Ref.\,\cite{ChenChen:2022qpy} was able to calculate and contrast the $u$- and $d$-quark contributions to the associated light-front transverse spatial density profiles:
\begin{equation}
\label{density}
\hat\rho_A^f(|\hat b|) = \int\frac{d^2 \vec{q}_\perp}{(2\pi)^2}\,{\rm e}^{i \vec{q}_\perp \cdot\hat b}G_A^f(x)\,,
\end{equation}
with $G_A^f(x)$ interpreted in a frame defined by $Q^2 = m_N^2 q_\perp^2$, $m_N q_\perp = (\vec{q}_{\perp 1},\vec{q}_{\perp 2},0,0)=(Q_1,Q_2,0,0)$.  These profiles are depicted in Fig.\,\ref{FigProfiles}.
We note that $|\hat b|$ and $\hat\rho_A^f $ are dimensionless; so, the images drawn in Fig.\,\ref{FigProfiles} can be mapped into physical units using:
\begin{equation}
\rho_A^f(|b| =| \hat b|/m_N) = m_N^2 \,  \hat \rho_A^f(|\hat b|)\,,
\end{equation}
in which case $|\hat b| = 1$ corresponds to $|b| \approx 0.2\,$fm and $\hat\rho_1^f=0.1 \Rightarrow \rho_1^f\approx 2.3/{\rm fm}^2$.

The top row of Fig.\,\ref{FigProfiles} provides two-dimensional renderings of the flavour-separated transverse density profiles calculated from a proton wave function which does \underline{not} include axialvector diquarks, \emph{i.e}., a scalar-diquark-only proton.  In this case, the $u$ quark profile is far more diffuse than that of the $d$ quark, \emph{viz}., its extent in the light-front transverse spatial plane is much greater.  One may quantify this by reporting the associated radii: $r_{A^d}^\perp = 0.24\,$fm, $r_{A^u}^\perp = 0.48\,$fm, so the $d/u$ ratio of radii is $\approx 0.5$.

The bottom row of Fig.\,\ref{FigProfiles} was obtained using a realistic proton wave function, in which both scalar and axialvector diquarks are present with the strength determined by the Faddeev equation in Fig.\,\ref{FigFaddeev}.  In this realistic case, the $d$ quark profile is not very different from that of the $u$ quark: relative to the $u$-quark profile, the intensity peak is only somewhat broader for the $d$ quark; and comparing radii,
\begin{equation}
r_{A^d}^\perp = 0.43\,{\rm fm}, \quad r_{A^u}^\perp = 0.49\,{\rm fm}, \quad r_{A^d}^\perp/r_{A^u}^\perp \approx 0.9\,.
\end{equation}

The CSM predictions for nucleon axial and pseudoscalar form factors discussed in this section complement those for the large $Q^2$ behaviour of nucleon electromagnetic elastic and transition form factors reported, \emph{e.g}., in Refs.\,\cite{Cui:2020rmu, Chen:2018nsg}.  One may now anticipate that predictions for form factors characterising weak interaction induced $N\to \Delta(1232)$ and $N\to N^\ast(1535)$ transitions will soon become available.  Each will shed new light on nucleon structure; and the former, calculated on a domain that stretches from low-to-large $Q^2$, will likely prove valuable in developing a more reliable understanding of neutrino scattering from nucleons and nuclei.

\section{Transition Form Factors of Heavy+Light Mesons}
\label{SecTransition}
%
Heavy mesons ($B_{q=c,s,u}$, $D_s$, $D$) are special for many reasons; and their mass budgets and role in exposing constructive interference between Nature's two known sources of mass are of particular interest herein.  Consider, therefore, Fig.\,\ref{Hmassbudget} and contrast the images with those in Fig.\,\ref{Fmassbudget}.  Evidently, for heavy meson masses:
(\emph{i}) the HB component is largest in each case and its relative size grows as the current-masses of the valence constituents increase;
(\emph{ii}) all receive a significant EHM+HB interference component, but its relative strength diminishes with increasing current-masses;
and (\emph{iii}) for vector heavy mesons, but not pseudoscalar mesons, there is an EHM component, but its relative strength drops as the HB component grows.

\begin{figure}[!t]
\hspace*{-1ex}\begin{tabular}{ll}
{\sf A} & {\sf B} \\[-2ex]
\includegraphics[clip, width=0.46\textwidth]{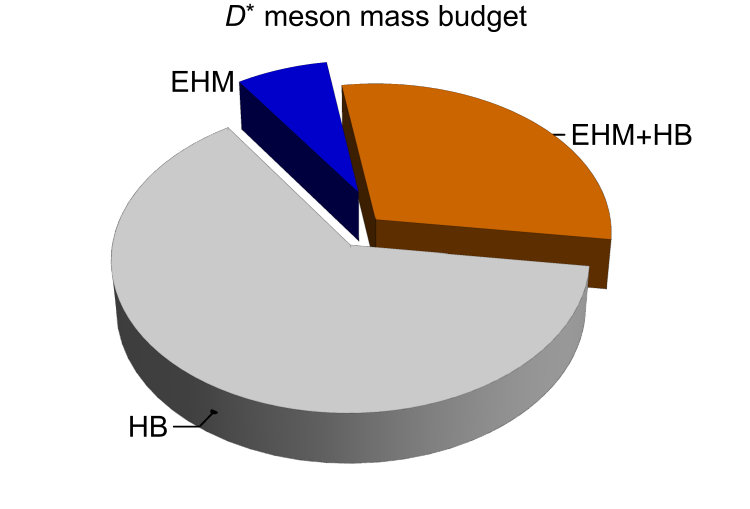} &
\includegraphics[clip, width=0.46\textwidth]{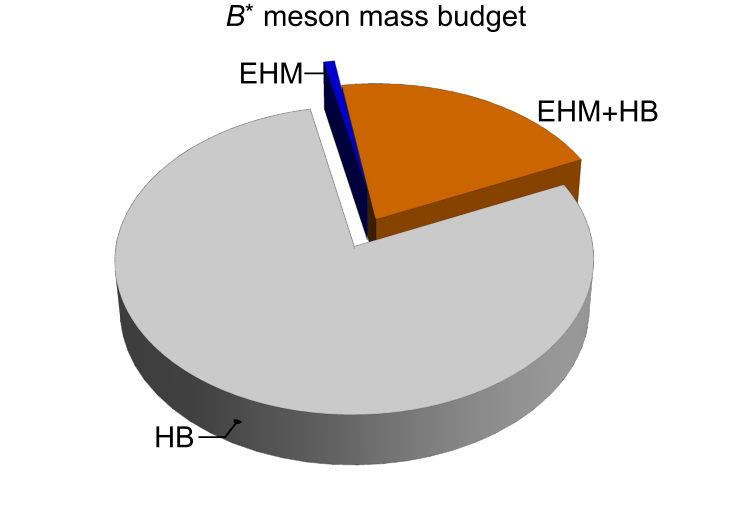}
\end{tabular}

\hspace*{-1ex}\begin{tabular}{ll}
{\sf C} & {\sf D} \\[-2ex]
\includegraphics[clip, width=0.5\textwidth]{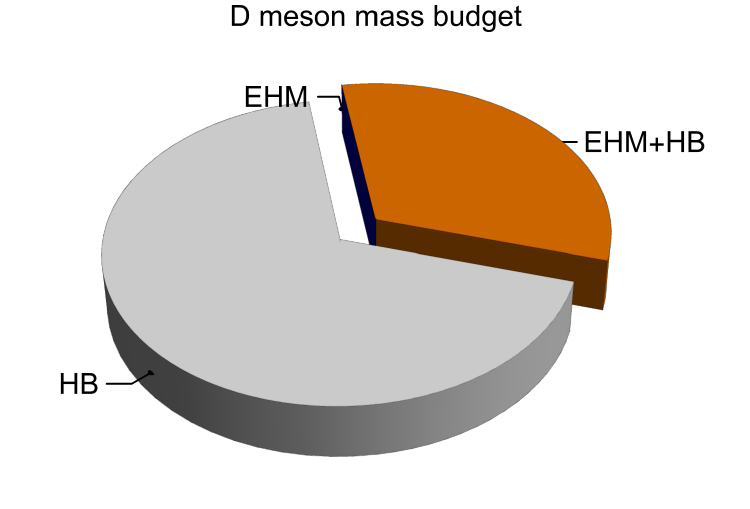}&
\hspace*{-1em}\includegraphics[clip, width=0.5\textwidth]{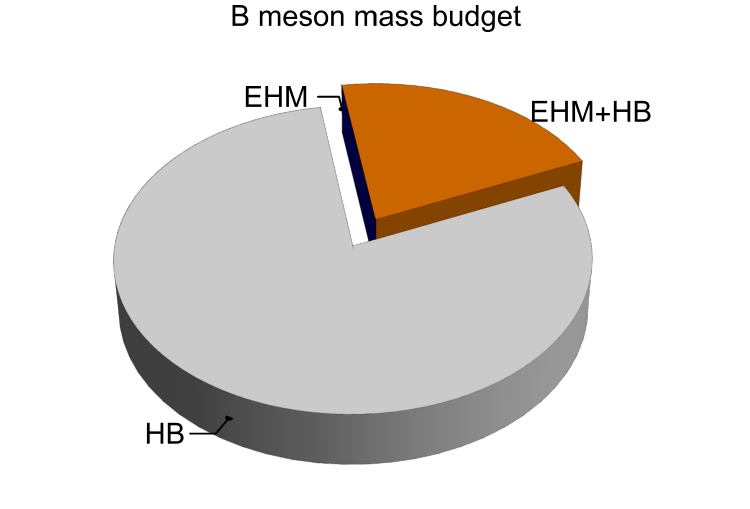}
\end{tabular}
\caption{\label{Hmassbudget}
Mass budgets:
\mbox{\sf A}\,--\,$D^\ast$ meson;
\mbox{\sf B}\,--\,$B^\ast$-meson;
\mbox{\sf C}\,--\,$D$ meson; and
\mbox{\sf D}\,--\,$B$ meson.
Each is drawn using a Poincar\'e invariant decomposition and the numerical values listed in Table~\ref{massbudget}.
(Separation at $\zeta = 2\,GeV$, calculated using information from Refs.\,\cite{Roberts:2021nhw, Workman:2022ynf}.)
 }
\end{figure}

Next consider semileptonic weak-interaction transitions between heavy and light mesons.  Comparing Figs.\,\ref{Hmassbudget} and \ref{Fmassbudget}, it is apparent that heavy-pseudoscalar to light-pseudo\-scalar transitions serve to probe the relative impacts of the strength of EHM+HB interference in the initial and final states, whereas heavy-pseudoscalar to light-vector transitions overlap systems in which HB mass is dominant with those whose mass owes almost entirely to EHM.  Both classes of transitions, therefore, present excellent opportunities for exposing the influence of Nature's two known sources of mass on physical observables.  These cases are of special interest, of course, because the transitions have long been used to place constraints on the values of the elements of the Cabibbo-Kobayashi-Maskawa (CKM) matrix, which parametrises quark flavour mixing in the SM.

A unifying analysis of both classes of transitions was recently completed using the SCI \cite{Xu:2021iwv, Xing:2022sor}, yielding results that compare favourably with other reliable experimental or independent theory analyses.  The SCI branching fraction predictions should therefore be a reasonable guide.  This is important because predictions were made for several branching fraction ratios -- $R_{D_{(s)}^{(\ast)}}$, $R_{J/\psi}$, $R_{\eta_c}$ -- whose values are direct tests of lepton universality in weak interactions: the SCI values confirm SM predictions, hence, speak for universality, as we discuss below.  The analyses also used $B_{(s)}\to D_{(s)}^\ast$ transitions to predict the precursor functions which evolve into the universal Isgur-Wise function \cite{Isgur:1989ed}, obtaining results in agreement with empirical inferences -- Refs.\,\cite[Eqs.\,(177), (181)]{HFLAV:2019otj}, \cite[Belle]{Glattauer:2015teq}.

The SCI's successes in these applications highlight the need for kindred studies using an interaction with a closer connection to QCD.  The impediment has always been the large disparity in masses that typically exists between initial and final states.  That mass imbalance requires, \emph{inter alia}, that any approach to the problem be simultaneously able to deal with both chiral and heavy-quark limits in quantum field theory.  To date, compared with the SCI, no framework with a better link to QCD can directly surmount this difficulty.  Nevertheless, following Ref.\,\cite{Yao:2020vef}, practicable and effective algorithms for continuum studies do now exist.  They exploit the strengths of the statistical SPM as a tool for interpolating data (broadly defined) and therefrom delivering extrapolations with a rigorously defined and calculable uncertainty \cite{Cui:2022fyr}.  Namely, results are calculated on domains of current-quark mass for which transition form factors may straightforwardly be obtained.  The SPM is then used to extrapolate those results and arrive at predictions for the physical processes of interest.

At present, CSM RL truncation predictions are available for the following transitions \cite{Yao:2021pyf, Yao:2021pdy}:
$B_c \to J/\psi, \eta_c$; $B_{(s)}\to \pi (K)$; $D_s \to K$; $D\to \pi, K$; and $K\to \pi$.  The last process is something of a test for the approach because such $K_{\ell 3}$ transitions have long been of experimental and theoretical interest \cite[Sec.\,62]{Workman:2022ynf}.
The calculated branching fractions are gathered in Table~\ref{TabBranch}, from which it will be seen that CSMs deliver sound results.

\begin{table*}[t]
\caption{\label{TabBranch}
CSM predictions for pseudoscalar meson semileptonic branching fractions \cite{Yao:2021pyf, Yao:2021pdy} -- each such fraction is to be multiplied by $10^{-3}$.  The column labelled ``ratio'' is the ratio of the preceding two entries in the row, so \emph{no} factor of $10^{-3}$ is applied in this column.  (A $1\sigma$ SPM uncertainty is listed for the CSM predictions.)
Ref.\,\cite[PDG]{Workman:2022ynf} lists the following values for the CKM matrix elements:
$|V_{us}| =0.2245(8)$, $|V_{cd}| = 0.221(4)$, $|V_{cs}|= 0.987(11)$
$|V_{ub}| = 0.00382(24)$; 
and the following lifetimes (in seconds):
$\tau_{K^+}=1.2379(21)\times 10^{-8}$,
$\tau_{D^0} = 4.10 \times 10^{-13}$,
$\tau_{D_s^\pm} = 5.04 \times 10^{-13}$,
$\tau_{\bar B^0} = 1.519 \times 10^{-12}$,
$\tau_{\bar B_s^0} = 1.515 \times 10^{-12}$,
$\tau_{B_c^\pm} = 0.51 \times 10^{-12}$.
%
}
\begin{tabular*}
{\hsize}
{
l@{\extracolsep{0ptplus1fil}}
|c@{\extracolsep{0ptplus1fil}}
c@{\extracolsep{0ptplus1fil}}
c@{\extracolsep{0ptplus1fil}}
|c@{\extracolsep{0ptplus1fil}}
c@{\extracolsep{0ptplus1fil}}
c@{\extracolsep{0ptplus1fil}}}\hline
${\mathpzc B}_{I\to F(\ell\nu_\ell)}$
& \multicolumn{3}{c|}{Refs.\,\cite{Yao:2021pyf, Yao:2021pdy}} & \multicolumn{3}{c}{PDG \cite{Workman:2022ynf} or other, if indicated} \\\hline
& $e^+ \nu_{e}\ $\phantom{[51]} & $\mu^+ \nu_\mu\ $ & ratio$\ $  & $e^+ \nu_{e}\ $ & $\mu^+ \nu_\mu\ $ & ratio$\ $ \\
$K^+\to\pi^0\ $ & $50.0(9)\phantom{00}\ $ & $33.0(6)\phantom{00}\ $ & $0.665\phantom{(70)}\ $ & $50.7(6)\phantom{00}\ $\phantom{[51]} & $33.5(3)\phantom{22}\ $ & $0.661(07)\ $ \\
$D^0\to\pi^-\ $ & $\phantom{1}2.70(12)\ $ & $\phantom{2}2.66(12)\ $ & $0.987(02)\ $ & $\phantom{0}2.91(4)\phantom{1}\ $\phantom{[51]} & $\phantom{0}2.67(12)\ $ & $0.918(40)\ $ \\
$D_s^+\to K^0\ $ & $\phantom{1}2.73(12)\ $ & $\phantom{2}2.68(12)\ $ & $0.982(01)\ $ & $\phantom{00}3.25(36)\ $\cite{Ablikim:2018upe} &  & \\
$D^0\to K^-\ $ & $39.0(1.7)\ $ & $38.1(1.7)\ $ & $0.977(01)\ $ & $\phantom{0}35.41(34)\phantom{1}\ $\phantom{[51]} & $34.1(4)\phantom{22}\ $ & $0.963(10)\ $ \\\hline
%
& $\mu^- \bar\nu_{\mu}\ $\phantom{[51]} & $\tau^- \bar\nu_\tau\ $ & ratio$\ $  & $\mu^- \bar\nu_{\mu}\ $ & $\tau^- \bar\nu_\tau\ $ & ratio$\ $ \\
$\bar B^0\to \pi^+\ $ & $\phantom{11}0.162(44)\ $ & $\phantom{11}0.120(35)\ $ & $0.733(02)\ $ & $\phantom{01}0.150(06)\ $\phantom{[51]} &  &  \\
$\bar B_s^0\to K^+\ $ & $\phantom{11}0.186(53)\ $ & $\phantom{11}0.125(37)\ $ & $0.667(09)\ $ & &  &  \\
$B_c\to\eta_c\ $ & $\phantom{00}8.10 \, (45)\phantom{0}\ $ & $\phantom{00}2.54(10)\phantom{0}\ $ & $0.31(2)\phantom{00}\ $ & &  &  \\
$B_c\to J/\psi\ $ & $\phantom{0}17.2 \, (1.9)\phantom{0}\ $ & $\phantom{0}4.17(66)\ $ & $0.24(5)\phantom{00}\ $ & & & \\\hline
\hline
\end{tabular*}
\end{table*}

\begin{figure}[t]
\centering
\includegraphics[width=0.66\textwidth]{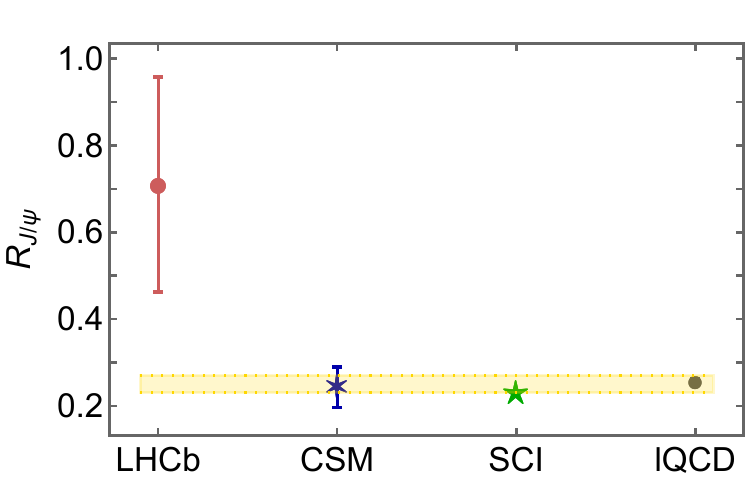}
\caption{\label{RJpsi}
Ratio $R_{J/\psi}$ in Eq.\,\eqref{eqRjpsi} -- red circle, empirical result \cite[LHCb]{Aaij:2017tyk};
blue asterisk -- CSM prediction \cite{Yao:2021pyf};
green star -- SCI prediction \cite{Xing:2022sor};
grey circle -- lQCD result \cite{Harrison:2020nrv, Harrison:2020gvo};
and gold band -- unweighted mean of central values from several calculations \cite{Tran:2018kuv, Issadykov:2018myx, Wang:2018duy, Leljak:2019eyw, Hu:2019qcn, Zhou:2019stx}.
%
}
\end{figure}

\begin{figure}[t]
\centering
\includegraphics[width=0.66\textwidth]{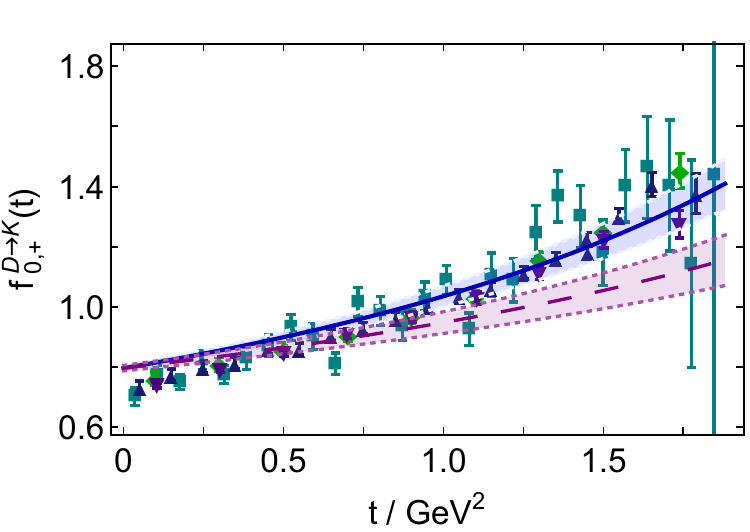}
\caption{\label{PDK}
$D\to K$ transition form factors.
$f_+$ -- solid blue curve; $f_0$ -- long-dashed purple curve; and like-coloured shaded bands -- associated SPM uncertainty in each case.
Data:
cyan squares \cite{Belle:2006idb}; green diamonds \cite{Besson:2009uv};
dark-blue up-triangles \cite{Ablikim:2015ixa}; and indigo down-triangles \cite{BESIII:2017ylw}.
}
\end{figure}

One of the key results in Ref.\,\cite{Yao:2021pyf} concerns a ratio of $B_c^+\to J/\psi$ branching fractions measured for the first time by the LHCb Collaboration fairly recently \cite{Aaij:2017tyk}:
\begin{equation}
\label{eqRjpsi}
R_{J/\psi} := \frac{{\mathpzc B}_{B_c^+\to J/\psi \tau \nu}}{{\mathpzc B}_{B_c^+\to J/\psi \mu \nu}} = 0.71 \pm 0.17 \,{\rm (stat)} \pm 0.18\,{\rm (syst)}\,.
\end{equation}
This value is plotted in Fig.\,\ref{RJpsi} and compared with the CSM prediction and other Standard Model calculations.  Evidently, the LHCb measurement lies approximately $2\sigma$ above the values predicted by reliable SM calculations.  If future, precision experiments do not deliver a markedly lower central value, then one might begin to judge that lepton flavour universality is violated in $B_c\to J/\psi$ semileptonic decays.  As yet, however, the experimental precision is insufficient to support such a claim.  Furthermore, a compelling case would need to include information on $B_c\to \eta_c$ semileptonic decays.  The CSM prediction is $R_{\eta_c}=0.313(22)$ -- see Table~\ref{TabBranch}; the SCI result is $R_{\eta_c}=0.25$ \cite{Xu:2021iwv}; and a mean value of $0.31(4)$ is obtained from modern continuum analyses \cite{Tran:2018kuv, Issadykov:2018myx, Wang:2018duy, Leljak:2019eyw, Hu:2019qcn, Zhou:2019stx}.  An experimental value is lacking.

The array of analyses in Ref.\,\cite{Yao:2021pdy} yield novel results in other areas.
Of particular interest are the discussions of $D^0\to K^-$ transition form factors and the value of $|V_{cs}|$.  Two independent form factors characterise $0^+ \to 0^+$ transitions, \emph{viz}.\ vector and scalar, $f_{+,0}(t)$, respectively, where $t$ is the Mandelstam variable whose value expresses the momentum transferred to the final state.  The CSM predictions are plotted in Fig.\,\ref{PDK} and compared with available data \cite{Belle:2006idb, Besson:2009uv, Ablikim:2015ixa, BESIII:2017ylw}.  The CSM result is largely consistent with this collection, although there may be a hint that it is too high at lower $t$ values.  Concerning branching fractions, form factor contributions from this domain are important.  It is therefore notable that, within mutual uncertainties, the CSM value for $f_+^{D\to K}(0)=0.796(9)$ agrees with the $N_f=2+1+1$ lQCD result in Ref.\,\cite{Lubicz:2017syv}: $0.765(31)$.

With the form factors in Fig.\,\ref{PDK}, Ref.\,\cite{Yao:2021pdy} obtained the $D^0\to K^-$  branching fractions listed in Table~\ref{TabBranch} when using the value of $|V_{cs}|$ listed in the caption: evidently, both the $e^+\nu_e$ and $\mu^+\nu_\mu$ fractions exceed their respective PDG values.  On the other hand, the ratio agrees within $1.4\sigma$; so, a common overall factor can remedy the mismatch.
Adopting this perspective, then the value $|V_{cs}| = 0.937(17)$
combined with the CSM form factors delivers branching fractions that match the PDG values, \emph{viz}.\ $3.52(18)$\% and $3.44(18)$\%, respectively.
Actually, referring to Ref.\,\cite[Sec.\,12.2.4]{Workman:2022ynf}, one sees that the inferred CSM value
is both commensurate with and more precise than one of the two used to arrive at the PDG average listed in the caption of Table~\ref{TabBranch}.  With $|V_{cs}| = 0.937(17)$ used instead to compute this average, one finds a slightly more precise central value that is $1\sigma$ lower:
\begin{equation}
\label{VcspredictionUpdate}
|V_{cs}| = 0.974(10)\,.
\end{equation}

\begin{figure}[t] 
\hspace*{-1ex}\begin{tabular}{lcl}
\large{\textsf{A}} & & \large{\textsf{B}}\\[-3ex]
%
\includegraphics[clip, width=0.47\textwidth]{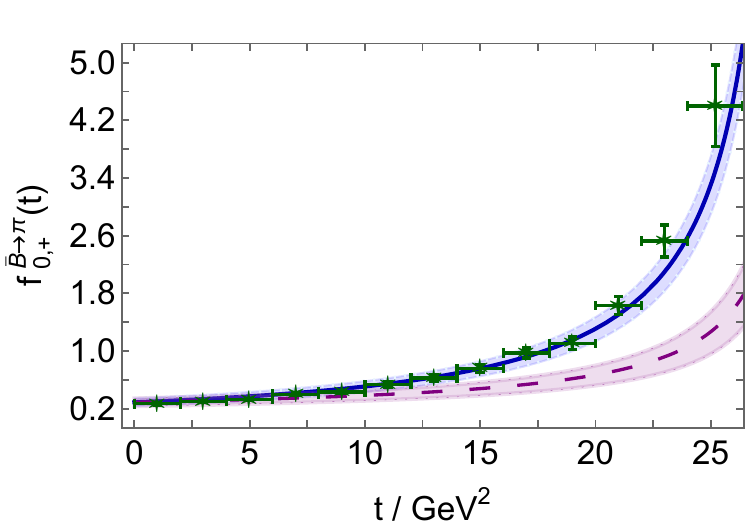} & \hspace*{0.em} &
\includegraphics[clip, width=0.47\textwidth]{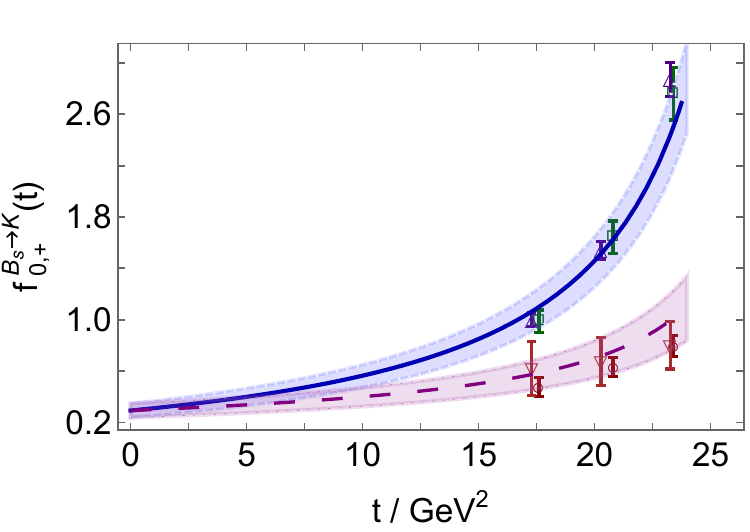}
\end{tabular}
\vspace*{2ex}
\caption{\label{PBinitial}
CSM predictions for $B_{(s)}$ semileptonic transition form factors: $f_+$ -- solid blue curve; $f_0$ -- long-dashed purple curve; and SPM uncertainty in each -- like-coloured shaded bands.
\emph{Left panel}\,--\,\mbox{\sf A}.  $\bar B^0 \to \pi^+$.
Data (green stars): reconstructed
from the average \cite[Tab.\,81]{HFLAV:2019otj} of data reported in Refs.\,\cite{BaBar:2010efp, Belle:2010hep, BaBar:2012thb, Belle:2013hlo}.
\emph{Right panel}\,--\,\mbox{\sf B}.  $\bar B_s^0 \to K^+$.  lQCD results:
$f_+$ -- indigo open up-triangles \cite{Bouchard:2014ypa} and green open boxes \cite{Flynn:2015mha};
$f_0$ --- brown open down-triangles \cite{Bouchard:2014ypa} and red open circles \cite{Flynn:2015mha}.
}
\end{figure}

Predictions for semileptonic $\bar B^0 \to \pi^+$, $\bar B_s^0 \to K^+$ transition form factors and branching fractions were also delivered in Ref.\,\cite{Yao:2021pdy}.  As emphasised above, such processes present challenges because $\pi$, $K$ are Nambu-Goldstone bosons and there is a huge disparity between the masses of the initial and final states.  Consequently, comparisons with data serve as a stringent test of the new CSM algorithms.

CSM predictions for the $\bar B^0 \to \pi^+$ transition form factors are depicted in Fig.\,\ref{PBinitial}A.  Regarding $f_+^{\bar B\to\pi}$, data have been collected by two collaborations \cite{BaBar:2010efp, Belle:2010hep, BaBar:2012thb, Belle:2013hlo}: within mutual uncertainties, the CSM predictions agree with this data.  The data support a value
\begin{equation}
f_+^{\bar B\to \pi}(t=0) = 0.27(2)\,,
\end{equation}
which is consistent with the CSM prediction \cite{Yao:2021pdy}: $0.29(5)$.

Using the form factors in Fig.\,\ref{PBinitial}A, one obtains the $\bar B^0 \to \pi^+$ branching fractions in Table~\ref{TabBranch}.  The PDG lists a result for the $\mu^-\nu_\mu$ final state, which matches the CSM prediction within mutual uncertainties.  Precise agreement is obtained using
\begin{equation}
|V_{ub}| = 0.00374(44)\,.
\end{equation}
This value is commensurate with other analyses of ${\mathpzc B}_{\bar B^0\to \pi^+ \mu^-\bar\nu_\mu}$ \cite[Sec.\,76.3]{Workman:2022ynf} and thus increases tension with the higher value inferred from inclusive decays.
No data is available on the $\tau^-\nu_\tau$ final state; so, the $\tau$:$\mu$ ratio is empirically unknown.  Here, a $N_f=2+1$-flavour lQCD study yields $0.69(19)$ \cite{Flynn:2015mha}, which, within its uncertainty, matches the CSM result:  $0.733(2)$.

Fig.\,\ref{PBinitial}B displays CSM predictions for the $\bar B_s\to K^+$ form factors.  Although the $B_s \to K^-$ transition was recently observed \cite{LHCb:2020ist}, with the measurement yielding the branching fraction
\begin{equation}
\label{LHCbmeasure}
{\mathpzc B}_{B_s^0\to K^- \mu^+ \nu_\mu} = [ 0.106 \pm 0.005_{\rm stat} \pm 0.008_{\rm syst}] \times 10^{-3}\,,
\end{equation}
no form factor data are yet available.  Comparisons are therefore made in Fig.\,\ref{PBinitial}B with results obtained using $N_f=2+1$-flavour lQCD \cite{Bouchard:2014ypa, Flynn:2015mha}.
Owing to difficulties encountered when using lattice methods to calculate form factors of heavy+light mesons, lQCD results are limited to a few points on the domain $t \gtrsim 17\,$GeV$^2$ -- see Fig.\,\ref{PBinitial}B.  Today, lattice analyses typically employ such results to construct a least-squares fit to the form factor points, using some practitioner-favoured functional form.  That fit is then employed to define the form factor on the whole kinematically accessible domain: $0\lesssim t \lesssim 25\,$GeV$^2$ in this case.
It is worth noting that at this time, given the small number of points and their limited precision, the SPM cannot gainfully be used to develop function-form unbiased interpolations and extrapolations of the lQCD output.

The CSM form factors in Fig.\,\ref{PBinitial}B yield the $\bar B_s^0 \to K^+$ branching fractions in Table~\ref{TabBranch}.  Fig.\,\ref{PBFBK}A compares the $\mu^-\bar\nu_\mu$ value with the measurement in Eq.\,\eqref{LHCbmeasure} and also results obtained via various other means.  Experiment and theory only agree because the theory uncertainty is large.
The unweighted theory average is $0.141(44)$\permil\ and the uncertainty weighted mean is $0.139(08)$\permil.
These values increase when entries V--VI \cite{Flynn:2015mha, FermilabLattice:2019ikx} are omitted:
unweighted $0.159(38)$\permil\ and uncertainty weighted $0.156(10)$\permil.
The extrapolations employed in V--VI \cite{Flynn:2015mha, FermilabLattice:2019ikx} lead to values of $f_+^{\bar B_s\to K}(0)$ that are $\sim 50$\% of those obtained in I--IV \cite{Wu:2006rd, Faustov:2013ima, Xiao:2014ana, Bouchard:2014ypa}: $0.148(53)$  vs.\ $0.299(86)$.  This can explain the difference in branching fractions: V--VI vs.\ I--IV in Fig.\,\ref{PBFBK}A.
Significantly, a different approach to fitting and extrapolating lQCD results, using the LHCb datum, Eq.\,\eqref{LHCbmeasure}, as an additional constraint, produces \cite{Gonzalez-Solis:2021awb}: $f_+^{\bar B_s\to K}(0)=0.211(3)$.

Ref.\,\cite{Gonzalez-Solis:2021awb} also infers $f_+^{\bar B\to \pi}(0)=0.255(5)$, leading to $f_+^{\bar B_s\to K}(0)/f_+^{\bar B\to \pi}(0)<1$.  This outcome conflicts with the CSM prediction, which has $f_+^{\bar B_s\to K}(0)/f_+^{\bar B\to \pi}(0)>1$ at the 85\% confidence level \cite[Eq.\,(9)]{Yao:2021pdy}, and the results in a raft of other studies, \emph{e.g}., Refs.\,\cite{Melikhov:2001zv, Faessler:2002ut, Ebert:2003wc, Ball:2004ye, Wu:2006rd, Khodjamirian:2006st, Lu:2007sg, Ivanov:2007cw, Faustov:2013ima}.  It is probable, therefore, that the Ref.\,\cite{Gonzalez-Solis:2021awb} value for $f_+^{\bar B_s\to K}(0)$ is too small.
It is worth remarking that the SCI is unclear on the value of this ratio.  It produces $f_+^{\bar B_s\to K}(0)/f_+^{\bar B\to \pi}(0)<1$, but the individual $t=0$ values are too large by a factor of two \cite[Table~3A]{Xu:2021iwv}.  On the other hand, the $t=0$ value of the kindred ratio of vector form factors in $\bar B_s\to K^\ast$, $\bar B \to \rho$ transitions is greater than unity \cite[Table~1A]{Xing:2022sor}.
Notably, the ratio $f_+^{\bar B_s\to K}(0)/f_+^{\bar B\to \pi}(0)$ is a marker for SU$(3)$-flavour symmetry breaking and its modulation by EHM; so it is worth reaching a sound conclusion on the value of the ratio.   It is here relevant to observe that $f_K/f_\pi = 1.2 > 1$.

\begin{figure}[t] 
\hspace*{-1ex}\begin{tabular}{lcl}
\large{\textsf{A}} & & \large{\textsf{B}}\\[-3ex]
%
\includegraphics[clip, width=0.47\textwidth]{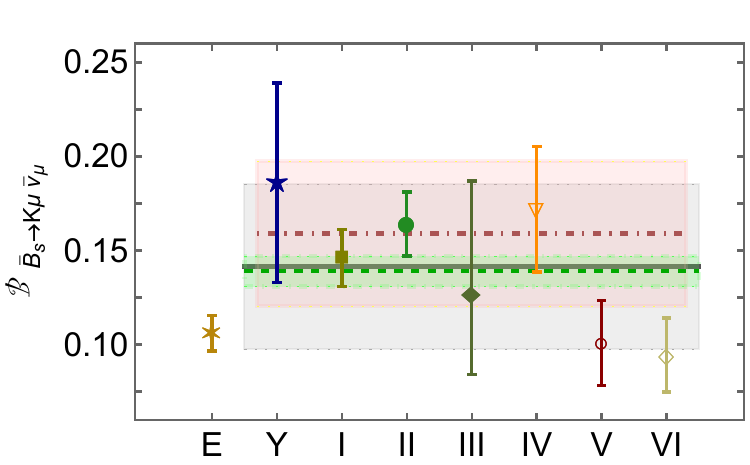} & \hspace*{0.em} &
\includegraphics[clip, width=0.47\textwidth]{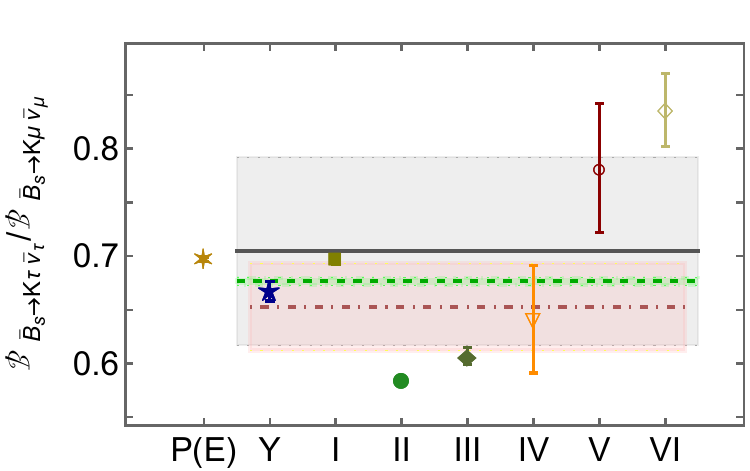}
\end{tabular}
\vspace*{2ex}
\caption{\label{PBFBK}
\emph{Left panel}\,--\,\mbox{\sf A}.  Branching fraction ${\mathpzc B}_{\bar B_s^0\to K^+ \mu^- \bar\nu_\mu}$ computed in Ref.\,\cite{Yao:2021pdy}, ``Y'', compared with
the value in Eq.\,\eqref{LHCbmeasure}, ``E'', \emph{viz}.\ a measurement of ${\mathpzc B}_{B_s^0\to K^- \mu^+ \nu_\mu}$ \cite{LHCb:2020ist}
and some results obtained using other approaches: continuum I -- III \cite{Wu:2006rd, Faustov:2013ima, Xiao:2014ana}; and lattice IV -- VI \cite{Bouchard:2014ypa, Flynn:2015mha, FermilabLattice:2019ikx}.
\emph{Right panel}\,--\,\mbox{\sf B}.
Branching fraction ratio ${\mathpzc B}_{\bar B_s^0\to K^+ \tau^- \bar\nu_\tau}/{\mathpzc B}_{\bar B_s^0\to K^+ \mu^- \bar\nu_\mu}$ computed in Ref.\,\cite{Yao:2021pdy} compared with some results obtained using other approaches.  Legend matches Panel~\mbox{\sf A}, except ``P(E)'' is the result from \cite{Gonzalez-Solis:2021awb}, \emph{i.e}., an estimate constrained by the datum in Ref.\,\cite{LHCb:2020ist}.
Both panels --
Grey line: unweighted mean of theory results.
Pink dot-dashed line: unweighted mean of theory results, omitting \mbox{\sf V--VI}.
Green dashed line: uncertainty-weighted average of theory results.
Like-coloured bands mark associated uncertainties in each case.
}
\end{figure}

Regarding the $\tau \nu_\tau$ final state in $\bar B_s\to K$ transitions, no empirical information is currently available; hence, none on the $|V_{ub}|$-independent ratio which would test lepton flavour universality.  In Fig.\,\ref{PBFBK}B, therefore, we compare the CSM prediction for this ratio, drawn from Table~\ref{TabBranch}, with results obtained via other means.
The unweighted average of theory results is
$0.705(87)$ and the uncertainty weighted mean is $0.678(03)$.
Omitting entries V--VI \cite{Flynn:2015mha, FermilabLattice:2019ikx}, these values are:
unweighted $0.653(41)$ and uncertainty weighted $0.677(03)$.
Within sound analyses, many uncertainties cancel in this ratio; so the results should be more reliable than any calculation of either fraction alone.  Nevertheless, the values are widely scattered, indicating that there is ample room for improving the precision of $\bar B_s^0 \to K$ theory.
Of course, measurements enabling extraction of $B_s^0\to K^-$ form factors would be very useful, too, for refining both (\emph{a}) comparisons with theory and between theory analyses and (\emph{b}) making progress toward a more accurate value of $|V_{ub}|$.

Such predictions for heavy-to-light meson electroweak transition form factors are a new branch of application for CSMs.  They are far more sophisticated and robust than the Schwinger-function parametrisation-based analyses in, \emph{e.g}.,  Refs.\,\cite{Ivanov:1998ms, Ivanov:2007cw, El-Bennich:2010uqs} and significantly improve upon earlier RL truncation studies of $\pi_{\ell 3}$ and $K_{\ell 3}$ transitions \cite{Ji:2001pj, Chen:2012txa}.  The keys to these advances are an improved understanding of RL truncation and the capacity to greatly expand its quark mass and mass-splitting domains of applicability using the SPM.  Fairly soon, one can expect these advances to be exploited in the study of kindred baryon transitions.

\section{Distribution Functions}
\label{SecDFs}
Hadron parton DFs are probability densities: each one describes the light-front fraction, $x$, of the hadron's total momentum carried by a given parton species within the bound-state \cite{Holt:2010vj}.  They are a much prized source of hadron structure information; and following the quark discovery experiments fifty years ago \cite{Taylor:1991ew, Kendall:1991np, Friedman:1991nq, Friedman:1991ip}, measurements interpretable in terms of hadron DFs have been awarded a high priority.  For much of this time, DFs were inferred from global fits to data, with the results viewed as benchmarks.  Such fitting remains crucial, providing input for the conduct of a huge number of experiments worldwide; but the past decade has seen the dawn of a new theory era.  Continuum and lattice studies of QCD are beginning to yield robust predictions for the pointwise behaviour of DFs; and these developments are exposing potential conflicts with the fitting results \cite{Roberts:2021nhw, Chang:2021utv, Cui:2021mom, Cui:2022bxn, Chang:2022jri, Lu:2022cjx, dePaula:2022pcb}.

Notwithstanding the enormous expense of time and effort, much yet remains to be learnt before proton and pion structure may be judged as understood in terms of DFs.  For instance and most simply, it is still unclear whether there are differences between the distributions of partons within the proton (Nature's most fundamental bound-state) and the pion (Nature's most fundamental (near) Nambu-Goldstone boson).  Plainly, if there are differences, then they must be explained.  As we have stressed above, answering the question of similarity/difference between proton and pion DFs is particularly important today as science seeks to expose and explain EHM \cite{Roberts:2020udq, Roberts:2020hiw, Roberts:2021xnz, Roberts:2021nhw, Binosi:2022djx, Papavassiliou:2022wrb}.

Regarding DFs measured in processes that do not resolve beam or target polarisation, practitioners experienced and involved with solving bound-state problems in QCD have learnt that, at the hadron scale, $\zeta_{\cal H}<m_p$, valence-quark DFs in the proton and pion behave as follows \cite{Brodsky:1994kg, Yuan:2003fs, Cui:2021mom, Cui:2022bxn, Chang:2022jri}:
\begin{equation}
\label{LargeX}
{\mathpzc d}^p(x;\zeta_{\cal H}), {\mathpzc u}^p(x;\zeta_{\cal H}) \stackrel{x\simeq 1}{\propto} (1-x)^3\,,
\quad
\bar {\mathpzc d}^\pi(x;\zeta_{\cal H}), {\mathpzc u}^\pi(x;\zeta_{\cal H})  \stackrel{x\simeq 1}{\propto} (1-x)^2\,;
\end{equation}
and it subsequently follows from the DGLAP equations \cite{Dokshitzer:1977sg, Gribov:1971zn, Lipatov:1974qm, Altarelli:1977zs} that the large-$x$ power on the related gluon DF is approximately one unit larger; and that for sea quark DFs is roughly two units larger.
%
Moreover, as the resolving scale increases to $\zeta > \zeta_{\cal H}$, all these exponents grow logarithmically.  However, fueling controversy and leading some to question the veracity of QCD \cite{Aicher:2010cb, Cui:2021mom, Cui:2022bxn}, these constraints are typically ignored in fits to the world's data on deep inelastic scattering (DIS) and kindred processes \cite{Ball:2016spl, Hou:2019efy, Bailey:2020ooq, Novikov:2020snp, Barry:2021osv}.
Furthermore, largely because pion data are scarce \cite[Table~9.5]{Roberts:2021nhw}, proton and pion data have never been considered simultaneously.  Therefore, the unified body of results in Ref.\,\cite{Lu:2022cjx}, which uses a single symmetry-preserving framework to predict the pointwise behaviour of all proton and pion DFs -- valence, glue, and four-flavour-separated sea -- is a significant advance.

In order to sketch this progress, it is necessary to recall that the modern approach to the CSM prediction of hadron DFs%
\footnote{Contemporary continuum methods for obtaining light-front amplitudes and density distributions from Euclidean space Schwinger functions are detailed, \emph{e.g}., in Refs.\,\cite{Chang:2013pq, Chang:2022jri}, \cite[Secs.\,3, 5]{Roberts:2021nhw}, \cite[Sec.\,IV]{Ding:2019lwe}, \cite[Secs.\,2, 5]{Cui:2020tdf}.}
is based on a single proposition \cite{Raya:2021zrz, Cui:2021mom, Cui:2022bxn, Chang:2022jri, Lu:2022cjx}:
%
\begin{description}
\item[\mbox{\sf P1}] ... There is an effective charge, $\alpha_{1\ell}(k^2)$, which, when used to integrate the one-loop perturbative-QCD DGLAP equations, defines a DF evolution scheme that is all-orders exact.  \label{P1item} 
\end{description}


\noindent As noted in connection with Fig.\,\ref{Falpha}, charges of this sort are discussed in Refs.\,\cite{Grunberg:1982fw, Grunberg:1989xf, Dokshitzer:1998nz}.  They need not be process-inde\-pen\-dent (PI); hence, not unique.  Moreover, the results delivered are independent of the explicit form of $\alpha_{1\ell}(k^2)$.
Notwithstanding these things, a suitable PI charge is available, \emph{viz}.\ the coupling discussed in Sec.\,\ref{SecPICharge}, which has proved efficacious.  In being defined by an observable -- in this instance, structure functions -- each such $\alpha_{1\ell}(k^2)$ is \cite{Deur:2016tte}:
consistent with the renormalisation group and renormalisation scheme independent;
everywhere analytic and finite;
and, crucially, provides an infrared completion of any standard perturbative running coupling.

\mbox{\sf P1} was used in Refs.\,\cite{Cui:2020dlm, Cui:2020tdf, Raya:2021zrz, Han:2020vjp, dePaula:2022pcb} to deliver meson DFs with a flavour symmetric sea.  A generalisation, which expresses key quark current-mass effects in the evolution kernels, was introduced in Ref.\,\cite{Lu:2022cjx} and used for the proton and pion.
It features a threshold function ${\cal P}_{qg}^\zeta \sim \theta(\zeta - \delta_{\mathpzc f})$, which ensures that a given quark flavour only becomes active in DF evolution when the energy scale exceeds a value determined by the quark's mass \cite[Fig.\,2.5]{Roberts:2021nhw}: $\delta_{u,d} \approx 0$,  $\delta_{s} \approx 0.1\,$GeV, $\delta_{c} \approx 0.9\,$GeV.
%
%
The impact of this modification is readily anticipated.
Supposing that all quark flavours are light, then each would be emitted with equal probability on $\zeta>\zeta_{\cal H}$; so, evolution would produce a certain gluon momentum fraction in the hadron plus a sea-quark fraction shared equally between all quark flavours.
Considering mass differences between the quarks, with some flavours being heavier than the light-quark threshold, then evolution on $\zeta>\zeta_{\cal H}$ will generate a gluon momentum fraction that is practically unchanged from the all-light quark case and a sea-quark fraction divided amongst the quarks in roughly inverse proportion to their mass.

It is worth reiterating here that $\zeta_{\cal H}$ is the scale at which the valence quasiparticle degrees-of-freedom carry all properties of a given hadron \cite{Ding:2019qlr, Ding:2019lwe, Cui:2020dlm, Cui:2020tdf, Han:2020vjp, Chang:2021utv, Xie:2021ypc, Raya:2021zrz, Cui:2021mom, Cui:2022bxn, Chang:2022jri, Lu:2022cjx, dePaula:2022pcb}.  Moreover, the value of this scale is a prediction.  Using the PI charge discussed in Sec.\,\ref{SecPICharge} to construct bound-state kernels informed by Refs.\,\cite{Qin:2011dd, Binosi:2014aea}, then \begin{equation}
\label{EqzetaH}
\zeta_{\cal H} = 0.331(2) \, {\rm GeV}.
\end{equation}

The value in Eq.\,\eqref{EqzetaH} is the same for all hadrons.

Furthermore, combined with evolution according to \mbox{\sf P1}, the character of $\zeta_{\cal H}$ ensures that all hadron DFs are intertwined at every scale $\zeta$.  Hence, this perspective suggests that it is incorrect to choose independent, uncorrelated functions to parametrise the DFs of different parton species when fitting data at any scale $\zeta > \zeta_{\cal H}$.  If one nevertheless chooses to ignore the innate associations, then DFs with unphysical features may be obtained -- see, e.g., Ref.\,\cite[Fig.\,6]{Cui:2021mom}.

\begin{figure}[t] 
\hspace*{-1ex}\begin{tabular}{lcl}
\large{\textsf{A}} & & \large{\textsf{B}}\\[-3ex]
%
\includegraphics[clip, width=0.47\textwidth]{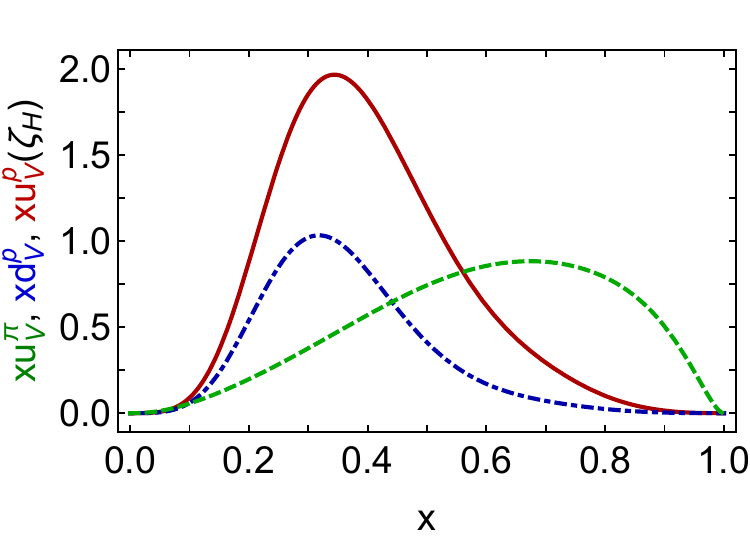} & \hspace*{0.em} &
\includegraphics[clip, width=0.47\textwidth]{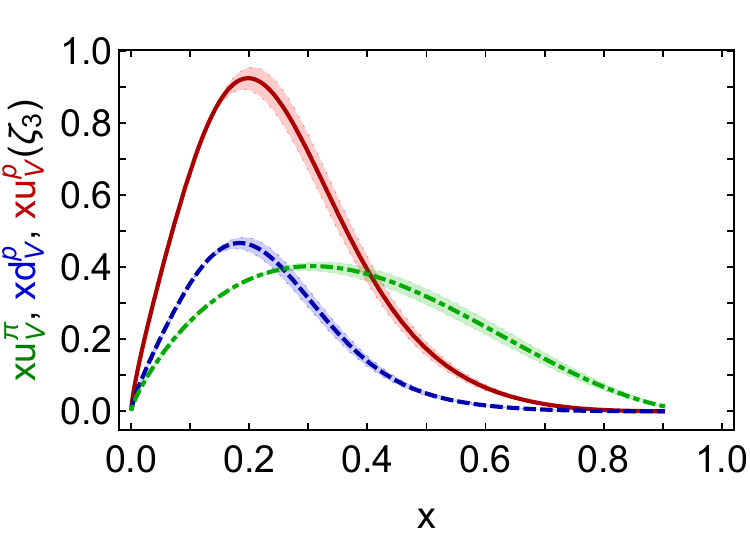}
\end{tabular}
\vspace*{2ex}
\caption{\label{ImageValence}
\emph{Left panel}\,--\,\mbox{\sf A}.
Proton and pion hadron scale valence parton DFs:
$x {\mathpzc u}^p(x;\zeta_{\cal H})$ -- solid red curve;
$x {\mathpzc d}^p(x;\zeta_{\cal H})$ -- dot-dashed blue curve;
and $x {\mathpzc u}^\pi(x;\zeta_{\cal H})$ -- dashed green curve.
\emph{Right panel}\,--\,\mbox{\sf B}.
Valence DFs in panel \mbox{\sf A} evolved to $\zeta_3=m_{J/\psi}=3.097\,$GeV according to \mbox{\sf P1}.
The band surrounding each CSM curve expresses the response to a $\pm 5$\% variation in $\zeta_{\cal H}$.
}
\end{figure}

CSM predictions for the $\zeta=\zeta_{\cal H}$ proton and pion valence DFs are drawn in Fig.\,\ref{ImageValence}A.
The following points are significant.
(\emph{i}) Each DF is consistent with the relevant large-$x$ scaling law in Eq.\,\eqref{LargeX}.  Hence, from the outset, whilst the $\zeta=\zeta_{\cal H}$ momentum sum rules for each hadron are necessarily saturated by valence degrees-of-freedom, \emph{viz}.\
\begin{equation}
\langle x \rangle_{{\mathpzc u}_p}^{\zeta_{\cal H}}=0.687\,,\;
\langle x \rangle_{{\mathpzc d}_p}^{\zeta_{\cal H}} = 0.313\,,\;
\langle x \rangle_{{\mathpzc u}_\pi}^{\zeta_{\cal H}} =0.5\,,
\end{equation}
the proton and pion valence DFs nevertheless have markedly different $x$-dependence.  (Nature's approximate ${\cal G}$-parity symmetry \cite{Lee:1956sw} entails $\bar {\mathpzc d}_\pi(x;\zeta)={\mathpzc u}_\pi(x;\zeta)$.)
(\emph{ii}) Owing to DCSB \cite{Lane:1974he, Politzer:1976tv, Pagels:1978ba, Pagels:1979hd, Higashijima:1983gx, Roberts:2000aa, Binosi:2016wcx}, an important corollary of EHM, QCD dynamics simultaneously produce a dressed light-quark mass function, $M_{u,d}(k^2)$, that is large at infrared momenta, $M_D:= M_{u,d}(k^2\simeq 0)  \approx 0.4\,$GeV and an almost massless pion, $m_\pi^2/M^2_D \approx 0.1$ -- see Ref.\,\cite[Sec.\,2]{Roberts:2021nhw}.  As a result, ${\mathpzc u}^\pi(x;\zeta_H)$ is Nature's most dilated hadron-scale valence DF.  This is highlighted by Fig.\,\ref{ImageValence}A and Refs.\,\cite{Cui:2021dlm, Cui:2020tdf}, and implicit in numerous other symmetry-preserving analyses, \emph{e.g}., Refs.\,\cite{Gao:2014bca, Binosi:2018rht, Ding:2018xwy, Lu:2021sgg}.

Evolving the DFs in Fig.\,\ref{ImageValence}A according to \mbox{\sf P1}, one obtains the $\zeta = m_{J/\psi}=:\zeta_3$ distributions in Fig.\,\ref{ImageValence}B.  Plainly, although the profiles change, the relative dilation of the DFs is preserved and is therefore a verifiable prediction of the EHM paradigm.

Given that Fig.\,\ref{ImageValence}B depicts the first CSM predictions for proton valence-quark DFs, one might question their reliability.  That issue can partly be addressed through a comparison with lQCD results.  The calculation of individual valence DFs using lQCD is problematic owing to difficulties in handling so-called disconnected contributions \cite{Alexandrou:2013cda}.  In the continuum limit, however, disconnected diagrams do not contribute to the isovector DF $[{\mathpzc u}^p(x;\zeta) - {\mathpzc d}^p(x;\zeta)]$, so computations of this difference are available \cite{Lin:2020fsj, Alexandrou:2021oih}.  Both analyses use the quasidistribution approach \cite{Ji:2020ect}, but the lattice algorithms and configurations are somewhat different.  Their comparison with CSM predictions is depicted in Fig.\,\ref{FigIsoVector}.  The  level of agreement is encouraging; especially because refinements of both continuum and lattice calculations may be anticipated.  For instance, the CSM predictions were obtained using a simplified proton Faddeev amplitude and the lattice studies must address issues with, \emph{inter alia}, the pion masses used, lattice artefacts and systematic errors, and convergence of the boost expansion in the quasidistribution approach.  (The last of these is a particular hindrance to lQCD extractions of DF endpoint behaviour \cite{Xu:2018eii}.)

\begin{figure}[t]
\centering
\includegraphics[width=0.66\textwidth]{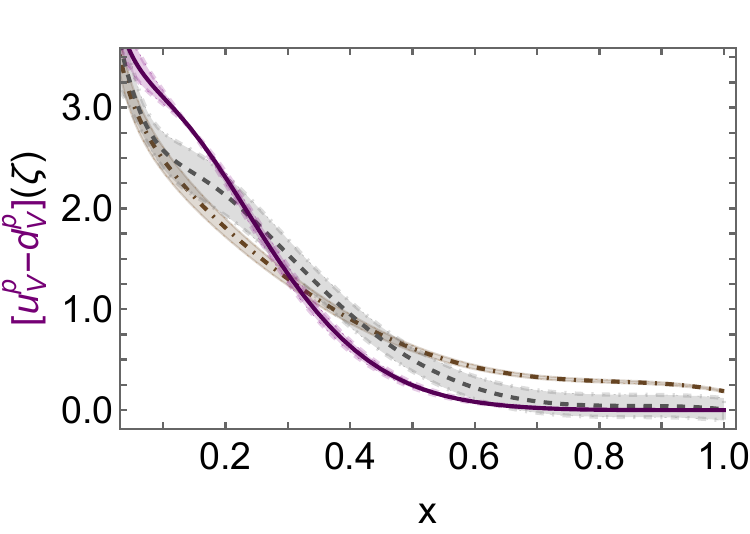}
\caption{\label{FigIsoVector}
Isovector distribution $[{\mathpzc u}^p(x;\zeta) - {\mathpzc d}^p(x;\zeta)]$.
CSM prediction -- solid purple curve, $\zeta = \zeta_3$;
lQCD result from Ref.\,\cite{Lin:2020fsj} -- dashed grey curve, $\zeta = \zeta_3$;
lQCD result from Ref.\,\cite{Alexandrou:2021oih} -- dot-dashed brown curve, $\zeta = \zeta_2$.
Like-coloured band bracketing each curve indicates associated uncertainty.
}
\end{figure}

In typical evolution kernels, gluon splitting yields quark+antiquark pairs of all flavours with equal probability.  However, it was long ago argued \cite{Field:1976ve} that because the proton contains two valence $u$ quarks and one valence $d$ quark, the Pauli exclusion principle should force gluon splitting to prefer $d+\bar d$ production over $u+\bar u$.
Consequently, when implementing evolution of proton singlet and glue DFs, Ref.\,\cite{Lu:2022cjx} followed Ref.\,\cite{Chang:2022jri} and introduced a small Pauli blocking factor into the gluon splitting function.  This correction preserves baryon number, but shifts momentum into $d+\bar d$ from $u+\bar u$, otherwise leaving the sum of sea-quark momentum fractions unchanged.  It vanishes with increasing $\zeta$, in order to express the declining influence of valence-quarks as the proton's sea and glue content increases.

The resulting CSM predictions for $\zeta=\zeta_3$ proton and pion glue DFs are drawn in Fig.\,\ref{ImageGlue}A.
The glue-in-$\pi$ DF is directly related to the $\zeta= 2\,{\rm GeV}=:\zeta_2$ result discussed in Ref.\,\cite{Chang:2021utv}, which is drawn in Fig.\,\ref{ImageGlue}B: evidently, it agrees with a recent lQCD calculation of the glue-in-$\pi$ DF \cite{Fan:2021bcr}.
Furthermore, reproducing the pattern seen with valence-quark DFs in Fig.\,\ref{ImageValence}, Fig.\,\ref{ImageGlue}A reveals that the glue-in-$\pi$ DF possesses significantly more support on the valence domain, $x\gtrsim 0.1$, than the glue-in-$p$ DF.  Once again, this feature is a measurable expression of EHM.

The $\zeta=\zeta_3$ light-quark sea DFs for the proton and pion are depicted in Fig.\,\ref{ImageGlue}C.  The EHM-induced pattern is also apparent here, \emph{viz}.\ the sea-in-$\pi$ DF possesses greater support on $x\gtrsim 0.1$ than the kindred sea-in-$p$ DFs.
DFs of the heavier sea quarks are also generated via evolution, with the results drawn in Fig.\,\ref{ImageGlue}D.  Interestingly, the $\zeta=\zeta_3$ $s$ and $c$ quark DFs are similar in size to those of the light-quark sea DFs; and, for these heavier quarks, too, the pion DFs have significantly greater support on the valence domain, $x\gtrsim 0.1$, than the related proton DFs.

\begin{figure}[t] 
\hspace*{-1ex}\begin{tabular}{lcl}
\large{\textsf{A}} & & \large{\textsf{B}}\\[-3ex]
%
\includegraphics[clip, width=0.47\textwidth]{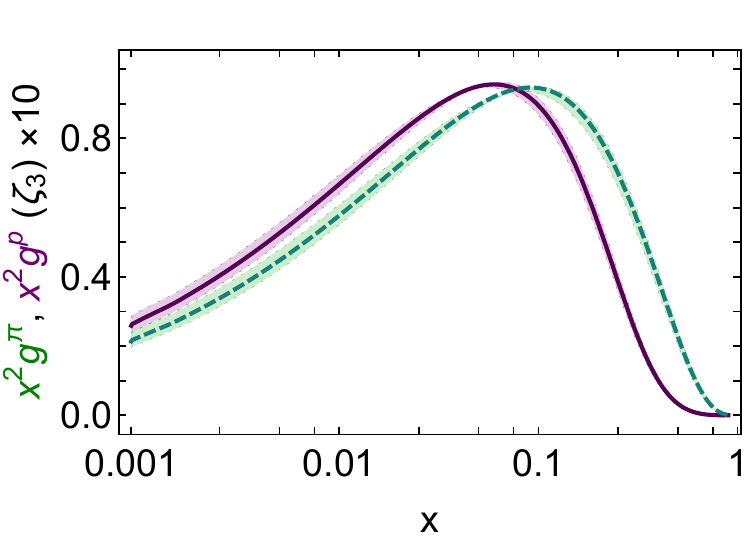} & \hspace*{0.em} &
\includegraphics[clip, width=0.47\textwidth]{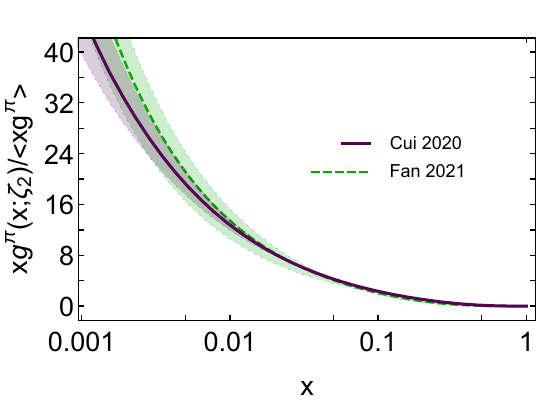}\\
\large{\textsf{C}} & & \large{\textsf{D}}\\[-2ex]
%
\includegraphics[clip, width=0.47\textwidth]{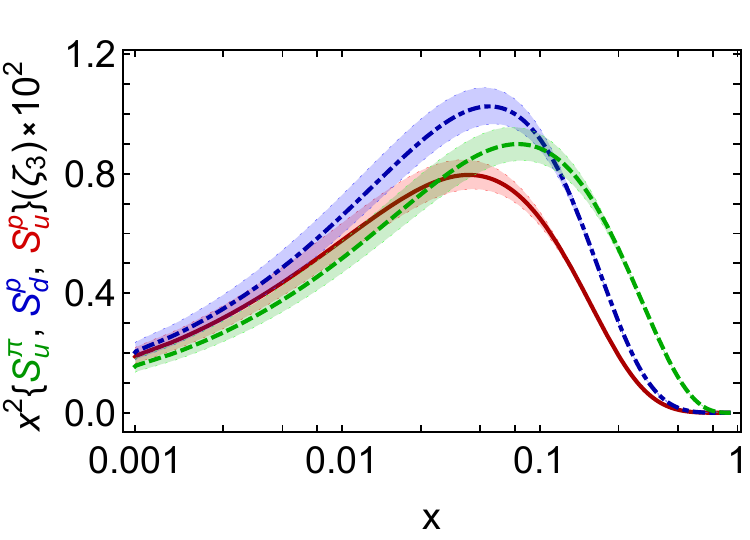} & \hspace*{0.em} &
\includegraphics[clip, width=0.47\textwidth]{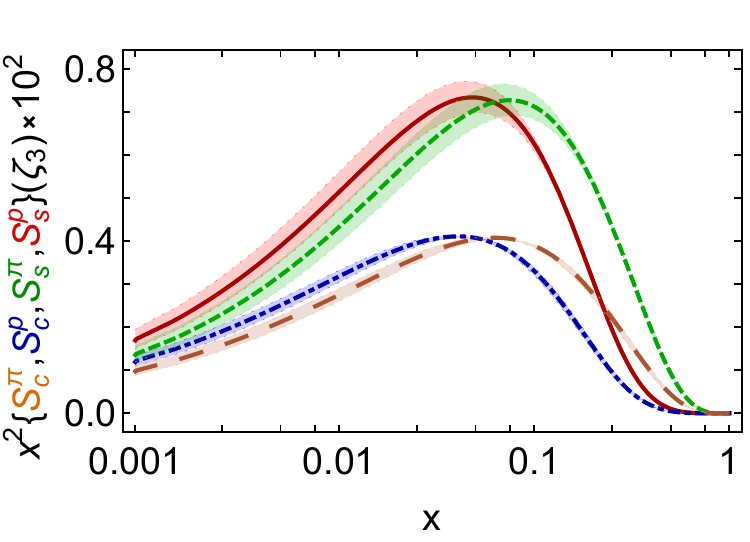}
\end{tabular}
\vspace*{2ex}
\caption{\label{ImageGlue}
\emph{Left upper panel}\,--\,\mbox{\sf A}.
Glue DFs -- $x^2 {\mathpzc g}$, in the proton (solid purple curve) and pion (dashed green curve) at $\zeta=\zeta_3$.  
\emph{Right upper panel}\,--\,\mbox{\sf B}.
Comparison between continuum \cite[Cui\,2020]{Cui:2020tdf} and lattice \cite[Fan\,2021]{Fan:2021bcr} results for the glue-in-pion DF at $\zeta=2\,$GeV.
\emph{Left lower panel}\,--\,\mbox{\sf C}.
Proton and pion light quark sea DFs:
$x^2 {\mathpzc S}_u^p(x;\zeta_{3})$ -- solid red curve;
$x^2 {\mathpzc S}_d^p(x;\zeta_{3})$ -- dashed blue curve;
and $x^2 {\mathpzc S}_u^\pi(x;\zeta_{3})$ -- dot-dashed green curve.
\emph{Right lower panel}\,--\,\mbox{\sf D}.
Proton and pion $c$- and $s$-quark sea DFs:
$x^2 {\mathpzc S}_s^p(x;\zeta_{3})$ -- solid red curve;
$x^2 {\mathpzc S}_s^\pi(x;\zeta_{3})$ -- dashed green curve;
$x^2 {\mathpzc S}_c^p(x;\zeta_{3})$ -- dot-dashed blue curve;
and
$x^2 {\mathpzc S}_c^\pi(x;\zeta_{3})$ -- long-dashed orange curve.
The band surrounding each CSM curve expresses the response to a $\pm 5$\% variation in $\zeta_{\cal H}$.  The uncertainty in the lQCD result is similarly indicated in Panel~\mbox{\sf B}.
}
\end{figure}

An analysis of the endpoint exponents of all $\zeta=\zeta_3$ DFs is also contained in Ref.\,\cite{Lu:2022cjx} along with simple interpolations of each DF that can readily be used by any practitioner -- Ref.\,\cite[Table~1]{Lu:2022cjx}.  It is worth reiterating the following remarks about the endpoint exponents.\\[-3.5ex]
\begin{description}
\item[(\emph{i})] The power laws express \emph{measurable} effective exponents, obtained from separate linear fits to $\ln[ x{\mathpzc p}(x)]$ on the domains $0<x<0.005$, $0.85<x<1$.  (Here, ${\mathpzc p}(x)$ denotes a generic DF.)
\item[(\emph{ii})] %
Within mutual uncertainties, proton and pion DFs have the same power-law behaviour on $x\simeq 0$:
\begin{equation}
\alpha^{p,\pi}_{\rm valence} \approx -0.22\,, \quad
\alpha^{p,\pi}_{\rm glue} \approx -1.6\,, \quad
\alpha^{p,\pi}_{\rm sea} \approx -1.5\,.
\end{equation}
\item[(\emph{iii})] On $x\simeq 1$, the following relationships exist for and between pion and proton DF exponents:
\begin{subequations}
\begin{align}
\beta^{\pi}_{\rm valence} & \approx 2.5\,,\quad
\beta^{p}_{\rm valence} \approx \beta^{\pi}_{\rm valence} + 1.6\,, \\
\beta^{p,\pi}_{\rm glue}&  \approx \beta^{p,\pi}_{\rm valence} + 1.4\,, \quad
\beta^{p,\pi}_{\rm sea} \approx \beta^{p,\pi}_{\rm valence} + 2.4\,.
\end{align}
\end{subequations}
\item[(\emph{iv})] Given (\emph{ii}), (\emph{iii}), then the CSM predictions are consistent with the QCD expectations discussed in connection with Eq.\,\eqref{LargeX}.
\item[(\emph{v})]  Existing phenomenological fits to relevant scattering data typically arrive at DFs which are inconsistent with (\emph{ii}), (\emph{iii}); hence, fail to meet many QCD-based expectations, \emph{e.g}., Refs.\,\cite{Accardi:2016qay, NNPDF:2017mvq, Hou:2019efy, Novikov:2020snp, Barry:2021osv}.  This point is also discussed elsewhere \cite{Courtoy:2020fex, Cui:2021mom, Cui:2022bxn}.
\end{description}

Owing to the Pauli blocking factor described above and as evident in Fig.\,\ref{ImageGlue}C, the DFs calculated in Ref.\,\cite{Lu:2022cjx} express an in-proton separation between $\bar d$ and $\bar u$ distributions.  This entails a violation of the Gottfried sum rule \cite{Gottfried:1967kk, Brock:1993sz}, which has been seen in experiments  \cite{NewMuon:1991hlj, NewMuon:1993oys, NA51:1994xrz, NuSea:2001idv, SeaQuest:2021zxb}.
Using the proton DFs in Fig.\,\ref{ImageGlue}C then, on the domain covered by the measurements in Refs.\,\cite{NewMuon:1991hlj, NewMuon:1993oys}, one obtains
\begin{equation}
\label{gottfried}
\int_{0.004}^{0.8} dx\,[\bar {\mathpzc d}(x;\zeta_3) - \bar {\mathpzc u}(x;\zeta_3)]
= 0.116(12)
\end{equation}
for the Gottfried sum rule discrepancy.  This value matches that inferred from recent fits to a large sample of high-precision data ($\zeta = 2\,$GeV) \cite[CT18]{Hou:2019efy}: 0.110(80); and is far more precise.
%

The result in Eq.\,\eqref{gottfried} corresponds to a strength for the Pauli blocking term in the gluon splitting function that shifts just $\approx 25$\% of the $u$ quark sea momentum fraction into the $d$ quark sea at $\zeta = \zeta_2$.  Changing the strength by $\pm 25$\% leads to the uncertainty indicated in Eq.\,\eqref{gottfried}.
Data from the most recent experiment focused on the asymmetry of antimatter in the proton \cite[E906]{SeaQuest:2021zxb} are presented in Fig.\,\ref{ImageSeaQuest}A.  They may be compared with the CSM result obtained using the proton DFs in Fig.\,\ref{ImageGlue}C.  
Evidently, a modest Paul blocking effect in the gluon splitting function is sufficient to explain modern data.

\begin{figure}[t] 
\hspace*{-1ex}\begin{tabular}{lcl}
\large{\textsf{A}} & & \large{\textsf{B}}\\[-3ex]
%
\includegraphics[clip, width=0.47\textwidth]{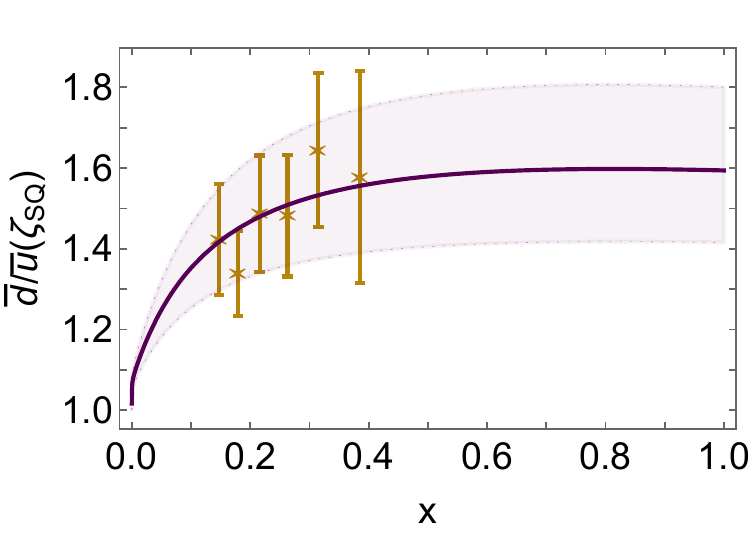} & \hspace*{0.em} &
\includegraphics[clip, width=0.47\textwidth]{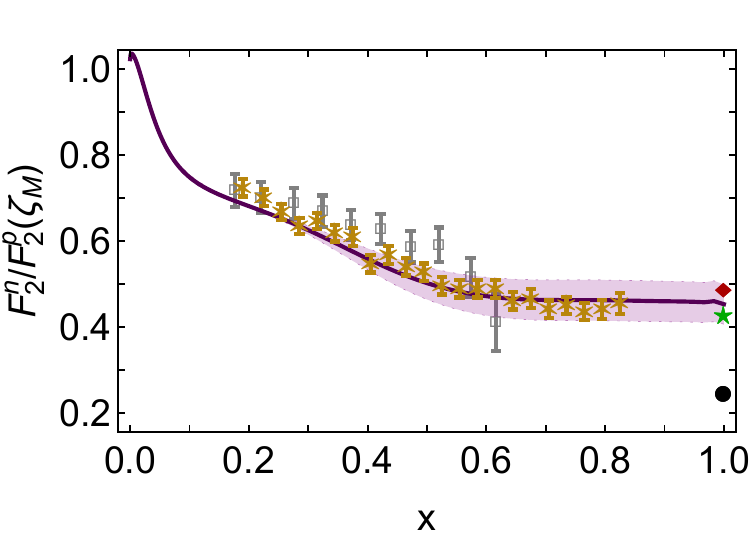}
\end{tabular}
\vspace*{2ex}
\caption{\label{ImageSeaQuest}
\emph{Left panel}\,--\,\mbox{\sf A}.
Ratio of light antiquark DFs.
Data: Ref.\,\cite[E906]{SeaQuest:2021zxb}.
CSM result (solid purple curve) obtained from valence-quark DFs in Fig.\,\ref{ImageValence} after evolution to $\zeta^2=\zeta_{\rm SQ}^2 = 30\,$GeV$^2$ \cite{Lu:2022cjx}.  The shaded band expresses the impact of a $\pm 25$\% variation in the strength of Pauli blocking.
\emph{Right panel}\,--\,\mbox{\sf B}.
Neutron-to-proton structure function ratio.
Data: \cite[BoNuS]{CLAS:2014jvt} -- open grey squares; and \cite[MARATHON]{Abrams:2021xum} -- gold asterisks.
Contemporary CSM result (solid purple curve) obtained from valence-quark DFs in Fig.\,\ref{ImageValence} after evolution to $\zeta=\zeta_{M}=2.7\,$GeV \cite{Chang:2022jri, Lu:2022cjx}.
Other predictions:
green star -- helicity conservation in the QCD parton model \cite{Farrar:1975yb, Brodsky:1979gy, Brodsky:1994kg};
red diamond -- large-$x$ estimate based on Faddeev equation solutions \cite{Roberts:2013mja};
and retaining only scalar diquarks in the proton wave function, which produces a large-$x$ value for this ratio that lies in the neighbourhood of the filled circle \cite{Close:1988br, Xu:2015kta}.
The band surrounding the CSM curve expresses the response to a $\pm 5$\% variation in the size of axialvector diquark contributions to the proton charge.  It is only noticeable on the valence quark domain.
}
\end{figure}

Proton and, in principle, neutron structure functions -- $F_2^{p,n}$ -- can be measured in the DIS of electrons from nucleons \cite{Taylor:1991ew, Kendall:1991np, Friedman:1991nq, Friedman:1991ip}.  The ratio $F_2^{n}(x)/F_2^{p}(x)$ is recognised as a sensitive measure of ${\mathpzc d}^p(x)/{\mathpzc u}^p(x)$ on $x\gtrsim 0.4$ \cite{Holt:2010vj} and the latter ratio is important because it is a keen discriminator between pictures of proton structure \cite{Roberts:2013mja, CLAS:2014jvt, Abrams:2021xum}.  The obstacle to an empirical result for $F_2^{n}(x)/F_2^{p}(x)$ is measurement of $F_2^{n}$: since isolated neutrons decay rather quickly, a suitable, effective ``free neutron target'' must be found.  Following Refs.\,\cite{Bodek:1973dy, Poucher:1973rg}, many experiments have used the deuteron.  However, despite this being a weakly bound system, the representation-dependence of proton-neutron interactions leads to large theory uncertainties in the extracted ratio on $x\gtrsim 0.7$ \cite{Whitlow:1991uw}.

A more favourable approach is provided by DIS measurements on $^3$H and $^3$He.  In this case, nuclear interaction effects cancel to a very large degree when extracting $F_2^n(x)/F_2^p(x)$ from the $^3$H:$^3$He ratio of scattering rates \cite{Afnan:2000uh, Pace:2001cm}.  Of course, $^3$H is highly radioactive; so, careful planning and implementation are required to deliver a safe target.  Recently, after years of development, all challenges were overcome and such an experiment was completed \cite{Abrams:2021xum}: the extracted data are drawn in Fig.\,\ref{ImageSeaQuest}B.  Importantly, within mutual uncertainties, the results from Ref.\,\cite{Abrams:2021xum} match those inferred from an analysis of nuclear DIS reactions, exploiting targets ranging from the deuteron to lead and accounting for the effects of short-range correlations in the nuclei \cite{Segarra:2019gbp}.  This speaks in support of the reliability of the analyses in both cases.

As described in Sec.\,\ref{SecBaryonWaveFunctions}, the Faddeev equation in Fig.\,\ref{FigFaddeev} makes firm statements about proton structure.  In particular, well-constrained studies predict that axialvector diquark correlations are responsible for approximately 40\% of the proton's charge -- see Ref.\,\cite[Fig.\,2]{Liu:2022ndb}; and, \emph{e.g}., this strength is confirmed in studies of nucleon axialvector and pseudoscalar currents -- see Sec.\,\ref{SecBaryonFormFactors}.  Consequently, this is the size of the axialvector diquark fraction in the nucleon Faddeev amplitudes used to calculate proton DFs in Refs.\,\cite{Chang:2022jri, Lu:2022cjx}.  Using the results therein, one may readily predict the neutron-proton structure function ratio:
\begin{align}
\label{F2nF2p}
\frac{F_2^n(x;\zeta)}{F_2^p(x;\zeta)} =
\frac{
{\mathpzc U}(x;\zeta) + 4 {\mathpzc D}(x;\zeta) + \Sigma(x;\zeta)}
{4{\mathpzc U}(x;\zeta) + {\mathpzc D}(x;\zeta) + \Sigma(x;\zeta)}\,,
\end{align}
where, in terms of quark and antiquark DFs,
${\mathpzc U}(x;\zeta) = {\mathpzc u}^p(x;\zeta)+\bar {\mathpzc u}^p(x;\zeta)$,
${\mathpzc D}(x;\zeta) = {\mathpzc d}^p(x;\zeta)+\bar {\mathpzc d}^p(x;\zeta)$,
$\Sigma(x;\zeta) = {\mathpzc s}^p(x;\zeta)+\bar {\mathpzc s}^p(x;\zeta)
  +{\mathpzc c}^p(x;\zeta)+\bar {\mathpzc c}^p(x;\zeta)$.
Supposing that valence quarks dominate on $x\simeq 1$, then the limiting cases ${\mathpzc d}^p(x)\equiv 0$ and ${\mathpzc u}^p(x)\equiv 0$ yield the Nachtmann bounds \cite{Nachtmann:1973mr}:
\begin{equation}
\label{F2nF2pVNachtmann}
1/4 \leq F_2^n(x)/F_2^p(x) \leq 4 \quad {\rm on}\; x\simeq 1.
\end{equation}

The $\zeta=2.7\,{\rm GeV}=:\zeta_M$ CSM prediction for $ F_2^n(x)/F_2^p(x)$ is drawn in Fig.\,\ref{ImageSeaQuest}B.  Its comparison with modern data \cite[MARATHON]{Abrams:2021xum} may be quantified by noting that the central curve yields $\chi^2/$degree-of-freedom$\;=1.3$.
It is worth stressing that the $x$-dependence of the CSM prediction in Fig.\,\ref{ImageSeaQuest}B was made without reference to any data.  Consequently, the agreement with the results published in Ref.\,\cite[MARATHON]{Abrams:2021xum} is meaningful and should serve to allay any concerns that the associated data analysis omitted some systematic effect deriving from nuclear structure modelling.

Such heightened confidence in the MARATHON data adds impact to the model-independent SPM analysis of that data described in Ref.\,\cite{Cui:2021gzg}; so, it is worth recapitulating some of the material therein.
The final results are highlighted by Fig.\,\ref{Fconclusion}, which compares the MARATHON-based SPM prediction for $\left. F_2^n/F_2^p\right|_{x\to 1}$ with: the nuclear DIS value \cite{Segarra:2019gbp}; theory predictions \cite{Segovia:2014aza, Xu:2015kta, Farrar:1975yb, Brodsky:1994kg}; and the phenomenological fit result in Ref.\,\cite{Accardi:2016qay}.
The figure also marks the Nachtmann lower bound, Eq.\,\eqref{F2nF2pVNachtmann}, which is saturated if valence $d$-quarks play no significant role at $x=1$; namely, when there are practically no valence $d$-quarks in the proton: $\left. {\mathpzc d}^p/{\mathpzc u}^p\right|_{x\to 1}=0$.  This outcome is characteristic of proton wave function models in which the valence $d$-quark is (almost) always paired with one of the valence $u$-quarks inside a scalar diquark \cite{Close:1988br, Anselmino:1992vg, Barabanov:2020jvn}.  Even allowing for the quark-exchange dynamics in Fig.\,\ref{FigFaddeev}, one still finds $\left. {\mathpzc d}^p/{\mathpzc u}^p\right|_{x\to 1} \approx 0$ if only scalar diquarks are retained \cite{Xu:2015kta}.

\begin{figure}[t]
\centering
\includegraphics[width=0.6\textwidth]{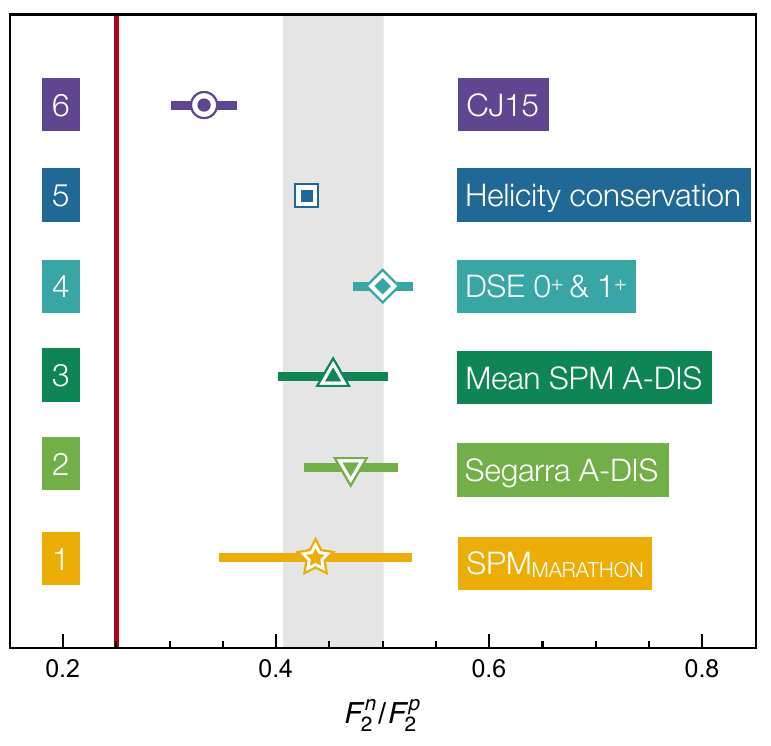}
\caption{\label{Fconclusion}
$\lim_{x\to 1}F_2^n(x)/F_2^p(x)$.
SPM prediction derived from Ref.\,\cite[MARATHON]{Abrams:2021xum} compared with results inferred from: nuclear DIS \cite{Segarra:2019gbp};
large-$x$ estimate from CSM studies \cite{Segovia:2014aza, Xu:2015kta};
quark counting (helicity conservation) \cite{Farrar:1975yb};
and a phenomenological fit (CJ15) \cite{Accardi:2016qay}.
The red vertical line indicates the Nachtmann lower-limit, Eq.\,\eqref{F2nF2pVNachtmann}, which is saturated if valence $d$-quarks play no material role at $x=1$;
and row~3 is the average in Eq.\,\eqref{mean}.
}
\end{figure}

The following observations serve as a summary of the analyses in Ref.\,\cite{Cui:2021gzg}.\\[-3.5ex]
\begin{description}
\item[Observation A] \ldots\
Applied to MARATHON data, the SPM yields $\left. F_2^n/F_2^p\right|_{x\to 1} =  0.437(85)$ $\Rightarrow$ $\left. {\mathpzc d}^p/{\mathpzc u}^p\right|_{x\to 1}=0.227(100)$.%
\footnote{
Extrapolations based on $[1,1]$ Pad\'e fits to MARATHON data, obtained using a one-point jackknife procedure, yield $F_2^n/F_2^p=0.395(3)$ on $x\simeq 1$ $\Rightarrow$ ${\mathpzc d}^p/{\mathpzc u}^d=0.169(3)$ \cite{Chen:2020ijn}.
Another analysis \cite{Pace:2022qoj}, employing practitioner-chosen polynomials as the basis for extrapolation, obtains $F_2^n/F_2^p=0.37(7)$ on $x\simeq 1$ $\Rightarrow$ ${\mathpzc d}^p/{\mathpzc u}^d=0.13(8)$.  The latter is less precise, but both results are consistent with the function form unbiased SPM prediction.}
The possibility $\left. {\mathpzc d}^p/{\mathpzc u}^p\right|_{x\to 1}=0$ is thus excluded with a 98.7\% level of confidence; hence, scalar-diquark-only models of proton structure are excluded with equal likelihood.
On the other hand, with this same 98.7\% level of confidence, the SPM analysis confirms the QCD parton model prediction \cite{Farrar:1975yb, Brodsky:1994kg}: ${\mathpzc d}^p(x) \propto {\mathpzc u}^p(x)$ on $x\simeq 1$.
\item[Observation B] \ldots\
The value of $\left. F_2^n/F_2^p\right|_{x\to 1}$ inferred from nuclear DIS \cite{Segarra:2019gbp} agrees with the SPM prediction; hence, they may be averaged to yield
\begin{equation}
\label{mean}
\left. F_2^n/F_2^p\right|_{x\to 1}^{\rm SPM\, \& \, DIS-A} = 0.454 \pm 0.047\,.
\end{equation}
This result is drawn at row 3 in Fig.\,\ref{Fconclusion}.  It corresponds to
\begin{equation}
\label{Final}
\lim_{x\to 1}\frac{{\mathpzc d}^p(x)}{{\mathpzc u}^p(x)} = 0.230 \pm 0.057
\end{equation}
and entails that the
probability that scalar-diquark-only models of proton structure might be consistent with available data is $1/141,000$.  In fact, with a high level of confidence, one may discard any proton structure model that delivers a result for $\left. F_2^n/F_2^p\right|_{x\to 1}$ that differs significantly from Eq.\,\eqref{mean}.  (As reviewed in Sec.\,\ref{SecBaryonFormFactors}, the ratio $g_A^d/g_A^u$ places an even harder exclusion bound on scalar-diquark-only models.)
\item[Observation C] \ldots\
Within uncertainties, the result in Eq.\,\eqref{mean} agrees with both:
(\emph{i}) the value obtained by assuming an SU$(4)$-symmetric spin-flavour wave function for the proton and helicity conservation in high-$Q^2$ interactions \cite{Farrar:1975yb, Brodsky:1994kg};
and (\emph{ii}) the prediction developed from proton Faddeev wave functions that contain both scalar and axialvector diquarks, with the axialvector contributing approximately 40\% of the proton charge \cite{Segovia:2014aza, Xu:2015kta}.  (Recall -- this was the axialvector diquark fraction built into the analyses in Refs.\,\cite{Chang:2022jri, Lu:2022cjx}.)
\end{description}

Following common practice, Ref.\,\cite[Table~2]{Lu:2022cjx} lists low-order $\zeta=\zeta_2, \zeta_3$ Mellin moments of all proton and pion DFs.  Given \mbox{\sf P1} and the character of the hadron scale, then comparable momentum fractions in the proton and pion are necessarily identical and, as anticipated above, the total sea-quark momentum fraction is shared between the quark flavours in roughly inverse proportion to their infrared dressed-mass, $M_f$.
Significantly, given that they are \emph{predictions}, the calculated values of the proton DF moments in Ref.\,\cite[Table~2]{Lu:2022cjx} are in fair agreement with those produced by phenomenological fits -- see \emph{e.g}., Ref.\,\cite[Table~VI]{Hou:2019efy}: referred to the CT18 column, the CSM results match at the level of $1.7(1.5)\,\sigma$.
Furthermore, within mutual uncertainties, the pion valence-quark DF moments agree with recent lQCD results \cite{Joo:2019bzr, Alexandrou:2021mmi}.

It is worth emphasising that the quantitative similarities also extend to the $c$ quark: Ref.\,\cite{Lu:2022cjx} predicts
$\langle x\rangle_{{\mathpzc c}_p}^{\zeta_2}=1.32(5)$\%, $\langle x\rangle_{{\mathpzc c}_p}^{\zeta_3}=1.82(6)$\%,
\emph{cf}.\ $1.7(4)$, $2.5(4)$\% in Ref.\,\cite[Fig.\,60]{NNPDF:2017mvq}.
Moreover, the CSM study predicts $\langle x\rangle_{{\mathpzc c}}^{\zeta=1.5\,{\rm GeV}}=0.64(3)$\% in both the pion and proton.  Regarding the pion, nothing is known about this momentum fraction; and in the proton, phenomenological estimates are inconclusive, ranging from $0$-$2$\% \cite[Fig.\,59]{NNPDF:2017mvq}.
%
Plainly, a significant $c$ quark momentum fraction is obtained under \mbox{\sf P1} evolution without recourse to ``intrinsic charm'' \cite{Brodsky:1980pb}.  This outcome, which is independent of the explicit form of $\alpha_{1\ell}(k^2)$, potentially challenges the findings in Ref.\,\cite{Ball:2022qks}.
Notwithstanding the size of these calculated fractions, we stress that, as apparent in Fig.\,\ref{ImageGlue}, ${\mathpzc S}^{\pi,p}_c(x)$ have sea-quark profiles.

Given these observations, one is led to reevaluate what is meant by intrinsic charm in the proton or any other hadron.  With $\zeta_{\cal H}$ being the scale at which valence quasiparticle degrees-of-freedom carry all measurable properties of a given hadron, then the Fock-space components which might be interpreted as intrinsic charm or intrinsic strangeness, etc., are sublimated into the nonperturbatively-computed $\zeta=\zeta_{\cal H}$ Schwinger functions that completely express the bound-state's structure.  In this context, such Fock space components are interpreted as being members of the set of basis eigenvectors representing a free-field light-front Hamiltonian.  One may exemplify this by noting that any true QCD solution for the dressed-quark Schwinger function (propagator) must contain infinitely many and all possible contributing Fock space vectors.  The putative ``intrinsic'' components within a bound-state are then revealed by evolving the hadron-scale Schwinger functions to higher scales, whereat an interpretation of data in terms of a Fock space expansion is relevant and practicable.  In one study or another, the actual expressions of the characters of intrinsic charm, strangeness, etc., will depend on the sophistication of the kernels used to calculate the hadron scale Schwinger functions.  Nevertheless, as highlighted by Ref.\,\cite{Lu:2022cjx}, any reasonable kernels will predict that a measurable fraction of the proton's light-front momentum is carried by the charm quark sea at all resolving scales for which data may be interpreted in terms of DFs.

With a symmetry-preserving framework in hand that has the demonstrated ability to provide simultaneous predictions for the entire array of proton and pion DFs, one is potentially in a position to bring a new order to the study of hadron structure functions.  Stringent new tests of the approach, including \mbox{\sf P1} -- see page~\pageref{P1item}, and the general character of the hadron scale, Eq.\,\eqref{EqzetaH}, will be found in, amongst other things, studies of helicity-dependent DFs.  New insights into proton spin structure may then be forthcoming.

\section{Conclusion}
We have sketched some recent advances in the use of continuum Schwinger function methods (CSMs) to link QCD and hadron observables.  The connecting bridge from theory to observation is supported by the three pillars of emergent hadrons mass (EHM):
(\emph{i}) dynamical generation of a gluon mass scale, whose size is roughly one half the proton mass;
(\emph{ii}) existence of a unique, process-independent effective charge, $\hat\alpha(k^2)$, which runs to a finite value at infrared momenta, $\hat\alpha(0)/\pi\approx 1$;
(\emph{iii}) and emergence of a running quark mass in the chiral limit, whose infrared value matches that typically identified as the constituent quark mass.
Our subsequent commentary stressed that the single phenomenon of EHM manifests itself differently in the diverse array of measurable quantities which define hadron physics.  No single observable is alone sufficient to validate the EHM paradigm for understanding strong interactions within the Standard Model (SM).  Instead, theory should identify a broad range of empirical consequences of EHM so that the order brought to a collective body of experimental results -- existing and future -- can be recognised as the signature of EHM.

Our developing understanding of EHM suggests that QCD is unique amongst known fundamental theories of natural phenomena.  It might be the first well-defined four-dimensional quantum field theory ever contemplated.  If so, then QCD could provide the archetype for theories that take physics beyond the SM.

Science has delivered theories of many things.  The best of them remain a part of the grander theories developed in response to new observations.  The basic question yet remains unanswered, \emph{viz}.\ is there a theory of everything?  Hadron physics and QCD might be pointing us toward an answer in exposing the special qualities of strong-coupling non-Abelian quantum gauge field theories.



\vspace{6pt}




\funding{Work supported by:
National Natural Science Foundation of China (grant no.\,12135007);
and
Helmholtz-Zentrum Dresden Rossendorf under the High Potential Programme.
}

\dataavailability{Not applicable.}

\acknowledgments{
This contribution is based on results obtained and insights developed through collaborations with many people, to all of whom we are greatly indebted.
%
%
}

\conflictsofinterest{The authors declare no conflict of interest.}



\abbreviations{Abbreviations}{
The following abbreviations are used in this manuscript:\\

\noindent
\begin{tabular}{@{}ll}
ACM & anomalous chromomagnetic moment \\
AdS/CFT (duality) & anti-de Sitter/conformal field theory (duality) \\
$\overline{\mbox{ard}}$ & mean absolute relative difference \\
CKM & Cabibbo-Kobayashi-Maskawa (matrix) \\
CSMs & continuum Schwinger function methods \\
DCSB & dynamical chiral symmetry breaking \\
DF & (parton) distribution function \\
DIS & deep inelastic scattering \\
DSE & Dyson-Schwinger equation \\
EHM & emergent hadron mass \\
FF & (parton) fragmentation function \\
JLab & Thomas Jefferson National Accelerator Facility
\end{tabular}

\noindent
\begin{tabular}{@{}ll}
lQCD & lattice-regularised quantum chromodynamics \\
NG (mode/boson) & Nambu-Goldstone (mode/boson) \\
PD (charge) & process-dependent (charge) \\
PDG & Particle Data Group and associated publications \\
PI (charge) & process-independent (charge) \\
QCD & quantum chromodynamics \\
QED & quantum electrodynamics \\
RGI & renormalisation group invariant \\
RL & rainbow-ladder (truncation) \\
SCI & symmetry-preserving treatment of a vector$\times$vector contact interaction \\
SM & Standard Model of particle physics \\
SPM & Schlessinger point method \\
VMD & vector meson dominance
\end{tabular}
}

\reftitle{References}
\begin{adjustwidth}{-\extralength}{0cm}



\end{adjustwidth}


\end{document}